\documentclass{jfm}
\usepackage{amsmath}
\usepackage{newtxtext}
\usepackage{newtxmath}
\usepackage{mathrsfs}
\usepackage{xcolor}
\usepackage{graphicx}
\usepackage{bm}
\usepackage{natbib}
\usepackage{epstopdf}
\usepackage{comment}
\usepackage{tikz}
\usetikzlibrary{decorations.pathreplacing}
\usepackage{hyperref}
\hypersetup{colorlinks=true,
            linkcolor=red,
            anchorcolor=blue,
            citecolor=blue}

\newcommand{\ri}{\mathop{\rm i}\nolimits}
\newcommand{\re}{\mathop{\rm e}\nolimits}

\title[Prediction of hypersonic bypass transition]
 {Prediction of bypass transition in hypersonic  blunt-plate boundary layers  subject to noisy conditions}

\author[Q. Song, M. Dong and L. Zhao]
 {Qinyang Song\aff{1},
  Ming Dong   \aff{2}\corresp{\email{dongming@imech.ac.cn}}\and \ 
  Lei Zhao    \aff{1,3}\corresp{\email{lei\_zhao@tju.edu.cn}}}

\affiliation
 {\aff{1}Department of Mechanics, Tianjin University, Tianjin 300072, PR China
  \aff{2}State Key Laboratory of Nonlinear Mechanics, Institute of Mechanics, Chinese Academy of Sciences, Beijing 100190, PR China
  \aff{3}National Key Laboratory of Vehicle Power System, Tianjin 300350, PR China}

\begin{document}
\maketitle
\begin{abstract}
In hypersonic boundary‐layer flows over blunt bodies,  laminar–turbulent transition exhibits  two distinct regimes: for small nose radii, increased bluntness delays transition; beyond a critical radius, further increasing bluntness reverses this trend. The latter regime corresponds to a bypass transition route, whose onset remains challenging to predict. The primary difficulty lies in capturing the excitation of non-modal streaks in the nose region, which is strongly affected by the bow shock and entropy layer effects. 
Recently, Zhao \& Dong (J. Fluid Mech., 2025, 1013: A44) develops a high-efficient, high‐accuracy shock‐fitting harmonic linearised Navier–Stokes (SF-HLNS) approach to quantify the excitation of linear non-modal perturbations. 
In this paper, we present a predictive framework for bypass transition by integrating the SF-HLNS approach with the  nonlinear parabolised stability equations (NPSE) and the bi-global stability analysis (BSA). The NPSE  is employed to   track the nonlinear evolution of the streaky perturbations up to  nonlinear saturation, while the BSA approach is used to capture the high-growth secondary instabilities. By integrating the growth rates of these secondary instability from their neutral positions,  an amplitude amplification factor is obtained, enabling the prediction of transition onset.
Under slow-acoustic forcing conditions drawn from wind-tunnel experiments, the present hybrid framework successfully reproduces the transition-reversal phenomenon at large nose radii, and yields quantitative agreement with measured transition locations, thereby validating its predictive capability.

\end{abstract}


\section{Introduction}\label{sec:intro}
Accurate prediction of transition location is critical for aerodynamic designs of  hypersonic vehicles, as the wall friction and heat flux undergo remarkable increase during the transition process. This task becomes even more challenging when the boundary layer features a blunt leading edge, a common design choice for hypersonic vehicles to prevent damage from the extreme heat loads at the stagnation point. {Wind-tunnel experiments \citep{stetson1967shock,stetson1983nosetip,lysenko1990influence,jewell2017boundary,borovoy2022laminar}} have revealed that the effect of nose bluntness on transition location is not monotonic. For configurations with relatively small bluntness, an increase in nose radius leads to a delay in transition; however, the opposite effect occurs when the nose radius surpasses a certain threshold. This phenomenon, known as the transition reversal phenomenon, is influenced by several factors, including the Mach number, Reynolds number, and the level of background noise.

To elucidate  the physical mechanism underling the transition reversal phenomenon, instability analyses of hypersonic boundary layers over blunt cones or wedges were conducted, revealing that increasing nose bluntness suppresses the growth rates of Mack modes, and the unstable Mack modes completely vanish once the nose radius exceeds a certain threshold \citep{malik1990effect,lei2012linear,Zhong2012direct,jewell2017boundary,jewell2018transition,wan2023effects}. This scenario features a natural transition route, where normal instability modes dominate the accumulation of perturbations during the laminar phase. An integration of the growth rate along the streamwise direction from the neutral position confirms the delay in transition onset as the nose radius is increased from a very sharp tip configuration, effectively explaining the experimental observations for configurations with small nose radii.

Additionally, an entropy layer forming from the nose region, characterised by a high entropy or density gradient, can support {an inviscid entropy-layer instability} \citep{reshotko1980stability,dietz1999entropy,Fedorov2004evolution}. It was found that the growth rate of the entropy-layer mode intensifies with increase of the nose radius. Although this mode emerges early around the nose region, its growth rate is typically low, making it unlikely to reach a finite-amplitude state unless the background noise is {exceptionally high} \citep{zhao2025excitation}.

In addition to natural transition, bypass transition offers an alternative route that does not rely on normal instability modes. In this process, freestream perturbations, such as acoustic, vortical, and entropy perturbations, enter the boundary layer and excite longitudinal streaky perturbations. Although these perturbations are composed of stable normal modes, their non-orthogonal nature allows them to exhibit transient growth due to the lift-up mechanism \citep{brandt2004transition}. These perturbations are also known as non-modal perturbations. While the growth of these non-modal perturbations is not successive, they can still accumulate to a nonlinear saturation state if (i) the level of freestream forcing or (ii) the transient amplification rate is sufficiently high. Notably, these non-modal perturbations are generally of low-frequency nature, which alone is insufficient to trigger transition directly through nonlinear cascade mechanisms. However, they can support secondary instabilities that exhibit high frequencies and high growth rates, allowing them to rapidly reach a nonlinear phase and eventually trigger the bypass transition.

Several theoretical frameworks have been established to predict bypass transition. The optimal growth theory emerged as a pioneering study to describe the transient growth of {non-modal perturbations} \citep{andersson1999optimal,paredes2018blunt,quintanilha2022transient}. In this theory, the evolution of non-modal perturbations is formulated as an optimization problem, employing an adjoint method to maximize energy amplification over a specified time period or spatial range. Subsequently, this optimization process can be approached as an input-output computation using {resolvent analysis} \citep{Towne2022}. While the optimal growth theory effectively identifies the streaky structure from the output perturbation profile, it does not take into account the receptivity process. This often results in an unphysical input perturbation profile and an overestimation of the maximum amplification rate \citep{zhao2025excitation}.

The boundary-region equation (BRE) appears as a more rational framework for considering the receptivity and evolution of non-modal perturbations. This approach employs an asymptotic matching technique to describe perturbations from the sharp leading edge to the downstream boundary layers. The linear and nonlinear evolution of non-modal perturbations in incompressible boundary layers over {thin flat plates were studied by} \citet{Leib1999} and \citet{ricco2011evolution}, respectively. More recently, this framework was extended to include the receptivity and nonlinear evolution of Görtler instabilities in compressible concave boundary layers \citep{xu2024excitation}. While the BRE provides an effective means for formulating the early stages of bypass transition, it faces limitations when applied to boundary layers with blunt noses. In particular, the presence of the detached bow shock and the entropy layer in hypersonic boundary layers poses challenges in matching perturbations from the stagnation point to the downstream boundary layers.

To address this limitation, \citet{zhao2025excitation} developed an innovative shock-fitting   harmonic linearised Navier-Stokes (SF-HLNS) approach. This method offers two key advantages: (i) it employs a Fourier transform with respect to time and the spanwise coordinate, reducing the four-dimensional system (three-dimensions in space and one-dimension in time) to a two-dimensional one; and (ii) it employs the shock-fitting technique to accurately describe the interaction between freestream perturbations and the bow shock. With this high-efficient and high-accuracy approach, they systematically studied the receptivity of non-modal perturbations. Their findings revealed that an increase in nose bluntness leads to an enhanced amplification rate of non-modal perturbations, qualitatively aligning with experimental observations for cases with relatively large bluntness. However, the SF-HLNS method alone cannot quantitatively predict the location of bypass transition, as it does not account for nonlinearity.

Downstream of the rapidly distorting leading‐edge region, perturbations in a smooth‐surface boundary layer exhibit slowly varying shape functions and wavelengths. Under these conditions, the  viscous terms involving second‐order streamwise derivatives become negligible, justifying the use of the parabolised stability equations (PSE) \citep{Herbert1997parabolized}. By retaining triadic interaction terms among Fourier modes, one arrives at the nonlinear PSE (NPSE) formulation \citep{zhao2016improved,song2023effect,song2024influence}. Consequently, in the study of bypass transition, when the linear non-modal perturbations are excited in the boundary layer, NPSE provides an efficient and accurate tool for capturing their subsequent nonlinear evolution downstream.

Once the non-modal streaks saturate at finite amplitude, small-scale perturbations emerge and undergo rapid growth via secondary instability (SI) mechanisms \citep{herbert1988secondary,andersson2001breakdown,ricco2011evolution}. These disturbances amplify into localized turbulent spots that convect downstream, merge, and ultimately lead to breakdown of the streaks, eventually causing the transition to turbulence \citep{Zhang2018,Wu2023new}. This secondary instability can be predicted through bi-global stability analysis (BSA) of the streaky base flow \citep{han2021secondary,songrj2023effect,song2024influence}.

While direct numerical simulations (DNSs) can reproduce the transition reversal phenomenon (see, for instance, \citet{guo2025transition}), there is a need for theoretical transition prediction approaches that offer higher efficiency and clearer understanding of the underlying physical mechanisms. 
Unlike natural transition, where the laminar phase is dominated by the amplification of linear normal modes and can be predicted by the conventional e-N method, bypass transition  still lacks a complete predictive framework. Two key challenges are (1) the amplitude accumulation of low-frequency non-modal perturbations, which is highly sensitive to freestream forcing; and (2) the requirement to know the nonlinear streaky base flow before estimating secondary-instability growth. Addressing these processes thus demands a multi-method approach. In this paper, we present a comprehensive framework for predicting bypass transition over blunt bodies with relatively high nose radii, based on the experimental configuration of \citet{borovoy2022laminar}. We will validate our predictions through direct comparison with their measured transition locations.

The rest of this paper is organized as follows.
In $\S$\ref{sec:method}, we describe the physical model and numerical methods used to predict bypass transition. $\S$\ref{sec:results} introduces the numerical results, including the base flow in $\S$\ref{subsec:baseflow}, the entropy-layer instability in $\S$\ref{subsec:modal-mode}, the estimate of the background noise level in $\S$\ref{subsec:estimate_noise}, the excitation of non-modal perturbations for varying controlling parameters in $\S$\ref{subsec:streak_excitation}, the nonlinear evolution of streaks in $\S$\ref{subsec:streak_nonlinear}, the secondary instability analysis in $\S$\ref{subsec:streak_sia}, and the comparison with the experimental data in $\S$\ref{subsec:comparison}. Finally, conclusions and discussions are offered in $\S$\ref{sec:Conclusion}.

\section{Description of the prediction framework for bypass transition and the physical model}\label{sec:method}
\subsection{Outlines of the bypass-transition-prediction method}\label{subsec:prediction_model}
As illustrated above, the bypass transition in blunt-nose boundary layers occurs through three critical steps: (i) the receptivity of non-modal perturbations resulting from external forcing by the oncoming stream, (ii) the transient growth and nonlinear evolution of streaky non-modal perturbations within the boundary layer, and (iii) the accumulation of secondary instabilities that triggers the breakdown of the streaky flow. To establish a rational transition-prediction method, we will quantify these three key processes using the following approaches:

(i) The SF-HLNS approach will be employed to describe the entrainment of freestream perturbations into the boundary layers through their interaction with the bow shock, followed by quantifying the receptivity of the boundary-layer perturbations; see a detailed introduction in $\S$\ref{subsubsec:SF-HLNS}.

(ii) The NPSE approach will be used to calculate the nonlinear evolution of boundary-layer perturbations initiated by small-amplitude disturbances excited during the receptivity process; further details are available in $\S$\ref{subsubsec:NPSE}.

(iii) The BSA approach will be employed to determine the growth rates of secondary instabilities once the primary perturbations reach a finite-amplitude state, enabling us to predict the onset of transition based on the accumulated amplification factor of these instabilities. Additional details can be found in \ref{subsubsec:SIA}.

\subsection{Physical model}\label{subsec:model}
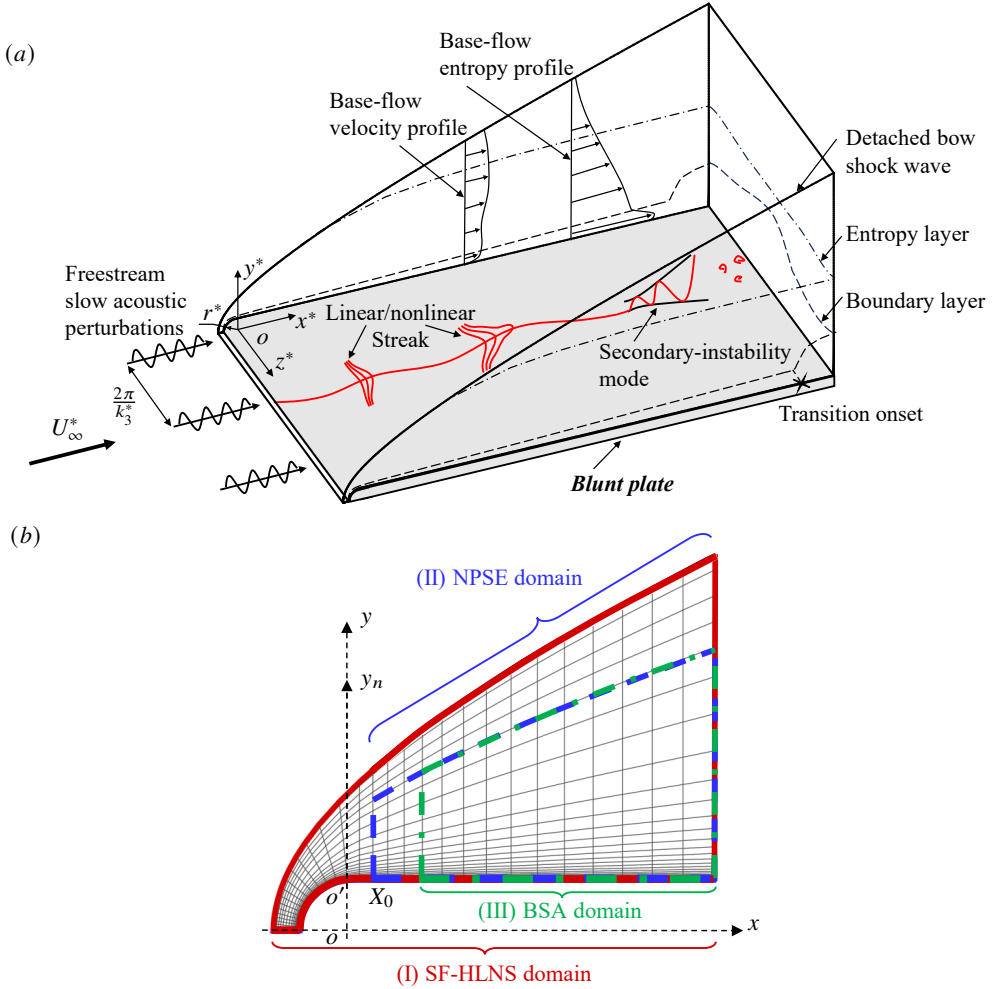
\begin{figure}
  \begin{center}
  \includegraphics[width = \textwidth]{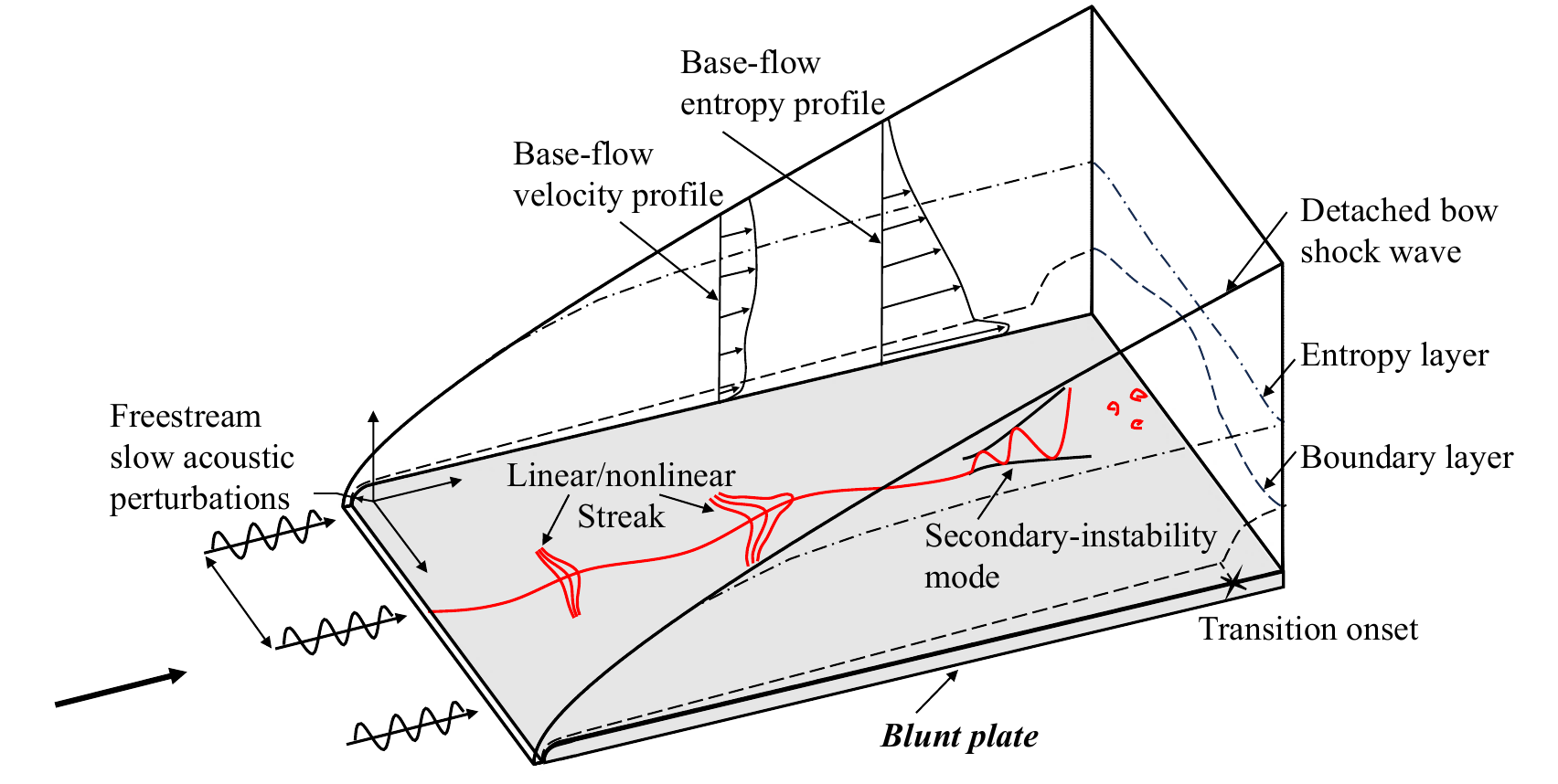}
  \put(-380,170){$(a)$}
  \put(-362,30){$U^*_{\infty}$}
  \put(-340,40){$\frac{2\pi}{k_3^*}$}
  \put(-305,73){$r^*$}
  \put(-285,64){$o$}
  \put(-270,70){$x^*$}
  \put(-290,90){$y^*$}
  \put(-278,55){$z^*$}\\
  \includegraphics[width = 0.49\textwidth]{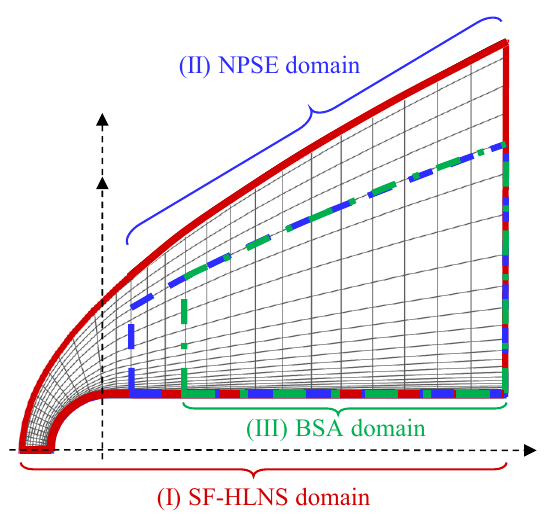}
  \put(-280,170){$(b)$}
  \put(-148,140){$y$}
  \put(-2,25){$x$}
  \put(-161,20){$o$}
  \put(-148,117){$y_n$}
  \put(-161,35){$o'$}
  \put(-145,35){${X_0}$}
  \caption{$(a)$ Sketch of  the physical model. $(b)$ Side view of computational domains for (I) SF-HLNS, (II) NPSE and (III) BSA. }
  \label{fig:model_sketch}
  \end{center}
\end{figure}

For demonstration, we choose a  hypersonic boundary-layer flow over a blunt plate as the physical model, as sketched in figure \ref{fig:model_sketch}$(a)$.
This model was also adopted in the experimental study in \citet{borovoy2022laminar}, whose experimental data will serve to validate our transition-prediction method. The mean flow field is characterised by three distinct layers: a shock layer detached from the body surface, an inviscid entropy layer exhibiting a significant rise in entropy, and a viscous boundary layer adjacent to the wall. Notably, in all the case studies we considered, the entropy layer remains consistently thicker than the boundary layer throughout the entire domain of interest. 
Following the experimental setup, we introduce a set of slow acoustic perturbations into the oncoming stream, which are assumed to be linearly independent across different Fourier components. Upon interacting with the detached shock wave, post-shock perturbations, including the acoustic, entropy, and vortical disturbances, are generated in the potential region, which in turn can excite perturbations within the boundary layer. In this study, we consider relatively large nose radii, thereby excluding the presence of Mack-mode instabilities in our domain of interest. Consequently, we will concentrate on the excitation and evolution of non-modal perturbations, which are typically characterised by low-frequency and three-dimensional nature. Due to the transient growth of these non-modal perturbations, even relatively weak freestream forcing can lead to the accumulation of finite-amplitude streaks in the boundary layers, which support secondary instabilities and ultimately result in a bypass transition to turbulence.

\subsection{Introduction of mathematical details}\label{subsec:mathe}
To describe the physical model, we select the nose radius $r^*$ as the reference length, and the physical quantities, including the density $\rho$, velocity field $\pmb{u} =(u, v, w)$, temperature $T$, pressure $p$, dynamic viscosity $\mu$ and time $t$, are normalised by the freestream quantities $\rho^*_{\infty}$, $U^*_{\infty}$, $T^*_{\infty}$, $\rho^*_{\infty}U^{*2}_{\infty}$, $\mu^*_{\infty}$ and $r^*/U^*_{\infty}$, respectively. In what follows, an asterisk represents a dimensional quantity.  Here, the velocity vector is defined in the Cartesian coordinate direction.
For convenience, we denote $\pmb\varphi=(\rho,\pmb{u},T,p)$ throughout this paper. The flow is governed by two dimensionless parameters, the Reynolds number and the Mach number, defined as 
\begin{equation}
  Re=\rho^*_{\infty}U^*_{\infty}r^*/\mu^*_{\infty}, \quad M=U^*_{\infty}/c^*_{\infty},
\end{equation}
where $c^*_{\infty}$ indicates the acoustic speed in the oncoming stream.

To facilitate a mathematical description of this system, we naturally introduce a dimensionless Cartesian coordinate system $(x,y,z)$, with its origin located at the center of the blunt nose shown in figure \ref{fig:model_sketch}(b). In the numerical process of SF-HLNS, the computational domain (indicated by the red region in figure  \ref{fig:model_sketch}(b)) is discretized into grid points. A corresponding computational coordinate system $(\xi, \eta, \zeta)$ aligned with these grid lines, is defined, and the dimensionless time is denoted as $\tau$. Detailed explanations concerning the numerical implementation are provided by \citet{zhao2025excitation}. 
To quantify the nonlinear evolution of the non-modal perturbations, we utilize {the NPSE approach} \citep{zhao2016improved}. For this purpose, we select a trapezoidal computational domain downstream of the nose region (the blue area marked in figure \ref{fig:model_sketch}(b)), with the inflow and upper-boundary perturbations obtained directly from the SF-HLNS calculations. The vertical grid lines are oriented perpendicular to the wall surface, and a body-fitted coordinate system $(x,y_n,z)$, defined  with $y_n=y-1$, is adopted.  As the non-modal streaky perturbations undergo transient growth, their amplitudes would reach the nonlinear phase, subsequently promoting the secondary instability with high growth rates. Thus, we select an analysis domain downstream, starting from the onset location of the secondary instability (the green region shown in figure \ref{fig:model_sketch}-(b)), and predict their amplitude accumulation using the BSA \citep{song2024influence}.

The governing equations are the three-dimensional (3-D) compressible Navier-Stokes (N-S) equations. We assume a perfect gas model with a constant specific heat ratio of $\gamma=1.4$ and a constant Prandtl number of $Pr=0.72$. Additionally, we employ Sutherland's law for viscosity.
The base flow ${\pmb\varphi}_B$ is two dimensional, and can be determined by solving the steady N-S equations using either shock-capturing or shock-fitting techniques, both of which yield comparably accurate numerical results, as confirmed by \citet{zhao2025excitation}.

\subsubsection{Slow acoustic wave in the oncoming stream}\label{subsubsec:gust}
In the conventional wind-tunnel experiments of \cite{borovoy2022laminar}, the dominant background disturbance is the slow acoustic wave radiated from the tunnel walls \citep{duan2019characterization}. Accordingly, we model the freestream perturbation as a series of harmonic acoustic waves. Each wave has a  frequency $\omega$, and the  streamwise, wall-normal and spanwise wavenumbers are given by the vector $\pmb k=(k_1, k_2, k_3)$. Accordingly, we can express it as
\begin{equation}\label{eq:gust}
 \frac{\varepsilon}{2}\hat{\pmb\varphi}_{\infty}\mathrm{exp}[\ri(k_1x+k_2y+k_3z-\omega t)]+~\mathrm{c.c.},
\end{equation}
where $\varepsilon$ denotes the  amplitude of the  static pressure perturbation in the oncoming stream, c.c. denotes the complex conjugate, and  $\hat{\pmb\varphi}_{\infty}=(\hat \rho_{\infty},\hat u_{\infty},\hat v_{\infty},\hat w_{\infty},\hat T_{\infty},\hat p_{\infty})$ denotes the relative magnitude of each quantity. For normalisation, we choose $\hat p_\infty=1$. Typically, the dispersion relation  of the  slow acoustic wave is $\omega=k_1- |\pmb{k}| / M $, and 
\begin{equation}
  \hat{\pmb\varphi}_{\infty}=\left( M^2, -\frac{k_1}{|\pmb k|}  M, -\frac{k_2}{|\pmb k|}  M, -\frac{k_3}{|\pmb k|} M,(\gamma-1)M^2,1 \right).
\end{equation}
The declination angle is defined by $\vartheta=\tan^{-1}(k_2/k_3)$.

\subsubsection{SF-HLNS approach}\label{subsubsec:SF-HLNS}
Given that the receptivity process is linear yet elliptic, we can formulate this process using a harmonic linearised framework. To simplify the challenges associated with calculating shock-perturbation interactions, we employ the shock-fitting technique. As a result, we employ the SF-HLNS approach, with the computational domain defined by the red lines in Figure \ref{fig:model_sketch}(b).

Forced by the acoustic perturbation (\ref{eq:gust}), the perturbation field $\pmb\varphi'$ and the shock-movement $H'$ are expressed in the harmonic form,
\begin{subequations}\label{eq:travelling}
  \begin{alignat}{1}
    \pmb\varphi'(\xi,\eta,\zeta,\tau)&=\frac{\varepsilon}{2} \hat{\pmb\varphi}(\xi,\eta)\mathrm{exp}[\ri(k_3\zeta-\omega \tau)]+~\mathrm{c.c.}, \\
    H'(\xi,\zeta,\tau)&=\frac{\varepsilon}{2} \hat{H}(\xi)\mathrm{exp}[\ri(k_3\zeta-\omega \tau)]+~\mathrm{c.c.}.
  \end{alignat}
\end{subequations}
Substituting  (\ref{eq:travelling}) into the linearised N-S equations, we derive the SF-HLNS equation system,
\begin{equation}\label{eq:HLNS}
    \mathcal L_1 \hat{\pmb \varphi}(\xi,\eta) +
    \mathcal L_2 \hat{H}(\xi)=0,
\end{equation}
where  $\mathcal L_1$ and $\mathcal L_2$ are the linear operators as outlined in \citet*{zhao2025excitation}.
The boundary conditions for system (\ref{eq:HLNS}) inlude a no-slip, isothermal condition at the wall,  and a  linearised Rankine–Hugoniot (R-H) relation at the shock, which is formulated in response to the   oncoming forcing defined in (\ref{eq:gust}). Additionally, the shock condition also incorporates the compatibility relation given by Eq. (2.39) in \citet*{zhao2025excitation}.
Detailed formulation and the corresponding algebraic  derivations can be found in  \citet*{zhao2025excitation}.

\subsubsection{NPSE approach}\label{subsubsec:NPSE}
In the boundary layer, non-modal perturbations experience transient growth due to the lift-up mechanism, potentially reaching a finite-amplitude state at some downstream location. Consequently, the linearised N-S equations no longer provide an adequate approximation for the evolution of these perturbations. Therefore, we need to employ the NPSE approach to address the nonlinear interactions among different Fourier components and the mean-flow distortion. As illustrated in figure \ref{fig:model_sketch}$(b)$, the NPSE calculations start  from a position downstream of the nose region (denoted by $X_{0}$), using the SF-HLNS solutions as the initial conditions. For the blunt-plate model, the body-fitted coordinate system $(x,y_n,z)$ employed  in the NPSE formulation is obtained by directly translating the original coordinate system $(x,y,z)$ via $y_n=y-1$. Therefore, the velocity vector  $\pmb{\hat u}$ remains consistent between the SF-HLNS and PSE frameworks.

In the NPSE approach, we use a marching scheme to compute the perturbation profile based on its upstream information, with the  numerical details outlined in our previous studies \citep{zhao2016improved,song2023effect,song2024influence}. The perturbation profile at  each streamwise position is expressed as 
\begin{equation}\label{eq:unknown_NPSE}
    \pmb\varphi'(x,y_n,z,t)=\sum_{m=-{\cal M},n=-{\cal N}}^{{\cal M},{\cal N}} \hat{\pmb\varphi}_{m,n}(x,y_n)\mathrm{exp}[\ri(nk_3 z-m\omega t)],
\end{equation}
where $m$ and $n$ denote the harmonic orders, {and $\cal M$ and $\cal N$ denote the truncation  orders.} Substituting (\ref{eq:unknown_NPSE}) into the N-S equations, and neglecting the $\partial_{x}^2$ terms, we derive the NPSE equations at each streamwise position,
\begin{equation}\label{eq:NPSE}
{\cal L}_3\hat{\pmb\varphi}_{m,n}=f_{m,n},
\end{equation}
where ${\cal L}_3$ is a parabolised operator, and $f_{m,n}$ denotes the nonlinear terms.  For convenience, we denote each Fourier component by $(m,n)$.

As will be demonstrated by the SF-HLNS calculations in Section \ref{subsec:streak_excitation}, the non-modal perturbations experience stronger amplification at lower frequencies, with stationary freestream acoustic perturbations exhibiting the highest receptivity efficiency. Consequently, in this paper, we set $\omega=0$ for the NPSE calculations, and set ${\cal M}=0$ in (\ref{eq:unknown_NPSE}) to remove the temporal harmonics. Therefore, the mean-flow distortion is denoted by (0,0),  the fundamental component is denoted by (0,1), and the harmonics are denoted by (0,2), (0,3), $\cdots$.
The truncation order of the spanwise-wavenumber harmonics is ${\cal N}=$5.

At the inlet boundary $x=X_0$ and upper boundary $y_n=Y_n$ of the computational domain, where the perturbation level is small, only the 
{(0,1)} component, obtained from the SF-HLNS calculations, is introduced. Again, the no-slip, isothermal boundary conditions are applied at the wall ($y_n=0$).

\subsubsection{BSA appraoch}\label{subsubsec:SIA}
Since low-frequency non-modal perturbations do not  breakdown themselves, the SI modes supported by the nonlinear streaky profile play a crucial role in triggering the bypass transition. To predict the transition onset, the accumulation of the SI amplitudes should be quantified. Thus, we perform the BSA based on the mean flow distorted by the strong non-modal streaks, as sketched in the green region in figure \ref{fig:model_sketch}$(b)$. The  streaky profile at a position $x$ is expressed as 
\begin{equation}\label{eq:SIA_baseflow}
  \bar{\pmb\varphi}(y_n,z;x)= {\pmb\varphi}_B(y_n;x) +\sum_{n=-{ \cal N}}^{ \cal N} \hat{\pmb \varphi}_{0,n}(y_n;x) \re^{\ri  n k_3 z},
\end{equation}
where $\hat{\pmb \varphi}_{0,n}$ is obtained from the NPSE calculation.

Since the length scale of the streaky base flow (\ref{eq:SIA_baseflow}) is much greater than that of the SI, we can introduce the parallel-flow assumption and express the SI mode as
\begin{equation}\label{eq:SIA_perturbation}
    \pmb\varphi'_{SI}(x,y_n,z,t)= \re^{\ri(\alpha {x}-\bar\omega t)} \re^{\ri \sigma_d k_3 z}\hat{\pmb\varphi}_{SI}(y_n,z) +~\mathrm{c.c.},
\end{equation}
where $\bar\omega$ and $\alpha$ denote the frequency and the streamwise wavenumber, respectively, and $\sigma_d\in[0,0.5]$ denotes the detuning parameter. Here, $\sigma_d=0$ denotes the fundamental SI modes, while $\sigma=0.5$ denotes the subharmonic SI modes. For a temporal mode, we specify $\alpha \in \mathbb{R}$, and calculate $\bar\omega=\bar\omega_r+\ri\bar\omega_i \in \mathbb{C}$, where $\bar\omega_i$ represents  the temporal growth rate. Conversely, for a spatial mode, we specify ${\bar\omega} \in \mathbb{R}$, and calculate $\alpha=\alpha_r+\ri\alpha_i \in \mathbb{C}$, where  $-\alpha_i$ represents  the spatial growth rate. It needs to be emphasized that the SI frequency $\bar\omega$ is much higher than that of the non-modal perturbation $\omega$.

Since the N-S equations involve only first-order time derivatives but second-order spatial derivatives, temporal mode calculations are more convenient. From the linearised N-S equations, we derive a generalized eigenvalue system,
\begin{equation}
    {\cal A} \hat{\pmb \varphi}_{SI}(y_n,z)=\bar\omega{\cal B} \hat{\pmb \varphi}_{SI}(y_n,z),
    \label{eq:SI}
\end{equation}
where ${\bar\omega}$ appears as the eigenvalue, and the coefficient matrices $\cal A$ and $\cal B$ are readily derived from the linear equations shown in \citet{song2024influence}. The Arnoldi iterative method is employed to search the eigenvalues initially, and the Rayleigh quotient iteration method is used to confirm its accuracy, as has been used in \citet{han2021secondary} and \citet{song2024influence}.

Noting that the spatial SI mode is more relevant for the transition prediction method, we convert the temporal SI solutions to a spatial mode using the following looping method:

(1) Provide an initial guess for $\alpha$.

(2) Solve for $\bar\omega(\alpha)$ from (\ref{eq:SI}).

(3) Calculate $d\bar\omega/d\alpha$ by introducing a small perturbation to $\alpha$, namely, $d\bar\omega/d\alpha=[\bar\omega(\alpha+\epsilon_\alpha)-\bar\omega(\alpha)]/\epsilon_\alpha$.

(4) Update $\alpha$ using the Newton iterative scheme and repeat step (2).
The iteration is terminated when the computed ${\bar\omega}$ is real and matches the specified value.

Once the spatial growth rate of SI at each position ${x}$, $-\alpha_i(x)$, is obtained, its amplitude amplification can be determined through spatial integration. For an SI mode with a frequency $\bar\omega$, we define an amplification factor  
\begin{equation}\label{eq:N_steady}
  N({x};\bar\omega)=\int_{x_n}^{{x}}- \alpha_i(\bar{x};\bar\omega) \mathrm{d}\bar x,
\end{equation}
where $x_n$ represents the neutral position of SI for frequency $\bar\omega$. The implication is that this SI is amplified $\mbox{e}^N$ times from the neutral position to ${x}$. According to the classical e-N method, the transition onset may be prescribed by the position where 
\begin{equation}
  \max_{\bar\omega}\{N({x},\bar \omega)\}=N_{\mathrm{tr}},  
\end{equation}
with $N_{\mathrm{tr}}$ being an empirical constant representing the transition threshold.

\section{Numerical results}\label{sec:results}
\subsection{Base flow}\label{subsec:baseflow}
\begin{table}
  \begin{center}
  \def~{\hphantom{0}}
  \begin{tabular}{ccccccc} \vspace{.2cm}
  Case  & 
  {$\begin{array}{c}  \mbox{Nose}\\\mbox{radius}\\r^*  \end{array}$} &
  {$\begin{array}{c}  \mbox{Reynolds}\\\mbox{number}\\Re  \end{array}$} &
  {$\begin{array}{c}  \mbox{Computational}\\\mbox{ domain outlet}\\x_L  \end{array}$} &
  {$\begin{array}{c}  \mbox{Transition }\\\mbox{onset}\\x_{\mbox{tr}}  \end{array}$} &
  {$\begin{array}{c}  \mbox{Transitional }\\\mbox{Reynolds number}\\Re_{\mbox{tr}} \end{array}$} \\ 
       A & 1.20 mm & 0.696$\times 10^5$ & 250 & 129 & 8.98$\times 10^6$\\
       B & 1.80 mm & 1.044$\times 10^5$ & 150 & 72 & 7.52$\times 10^6$\\ 
       C & 2.00 mm & 1.160$\times 10^5$ & 150 & 43 & 4.99$\times 10^6$\\
       D & 2.60 mm & 1.508$\times 10^5$ & 80  & 30 & 4.52$\times 10^6$\\
       E & 3.00 mm & 1.740$\times 10^5$ & 80  & 23 & 4.00$\times 10^6$\\ 
  \end{tabular}
  \caption{Parameters for case studies and experimental transition locations.}
  \label{tab:para_case}
  \end{center}
\end{table}

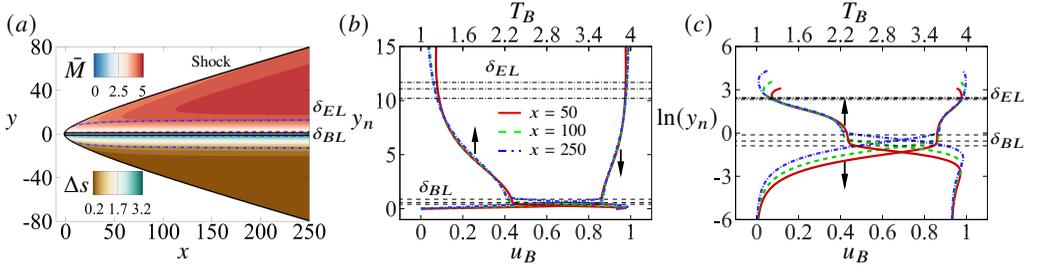
\begin{figure}
  \begin{center}
  \includegraphics[width = 0.33\textwidth]{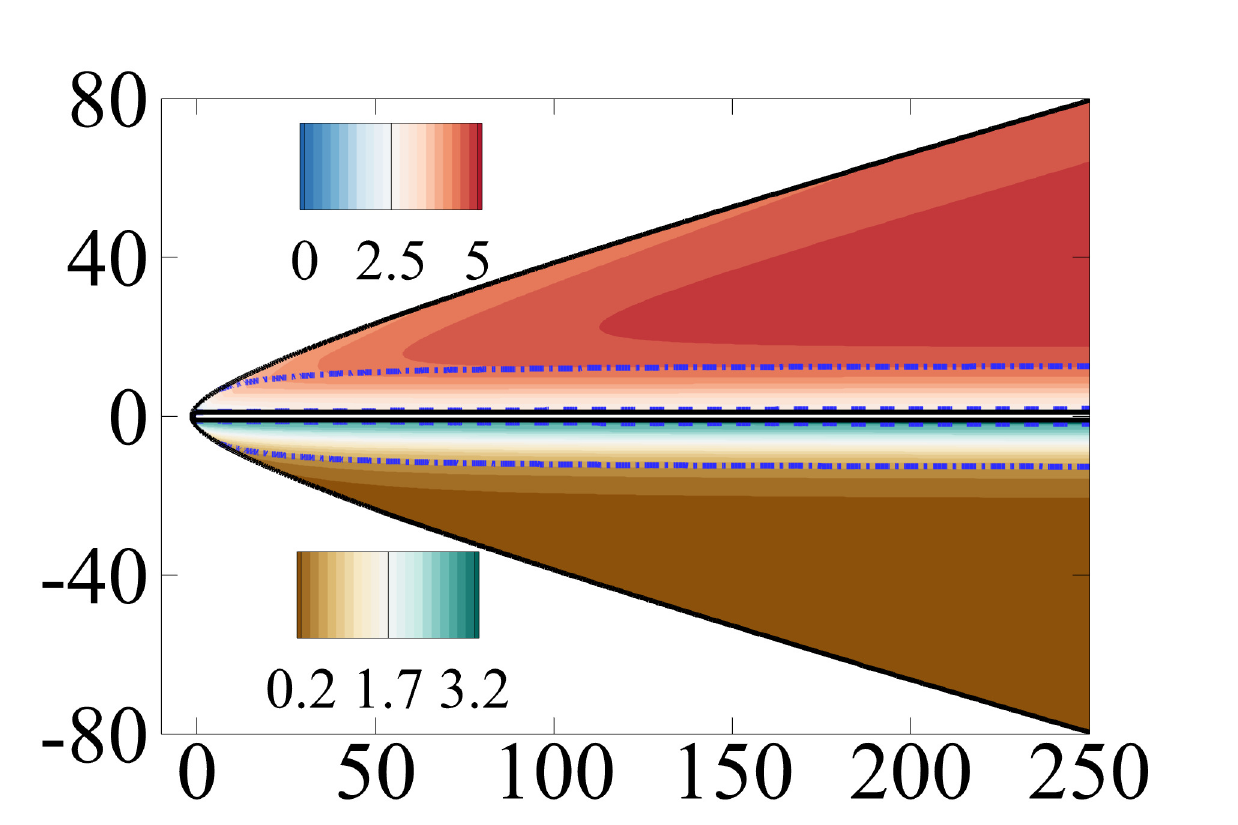}
  \includegraphics[width = 0.66\textwidth]{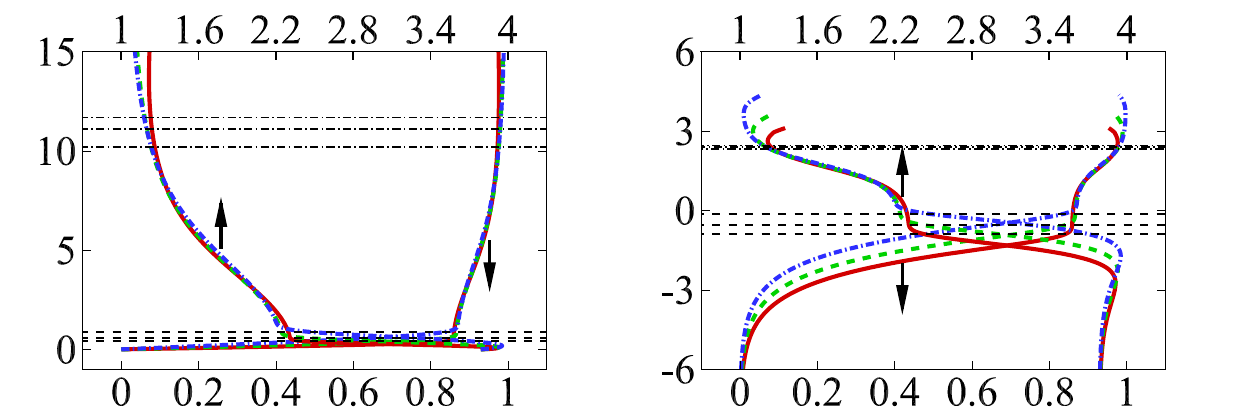}
  \put(-385,80){$(a)$}
  \put(-385,45){$y$}
  \put(-320,-5){$x$}
  \put(-363,63){$ \bar M$}
  \put(-363,20){$ \Delta s$}
  \put(-260,80){$(b)$}
  \put(-255,45){$y_n$}
  \put(-195,-5){$ u_B$}
  \put(-195,85){$ T_B$}
  \put(-130,80){$(c)$}
  \put(-140,45){ln$(y_n)$}
  \put(-70,-5){$ u_B$}
  \put(-70,85){$ T_B$}
  \put(-200,35) {\begin{tikzpicture}
    \draw[red,thick]   (0,0) -- (0.3,0);
    \draw[green,thick,dashed] (0,-0.25) -- (0.3,-0.25);
    \draw[blue,thick,dash dot]  (0,-0.5) -- (0.3,-0.5);
    \end{tikzpicture}}
  \put(-315,67) {\fontsize{6pt}{6pt}\selectfont  Shock}
  \put(-188,48) {\fontsize{6pt}{6pt}\selectfont $x=50$}
  \put(-188,41) {\fontsize{6pt}{6pt}\selectfont $x=100$}
  \put(-188,34) {\fontsize{6pt}{6pt}\selectfont $x=250$}
  \put(-270,40) {\fontsize{6pt}{6pt}\selectfont  $\delta_{BL}$}
  \put(-270,50) {\fontsize{6pt}{6pt}\selectfont  $\delta_{EL}$}
  \put(-230,20) {\fontsize{6pt}{6pt}\selectfont $\delta_{BL}$}
  \put(-205,65) {\fontsize{6pt}{6pt}\selectfont $\delta_{EL}$}
  \put(-15,38) {\fontsize{6pt}{6pt}\selectfont  $\delta_{BL}$}
  \put(-15,55) {\fontsize{6pt}{6pt}\selectfont  $\delta_{EL}$}
  \caption{($a$) Contours of the local Mach number $\bar M$ {and the entropy increment $\Delta s$} in the $x-y$ plane for case A (not to scale). ($b$) and ($c$) Wall-normal profiles of $u_B$ and $T_B$ in regular and logarithmic scales, respectively.  $\delta_{EL}$ and $\delta_{BL}$ mark the edges of the entropy and boundary layers, respectively.}
  \label{fig:baseflow_ma_r1.2}
  \end{center}
\end{figure}
Experimental data from \citet{borovoy2022laminar} exhibit two opposing trends in boundary‐layer transition: in one, increasing nose radii delays transition; in the other, it leads to earlier transition.  In the present study we focus on the latter behavior, examining five configurations listed in Table \ref{tab:para_case}. All cases employ an oncoming Mach number $M=5$, {unit Reynolds number $Re_{\infty}^*:=\rho_\infty^*U_{\infty}^*/\mu^*_\infty=5.80\times10^7$ /m, total temperature $T_{0}^*=464$ K and static temperature} $T_\infty^*=77.33$ K, with a wall temperature  $T_w^*=293$ K (corresponding to $T_w=T_w^*/T_{\infty}^*=3.79$).  The measured transition locations $x_{\mathrm{tr}}$ (second‐to‐last column) and corresponding transitional Reynolds numbers ${Re_{\mathrm{tr}}}:=\rho_\infty^*U_\infty^*{x_{\mathrm{tr}}} r^*/\mu_\infty^*$ (last column) confirm that increasing the nose radius shifts the transition position upstream, thereby reducing the streamwise extent $x_L$ needed for the base‐flow calculation.

To obtain the blunt-plate base flow, an in-house shock-fitting DNS code validated by \citet{zhao2025excitation} is employed. For each case, $551\times301$ grid points are employed, and a resolution study  for case A is presented in figure \ref{fig:gridcheck}(a) in Appendix \ref{app:A}. 
Figure \ref{fig:baseflow_ma_r1.2}-(a) presents contours of the local Mach number  $\bar M=M\sqrt{|\pmb{u}_B|^2/{{T}_B}}$ {and the normalised entropy increment $\Delta s=[\gamma/(\gamma-1)]\mathrm{ln} T_B-\mathrm{ln}(\gamma M^2 p_B)$} for case A.
In the nose region, the Mach number decreases significantly as the flow passes through the strong, nearly normal shock wave, corresponding to a notable entropy increase in this area. As the distance from the leading edge increases, however, the shock angle gradually decreases, causing a milder drop in Mach number in the downstream post-shock region. Meanwhile, the low-$\bar M$ region formed around the nose extends downstream, generating a distinct entropy layer characterised by a pronounced entropy rise beneath the high-$\bar M$ region. The outer edge of this entropy layer, marked as $\delta_{EL}$, is illustrated by the dot-dashed line in figure \ref{fig:baseflow_ma_r1.2}-(a). In fact, both the entropy layer itself  and the region  between its outer edge and the shock are inviscid. A viscous boundary layer emerges only in the immediate vicinity of the wall, with its outer boundary denoted by $\delta_{BL}$. Here, the means to determine the thicknesses of the entropy and boundary layers follow those in \cite{zhao2025excitation}.

Although the boundary layer is not clearly visible in the $\bar M$-contour plot, its presence is evident from the high gradients of $u_B$ and $T_B$ seen in their vertical profiles shown in figures \ref{fig:baseflow_ma_r1.2}-(b,c). Notably, the boundary-layer thickness expands as the flow moves downstream. By comparing these results with those from other cases (although not shown), it is observed that an increase in Reynolds number produces a thinner boundary layer. In contrast, the entropy layer remains largely unchanged due to its inviscid nature.

\subsection{Modal instability analysis}\label{subsec:modal-mode}
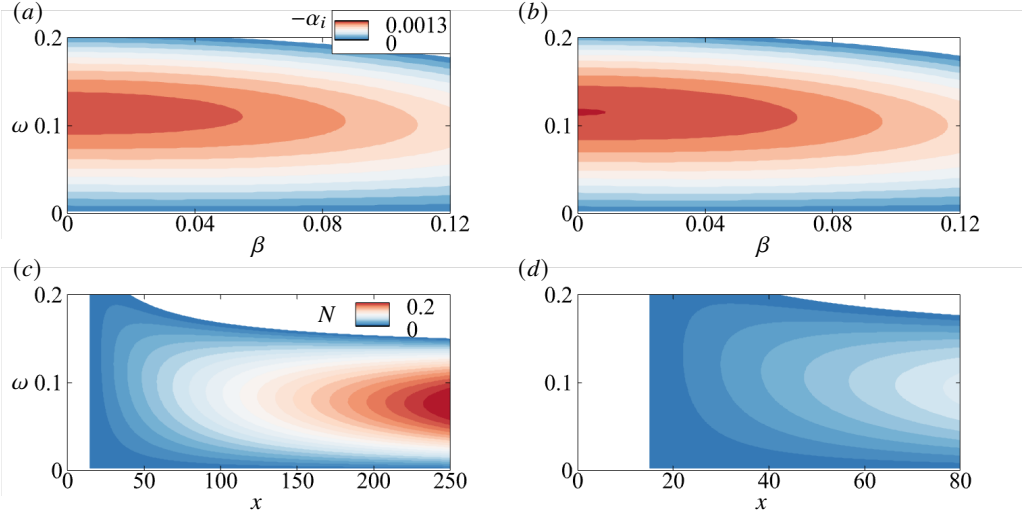
\begin{figure}
  \begin{center}
  \includegraphics[width = \textwidth]{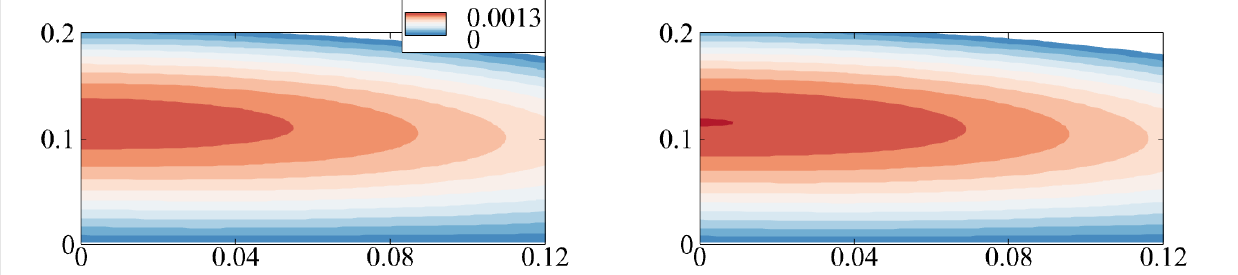}
  \put(-380,82) {$(a)$}
  \put(-190,82) {$(b)$}
  \put(-275,80) {$-\alpha_i$}
  \put(-380,40) {$\omega$}
  \put(-290,-5) {$\beta$}
  \put(-100,-5) {$\beta$}\\
  \includegraphics[width = \textwidth]{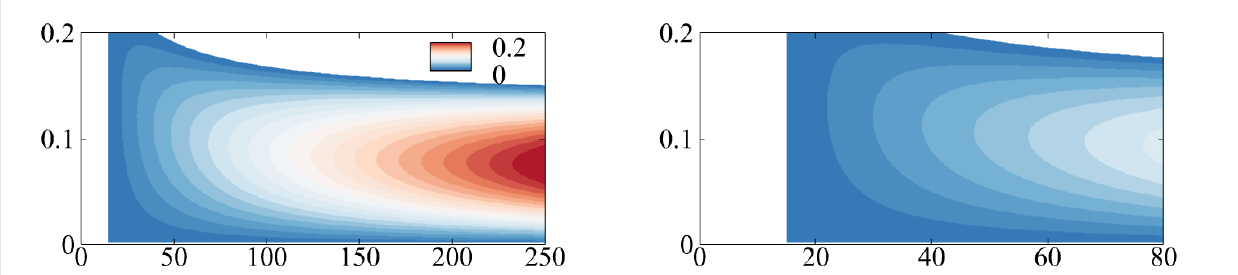}
  \put(-380,82) {$(c)$}
  \put(-190,82) {$(d)$}
  \put(-265,65) {$N$}
  \put(-380,40) {$\omega$}
  \put(-290,-5) {${x}$}
  \put(-100,-5) {${x}$}
  \caption{Properties of entropy-layer modes. $(a,b)$ Contours of $-\alpha_i$ in the $\beta$-$\omega$ plane at   ${x}=25$; $(c,d)$ contours of $N$-factor of the 2-D entropy-layer modes in the ${x}$-$\omega$ plane. Left column: case A; right column: case E.}
  \label{fig:entropymode}
  \end{center}
\end{figure}

For a given base-flow profile $u_B(y_n;x)$ and $T_B(y_n;x)$, stability analysis can be carried out by solving the compressible Orr–Sommerfeld equations, as detailed in \citet{Dong2020receptivity}. Here, we denote the frequency, streamwise wavenumber, and spanwise wavenumber by $\omega$, $\alpha$ and $\beta$, respectively, and the eigenfunctions by $\hat{\pmb\varphi}(y_n)$. For all cases considered in this study, no unstable Mack modes are detected within the computational domain. However, unstable entropy-layer modes are identified, as shown in figure \ref{fig:entropymode}. The entropy-layer mode is more unstable for two-dimensional configurations. While the entropy-layer mode exhibits exponential growth at relatively high frequencies, the accumulated $N$-factor remains small (at most approximately 0.2) up to the outlet of the computational domain. Although the growth rate increases with increasing $Re$,  the accumulated $N$-factor becomes smaller due to the reduced length of the computational domain at higher Reynolds numbers. Such  small amplification factors are insufficient to trigger transition. Therefore, we conclude that the experimentally observed transition is driven by non-modal perturbations via a bypass route.

\subsection{Estimate of background noise}\label{subsec:estimate_noise}
To enable a fair comparison with the experimental transition measurements, it is  necessary to first estimate the spectra of freestream perturbations upstream of the bow shock. However, the noise level and power spectra of the UT-1M wind tunnel employed in \cite{borovoy2022laminar} were not reported in the literature. Therefore, the perturbation spectra in this study are estimated by referring to the data from other conventional wind tunnels \citep{wagner2018combined,duan2019characterization,liu2022interaction}.

\begin{figure}
  \begin{center}
  \includegraphics[width = 0.8\textwidth]{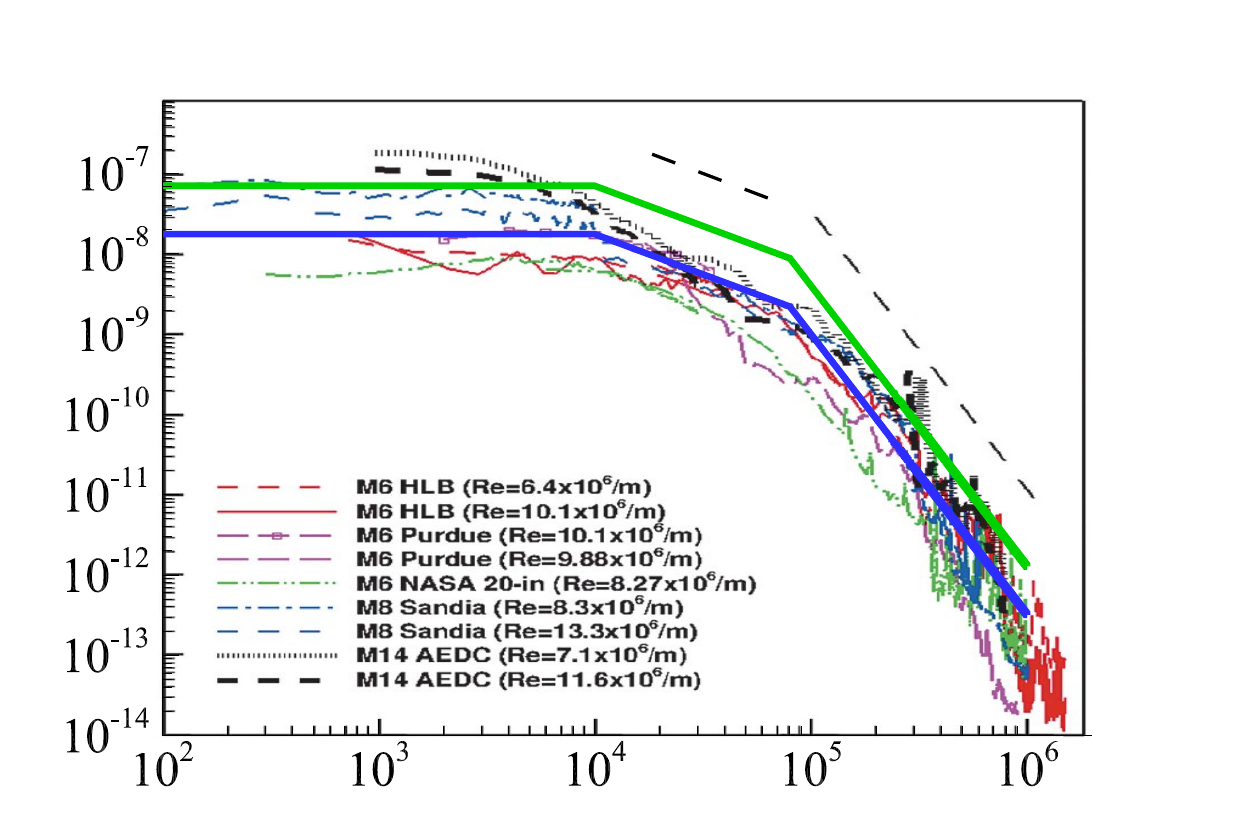}
  \put(-310,60) {{\rotatebox{90}{$\mathrm{PSD}\ [(p^{\ast\prime}_{0s}/p^*_{0s})^2/\mathrm{Hz}]$}}}
  \put(-170,0)  {$f^*$ (Hz)}
  \put(-253,100) {\begin{tikzpicture}
    \draw[green,line width=2pt]   (0,-0.35) -- (1,-0.35);
    \draw[blue,line width=2pt]   (0,-0.7) -- (1,-0.7);
    \draw[decorate, decoration={brace, amplitude=3pt}, thick](2.8,-0.2) -- (2.8,-0.7);
    \end{tikzpicture}}
  \put(-217,110) {$Q={3.5}\%$}
  \put(-217,100) {$Q={1.75}\%$}
  \put(-165,105) {Eq.(\ref{eq:PSD})}
  \put(-130,168) {{$f^{*-1}$}}
  \put(-78,125)  {{$f^{*-3.5}$}}
  \caption{Validation of  (\ref{eq:PSD}) by the wind-tunnel-noise measurements of PSD   \citep{duan2019characterization}, where {two} levels of $Q$ are selected. }
  \label{fig:line_PSD}
  \end{center}
\end{figure}
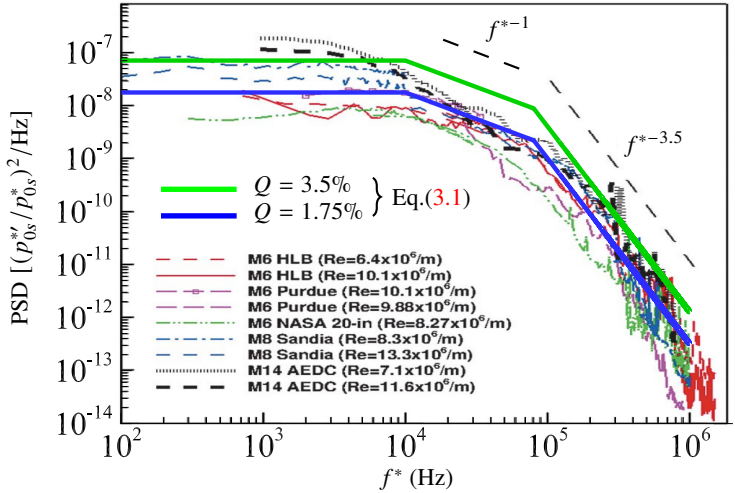
In wind tunnel experiments,  Pitot tubes are commonly used to measure the power spectral density (PSD) of the  total pressure perturbation signal, {$p^{\ast\prime}_{0s}$}, at the stagnation point. The PSD quantifies the frequency-dependent distribution of the total pressure perturbation power. 
We denote the normalised PSD of {$(p^{\ast\prime}_{0s}/p^*_{0s})^2$} by {$S^*(f^*)$}, where $f^{*}=\omega U^*_{\infty}/(2\pi r^*)$ represents the dimensional frequency. 

As shown in  figure 2 of \citet{duan2019characterization} (reproduced by the thin lines in figure \ref{fig:line_PSD}), the PSD measurements from multiple wind-tunnel runs show good overall agreement. The spectra are characterised by three distinct scaling regions:   a quasi-constant plateau at low frequencies, an intermediate region following an {$f^{*-1}$} decay, and a high-frequency region with a steeper  {$f^{*-3.5}$} decay. The transitions between these regions occur at approximately {10} kHz and {80} kHz. To model the unified spectral shape, we propose an empirical formula based on this observed behaviour,
\begin{equation}\label{eq:PSD}
{S^*}(f^*)= {A_S S^*_0}(f^*),\quad {S^*_0}(f^*)=\left\{\begin{aligned}
    &{a_0},         &f^* \le {10}\ \mathrm{kHz}, \\
    &a_1 f^{*-1},  &{10} < f^* \le {80}\ \mathrm{kHz}, \\
    &{a_2} f^{*-3.5},&f^* > {80}\ \mathrm{kHz},
  \end{aligned}\right.
\end{equation}
where $A$ characterizes the magnitude of the normalised pressure perturbation power, and {$S^*_0$} represents the unit distribution function, satisfying $\int_0^\infty {S^*_0}\mathrm{d}f^*=1$, with the parameters set ${(a_0,a_1,a_2)}=(0.0287\ \mathrm{kHz^{-1}},0.287,1.65\times10^{4}\ \mathrm{kHz^{2.5}})$. 
In conventional wind tunnels, the root-mean-square (RMS)  of the perturbation total pressure relative to the mean total pressure at the stagnation point,  {$p^{\ast\prime}_{0s,RMS}/p^*_{0s}$}, as measured by the Pitot tubes, is typically on the order of {$Q=1\% \sim 5\%$} \citep{casper2016hypersonic}. This corresponds to a PSD magnitude {$A_S=(\sqrt{2}Q)^2$}, where the factor $\sqrt{2}$ appears to convert the RMS value to the perturbation amplitude. The thick blue curve in  figure  \ref{fig:line_PSD} represents the evaluation of  (\ref{eq:PSD}) using {$Q=1.75\%$}, which agrees well with the experimental measurements. To further assess the  sensitivity of our transition prediction framework to background noise,   we also perform calculations with twice the nominal noise level ($Q=3.5\%$); the corresponding PSD is indicated  by the thick green line in figure \ref{fig:line_PSD}.

As reported in \cite{borovoy2022laminar},  the total  and static temperatures are {$T_{0}^*=464$ K} and {$T_{\infty}^*=77.3$ K}, respectively. For a Mach number $M=5$ and a unit Reynolds ${Re^*_{\infty}}=5.80\times10^7$ /m,  the corresponding sound speed, the freestream velocity and the density of the oncoming stream are calculated as $a_\infty^*={176\ \mathrm{m/s}}$, $U_\infty^*={881\ \mathrm{m/s}}$ and $\rho_\infty^*={0.348\ \mathrm{kg/m^3}}$, respectively. Using the equation of state, we obtain the static pressure of the oncoming stream $p_\infty^*={7.72\times10^3\ \mathrm{Pa}}$, which can then be converted to the mean total pressure 
\begin{equation}
p_{0\infty}^*=p_{\infty}^* \left[1+\frac{(\gamma-1)}{2}M^2\right]^{\gamma/(\gamma-1)}=4.08\times10^6\ \mathrm{Pa}.
\end{equation}
As the flow passes through the bow shock, its total pressure drops sharply and then remains nearly constant in the post-shock region all the way to the stagnation point. Therefore, we can estimate the mean total pressure at the stagnation point using the R-H relation of a normal shock,
\begin{equation}
    p_{0s}^*={p_{0\infty}^* \left[\frac{(\gamma+1)M^2}{(\gamma-1)M^2+2}\right]^{\frac{\gamma}{\gamma-1}} \cdot \left[\frac{(\gamma+1)}{2\gamma M^2-(\gamma-1)}\right]^{\frac{1}{\gamma-1}}=2.52\times10^5\ \mathrm{Pa}}. 
\end{equation}
Consequently, the RMS of the perturbation total pressure at the stagnation point, measured by the Pitot tube, is ${p^{\ast\prime}_{0s,RMS}\approx p^*_{0s}\times Q}$.

By integrating the power spectral density over the low- (0-{10} kHz), medium- ({10-80} kHz) and high- ($>${80} kHz) frequency bands, we can estimate the total pressure perturbation power for the three bands, ${S_{\mathrm{low}}}=\int_0^{{10}}S^* \mathrm{d}f^*={0.287A_S}$, ${S_{\mathrm{mid}}}=\int_{{10}}^{{80}}S^* \mathrm{d}f^*={0.598A_S}$ and ${S_{\mathrm{high}}}=\int_{{80}}^\infty{S^* \mathrm{d}f^*=0.115A_S}$, respectively. 
Thus, the contributions of these three bands to the power of the total‐pressure perturbation   are in the following proportions: 
\begin{equation}
{P_{\mathrm{low}}=\frac{S_{\mathrm{low}}}{A_S}=0.287},\quad 
{P_{\mathrm{mid}}=\frac{S_{\mathrm{mid}}}{A_S}=0.598},\quad
{P_{\mathrm{high}}=\frac{S_{\mathrm{high}}}{A_S}=0.115}.\label{eq:portion}
\end{equation}

In our theoretical model, the non-modal streaks are intrinsically low-frequency. Therefore, we restrict our attention to perturbations with $f^*\leq {10}$ kHz, for which the perturbation level is estimated to be (noting that the RMS is the square root of the power)
\begin{equation}
{\frac{p^{\ast\prime}_{0s,RMS,f^*_{\mathrm{low}}}}{p^*_{0s}}}={\frac{p^{\ast\prime}_{0s,RMS}}{p^*_{0s}}\times \sqrt{P_{\mathrm{low}}}}=\left\{\begin{array}{ll}
   {0.00938}  &\mbox{for }Q={1.75}\% \\
   {0.0188} & \mbox{for }Q={3.5}\%
\end{array}\right..
\end{equation}

Since our model also incorporates the spanwise spatial‐spectrum distribution, only a portion of the low‐frequency RMS pressure perturbation {$p^{\ast\prime}_{0s,RMS,f^*\leq 10\ \mathrm{kHz}}$} serves as the seed for non-modal excitation. Therefore, the power spectrum is further decomposed into low-, mid-, and high-wavenumber bands in each spatial direction. Given that the non-modal perturbations are characterised by short spanwise scales, the high-wavenumber components of the freestream perturbations are most relevant.  Lacking direct wavenumber‐spectrum measurements, we invoke Taylor’s frozen‐field hypothesis and adopt the band fractions given in (\ref{eq:portion}). Furthermore, the wavenumbers of slow acoustic waves radiated by wind-tunnel wall turbulence in different directions are interrelated through the dispersion relation. Based on these considerations, the effective seed perturbation level can be estimated  as  
\begin{equation}
{\frac{p^{\ast\prime}_{0s,RMS,f^*_{\mathrm{low}},k^*_{\mathrm{high}}}}{p^*_{0s}}}={\frac{p^{\ast\prime}_{0s,RMS,f^*_{\mathrm{low}}}}{p^*_{0s}}\times \sqrt{P_{\mathrm{high}}}}=\left\{\begin{array}{ll}
   {0.00318} &\mbox{for }Q={1.75}\% \\
   {0.00638} &\mbox{for }Q={3.5}\%
\end{array}\right..
\end{equation}

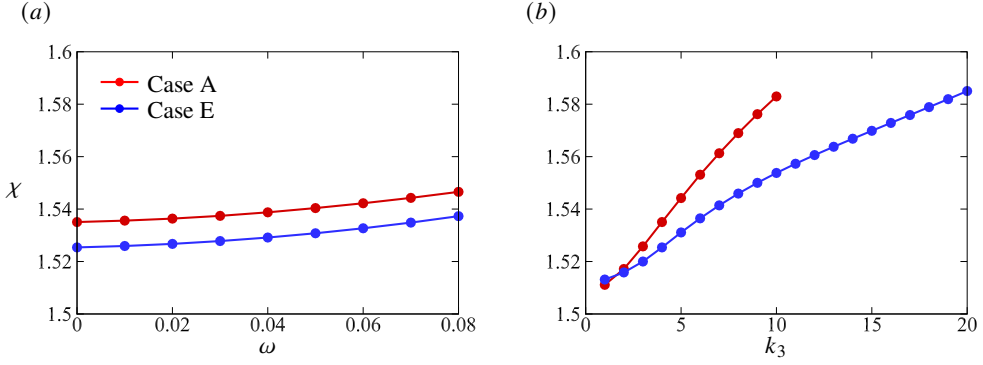
\begin{figure}
  \begin{center}
  \includegraphics[width = \textwidth]{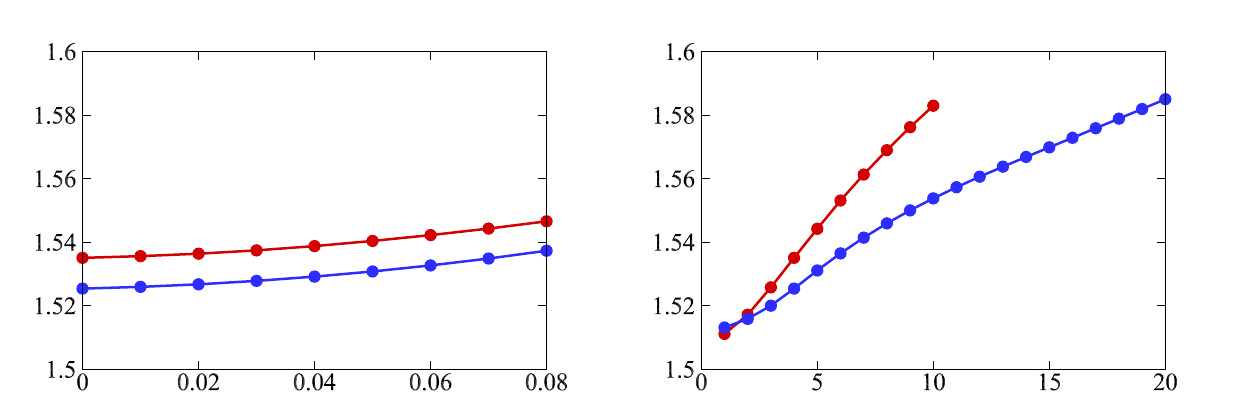}
  \put(-380,125){$(a)$}
  \put(-190,125){$(b)$}
  \put(-385,60) {{$\chi$}}
  \put(-290,0)  {$\omega$}
  \put(-100,0)  {$k_3$}
  \put(-350,90) {\begin{tikzpicture}
    \draw[red,thick]   (0,0) -- (0.5,0);
    \draw[blue,thick] (0,-0.35) -- (0.5,-0.35);
    \node[draw,red,circle,fill,inner sep=1pt] at (0.25,0) {};
    \node[draw,blue,circle,fill,inner sep=1pt] at (0.25,-0.35) {};
    \end{tikzpicture}}
  \put(-332,98){Case A}
  \put(-332,88){Case E}
  \caption{Dependence of {$\chi$} defined in Eq.(\ref{eq:sf-hlns_relation}) on ($a$) $\omega$ {(for $k_3=4$)} and ($b$) $k_3$ {(for $\omega=0$)}.}
  \label{fig:line_sfhlns_relation}
  \end{center}
\end{figure}
The next task is to convert the  total pressure perturbation {$p^{\ast\prime}_{0s}$} at the stagnation point to the static pressure perturbation {$p^{\ast\prime}_{\infty}$} in the oncoming stream. This can be implemented by the numerical results of SF-HLNS calculations. For simplicity, we introduce a relation characterizing the ratio of the relative perturbation level of the oncoming stream to that at the stagnation point,
\begin{equation}\label{eq:sf-hlns_relation}
  \chi = {\frac{p^{\ast\prime}_{\infty}/p^*_{\infty}}{p^{\ast\prime}_{\mathrm{0s}}/p^*_{\mathrm{0s}}}}.
\end{equation}
Choosing cases A and E, the cases with the lowest and highest Reynolds numbers of our study, we plot the dependence of $\chi$ on $\omega$ (for $k_3=4$) and $k_3$ (for $\omega=0$) in figures \ref{fig:line_sfhlns_relation}-(a) and (b), respectively. It is evident that  $\chi$ varies only slightly and remains approximately  1.55. Hence, the relative static‐pressure perturbation in the free stream can be estimated as
\begin{equation}
  {\frac{p^{\ast\prime}_{\infty,f^*_{\mathrm{low}},k^*_{\mathrm{high}}}}{p^*_{\infty}}}= \sqrt{2}\chi {\frac{p^{\ast\prime}_{0s,RMS,f^*_{\mathrm{low}},k^*_{\mathrm{high}}}}{p^*_{0s}}}= \left\{\begin{array}{ll}
  {0.00697}  &\mbox{for }Q={1.75}\% \\
  {0.0140}   &\mbox{for }Q={3.5}\%
\end{array}\right.,
\end{equation}
where the factor of $\sqrt{2}$ converts the RMS value to the peak value of the perturbation. In our study, the pressure is normalized by {$\rho^*_\infty U_\infty^{*2}$}, then we can estimate that
\begin{equation}
  \varepsilon={\frac{p^{\ast\prime}_{\infty,f^*_{\mathrm{low}},k^*_{\mathrm{high}}}}{\rho^*_\infty U_\infty^{*2}}}=\left\{\begin{array}{ll}
  {0.0002} &\mbox{for }Q={1.75}\% \\
  {0.0004} &\mbox{for }Q={3.5}\%
\end{array}\right..
\end{equation}

It should be noted that the estimate of $\varepsilon$ in this subsection is rather preliminary and subject to several uncertainties.  First, the assumed noise level of {$Q=1.75\%$ and $3.5\%$ are typical examples} of many conventional wind tunnels and may differ from that in the facility used  by \cite{borovoy2022laminar}. Second, the PSD curve fitting procedure itself introduces additional inaccuracy. Third, the receptivity coefficient depends on both frequency and wavenumber, and representing the broadband low-frequency, high-wavenumber spectrum with a single Fourier mode may lead to quantitative discrepancies. Despite these limitations, the following analyses will present bypass transition predictions for the two representative noise levels, demonstrating that the proposed prediction framework remains robust with respect to the specific choice of freestream disturbance amplitude. We emphasize, however, that direct measurement of the facility noise spectrum is essential for the reliable interpretation of wind-tunnel transition experiments.

\subsection{SF-HLNS calculations}\label{subsec:streak_excitation}

\begin{figure}
  \begin{center}
   \includegraphics[width = 0.49\textwidth]{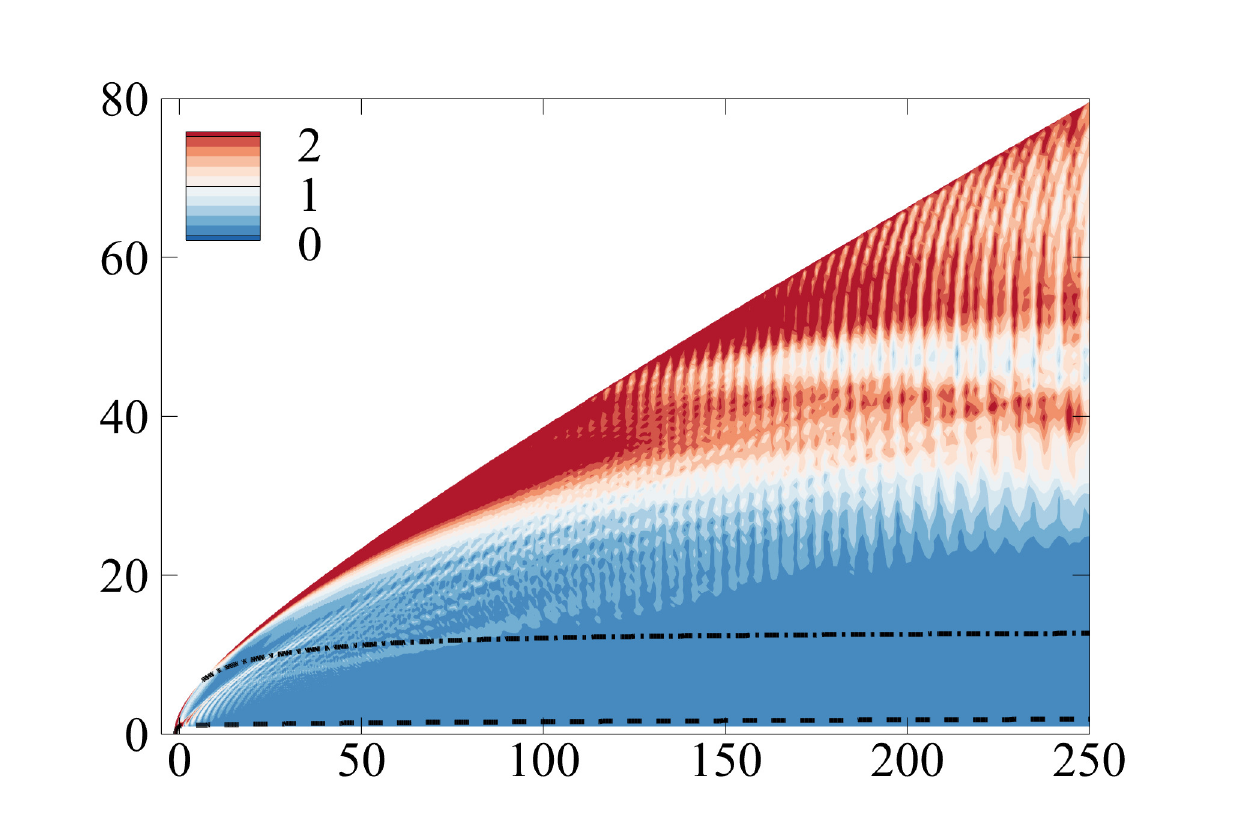}
  \includegraphics[width = 0.49\textwidth]{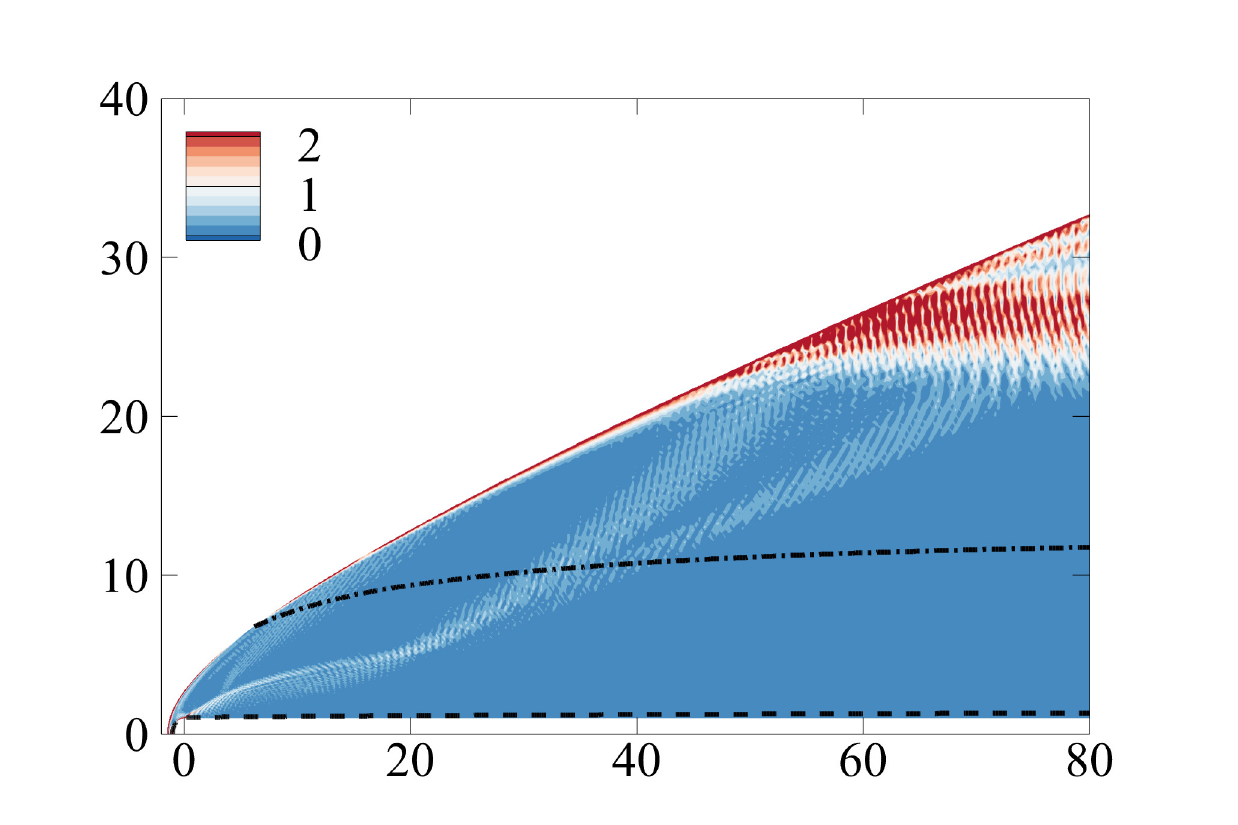}
  \put(-380,115){$(a)$}
  \put(-190,115){$(b)$}
  \put(-380,60){$y$}
  \put(-315,95){$|\hat p|$}
  \put(-125,95){$|\hat p|$}\\
  \includegraphics[width = 0.49\textwidth]{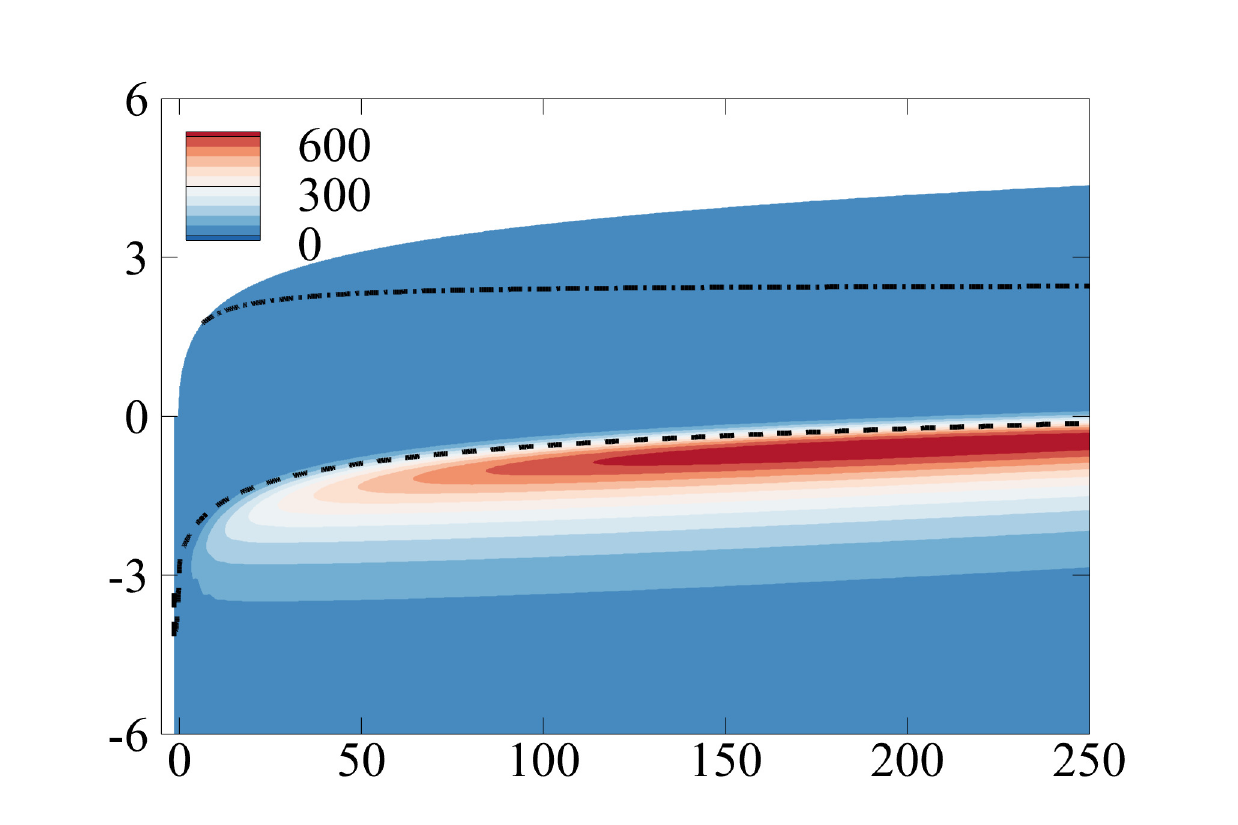}
  \includegraphics[width = 0.49\textwidth]{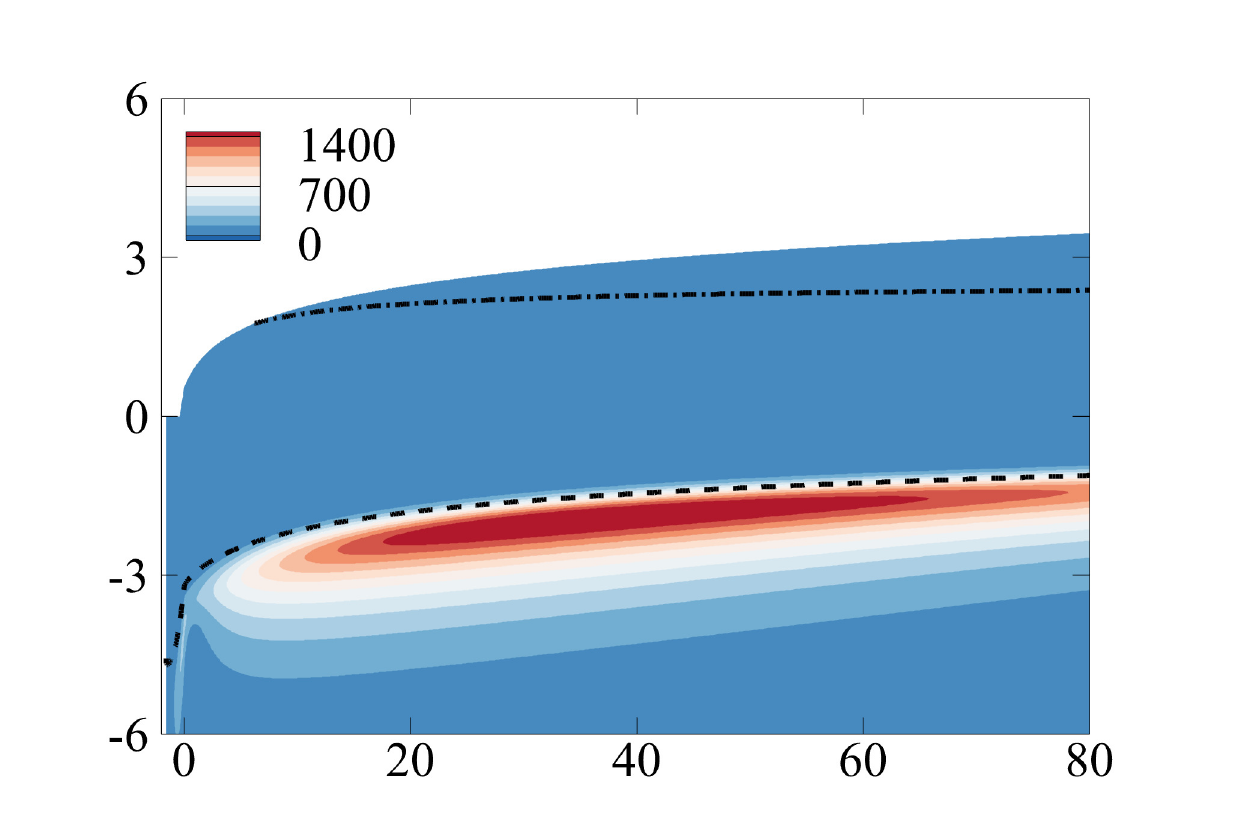}
  \put(-380,115){$(c)$}
  \put(-190,115){$(d)$}
  \put(-380,70){{$\ln( y_n)$}}
  \put(-315,95){$|\hat u|$}
  \put(-125,95){$|\hat u|$}\\
  \includegraphics[width = 0.49\textwidth]{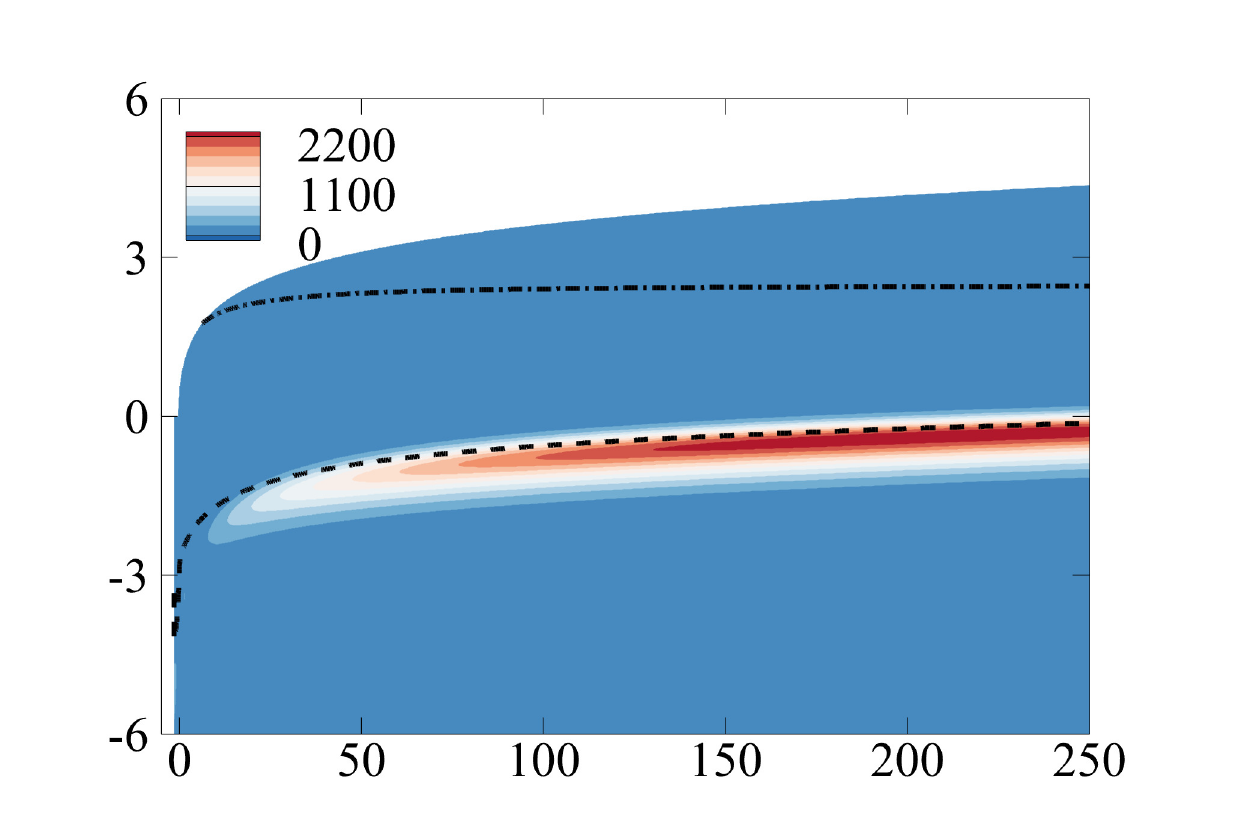}
  \includegraphics[width = 0.49\textwidth]{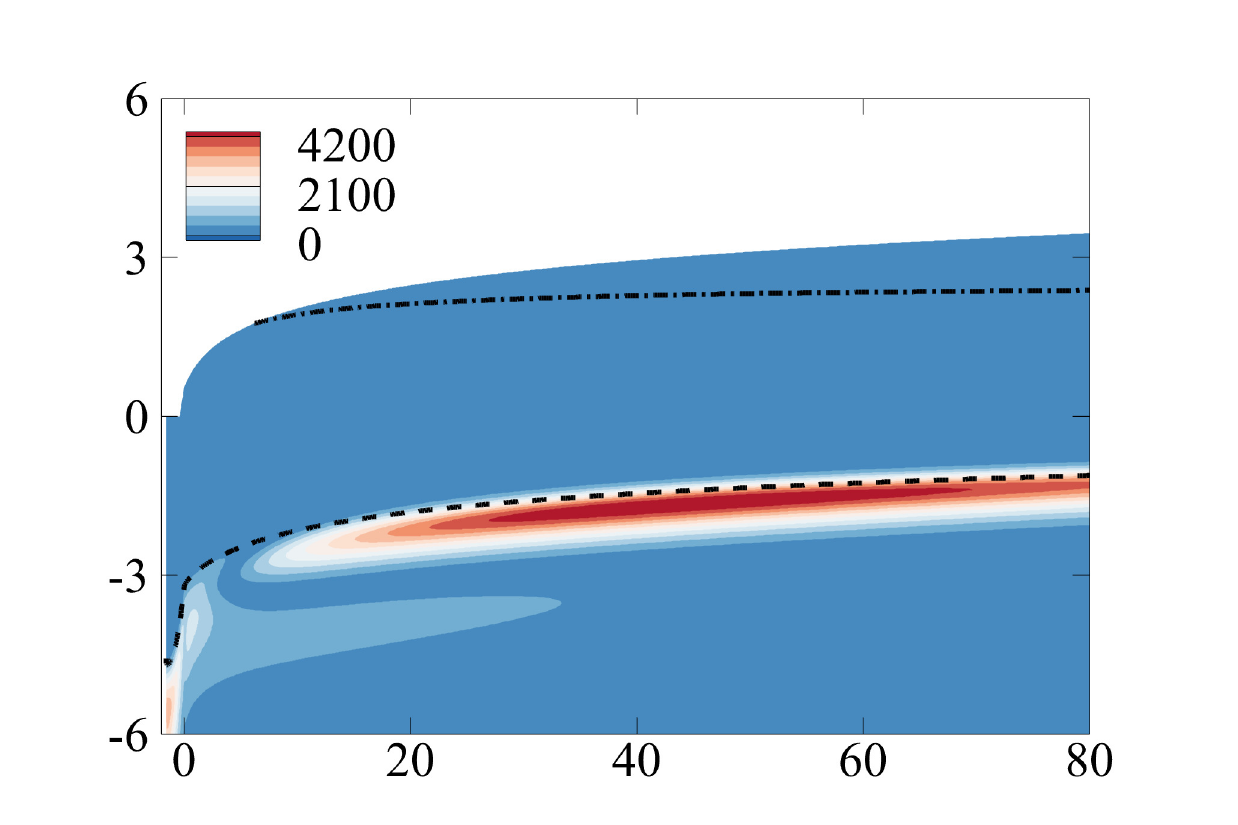}
  \put(-380,115){$(e)$}
  \put(-190,115){$(f)$}
  \put(-380,70){{$\ln( y_n)$}}
  \put(-285,0){${x}$}
  \put(-95,0){${x}$}
  \put(-315,95){$|\hat T|$}
  \put(-125,95){$|\hat T|$}
  \caption{Perturbation field obtained by SF-HLNS calculations for $\omega=0$ and $\vartheta=0^{\circ}$. Left column: case A with $k_3=4$; right column: case E with $k_3=16$. Top, middle and bottom rows: contours of  $\hat p$, $\hat u$ and $\hat T$. }
  \label{fig:linear_cont_pusT_r1.2}
  \end{center}
\end{figure}
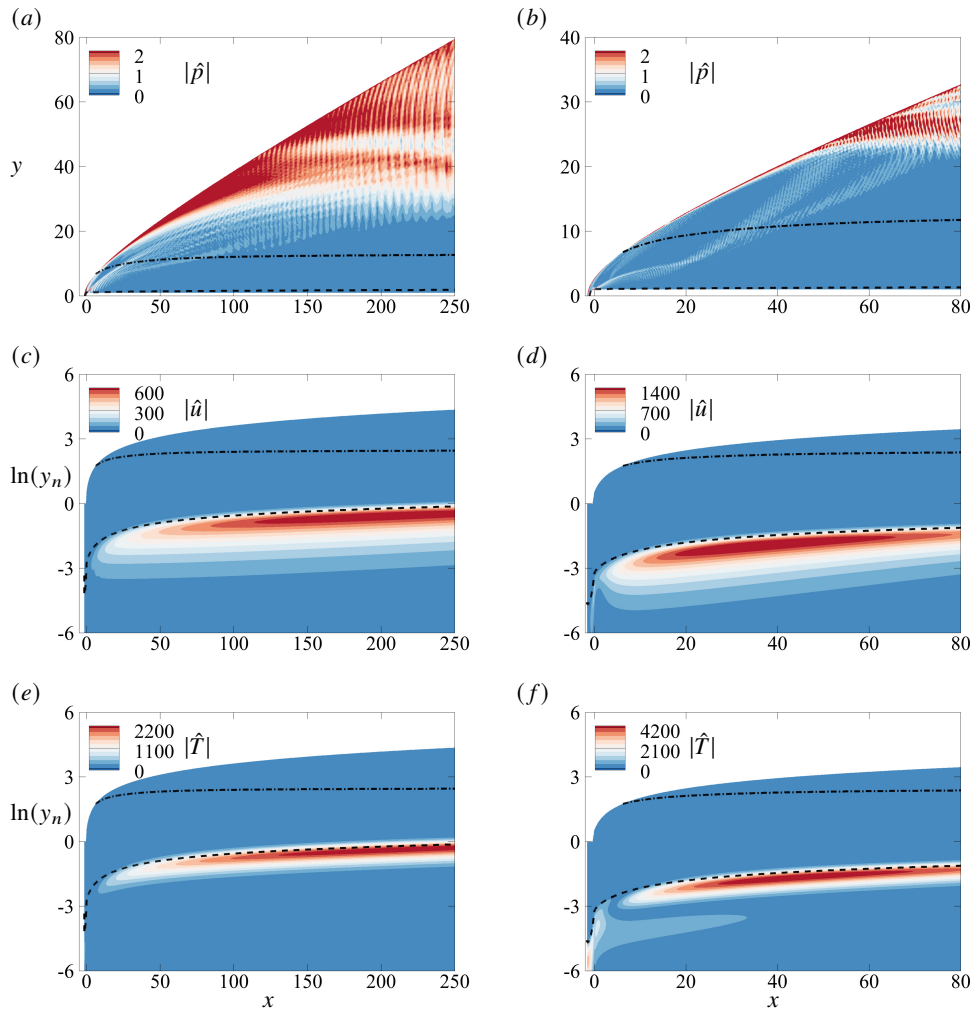
Using the SF-HLNS approach, we calculate the perturbation field within the region between the bow shock and the wall. As a linear approach, the resulting receptivity coefficients are independent of the background noise level as estimated in the last subsection.  However, they will determine the initial conditions for the NPSE computations to be  discussed in the following subsection.  Figures \ref{fig:linear_cont_pusT_r1.2}(a) and (b) present the pressure contours in the $x$-$y$
 plane for cases A and E, respectively, illustrating the dominant acoustic beam propagating through the post-shock potential region located above the entropy layer. To clearly visualize perturbations within the boundary-layer region, we plot contours of $|\hat u|$ and $|\hat T|$ in the ${x}$-$\ln(y_n)$ plane, as shown in panels (c) to (f). It is evident that both  $\hat u$ and $\hat T$ achieve their maxima near the edge of the boundary layer, with the magnitude of $\hat T$ being greater. Moreover, an increase in $Re$ strengthens the downstream amplitude amplification of these perturbations.

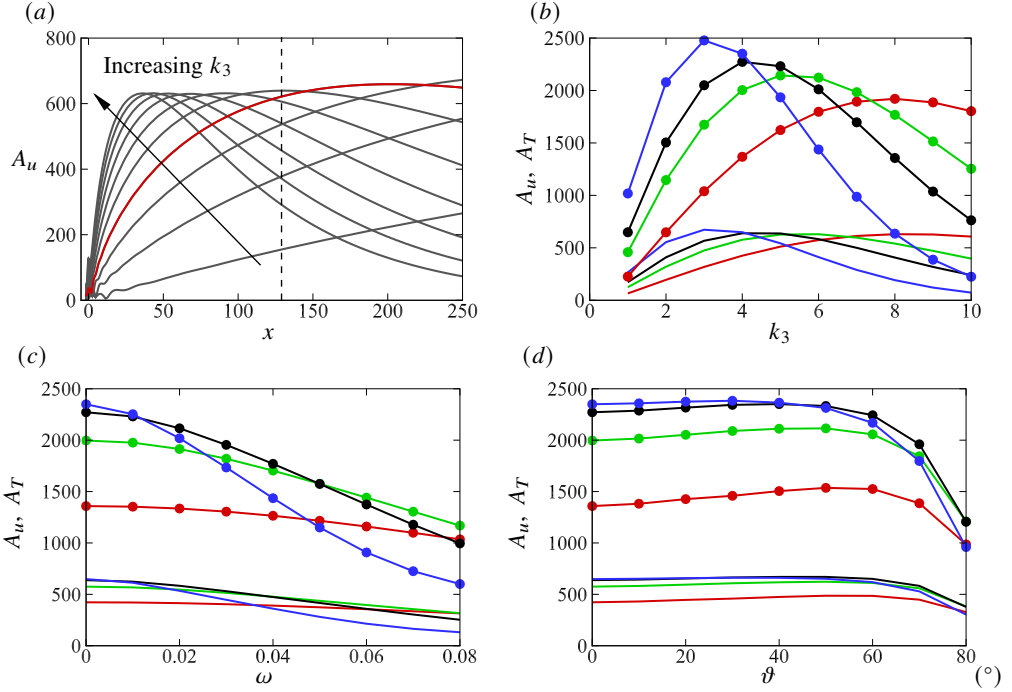
\begin{figure}
  \begin{center}
  \includegraphics[width = \textwidth]{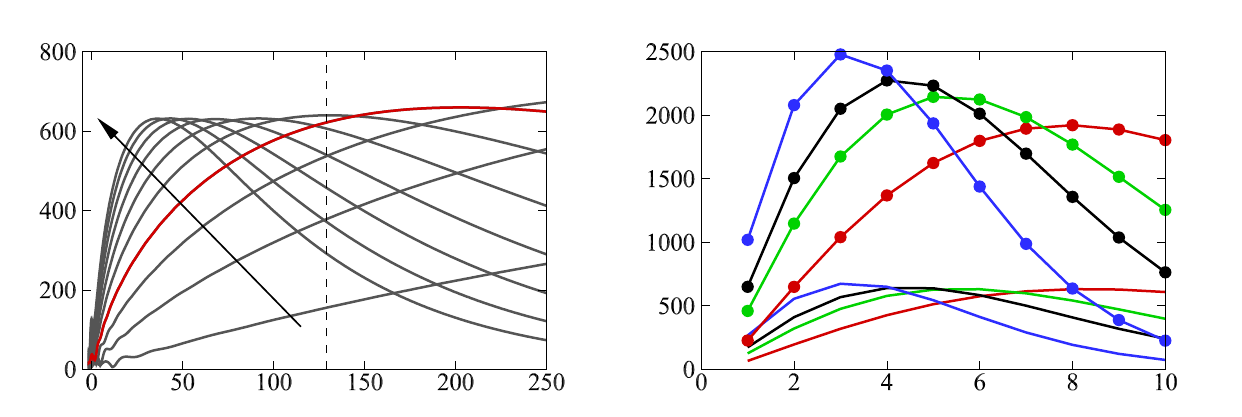}
  \put(-380,120){$(a)$}
  \put(-190,120){$(b)$}
  \put(-350,100){Increasing $k_3$}
  \put(-385,65){$A_u$}
  \put(-195,50){\rotatebox{90}{$A_u,~A_T$}}
  \put(-290,0){${x}$}
  \put(-100,0){$k_3$}\\
  \includegraphics[width = 0.49\textwidth]{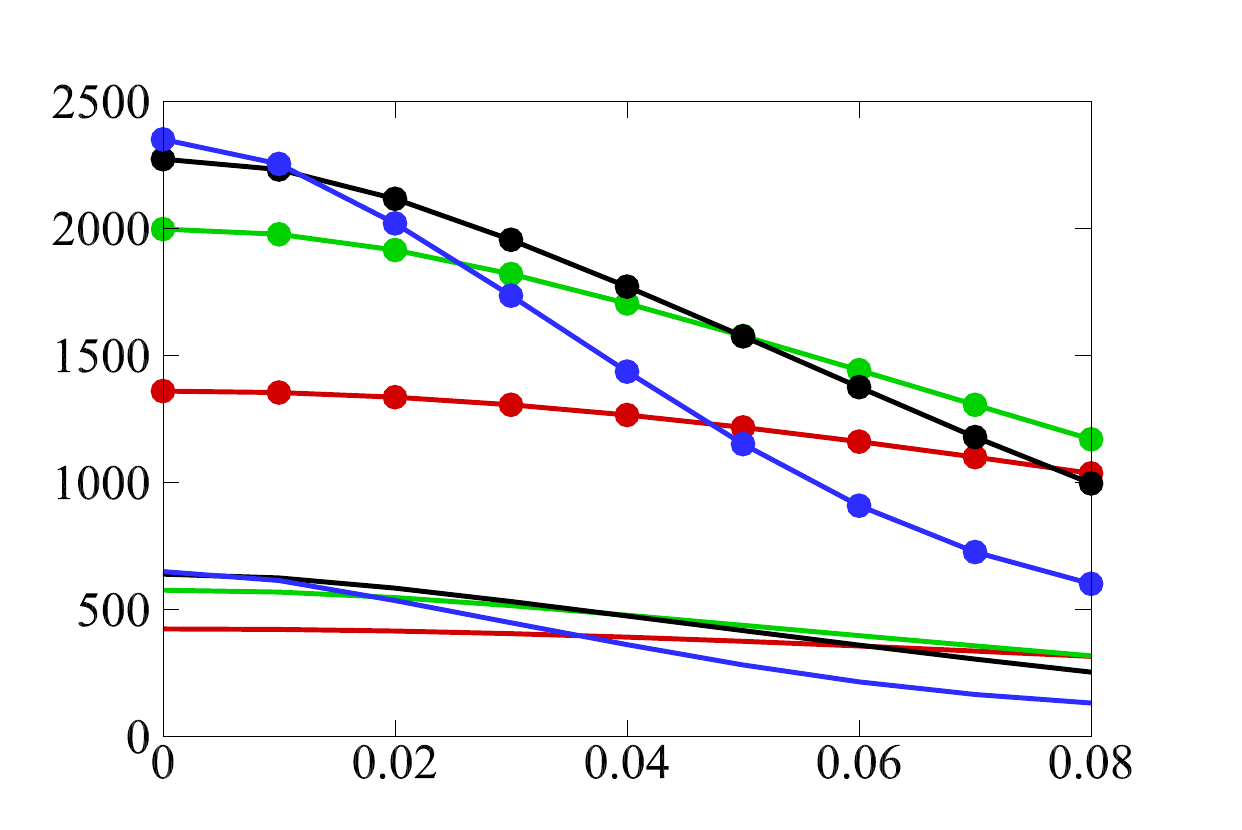}
  \includegraphics[width = 0.49\textwidth]{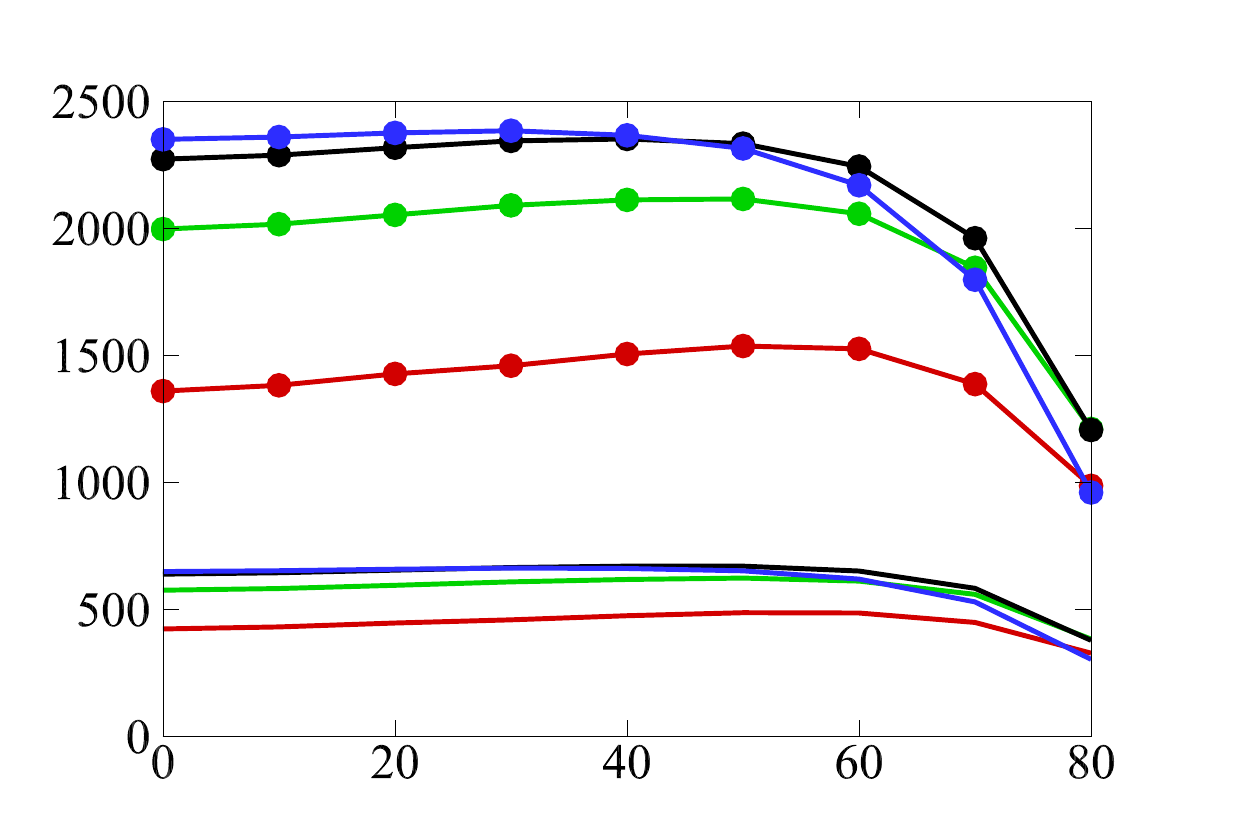}
  \put(-380,120){$(c)$}
  \put(-190,120){$(d)$}
  \put(-385,50){\rotatebox{90}{$A_u,~A_T$}}
  \put(-195,50){\rotatebox{90}{$A_u,~A_T$}}
  \put(-290,0){$\omega$}
  \put(-100,0){$\vartheta$}
  \put(-20,0){$(^{\circ}) $}
  \caption{Perturbation amplitude distributions for case A. ($a$) Streamwise evolution of  $A_u$  for  $k_3$ varying from 1 to 10 in increments  of 1, with $\omega=0$ and $\vartheta=0{^{\circ}}$. The vertical line indicates the experimentally observed transition onset from \citet{borovoy2022laminar}. 
  ($b$,$c$,$d$): Dependence of $A_u$ (solid lines) and $A_T$ (symboled lines) on $k_3$ (with $\omega=0$ and $\vartheta=0{^{\circ}}$), $\omega$ (with $k_3=4$ and {$\vartheta=0{^{\circ}}$}) and $\vartheta$ (with $\omega=0$ and $k_3=4$) at  ${x}=$ 50 (red), 100 (green), 146 (black) and 250 (blue).}
  \label{fig:linear_line_auat_k3k1k2_r1.2}
  \end{center}
\end{figure}
To clearly demonstrate the amplification behavior  of the non-modal perturbations, we define their amplitudes in terms of the maximum magnitude of either streamwise velocity or temperature,
\begin{equation}
A_u({x})=\max_{y_n}|\hat u({x},y_n)|,\quad A_T{(x)}=\max_{y_n}|\hat T({x},y_n)|.
\end{equation}
Figure \ref{fig:linear_line_auat_k3k1k2_r1.2}(a) shows the streamwise evolution of $A_u$ for ten different spanwise wavenumbers $k_3$ ranging from 1 to 10 for case A. At a relatively high $k_3$, the non-modal perturbation undergoes transient growth, after which it monotonically decays further downstream. As $k_3$ decreases,  the position corresponding to the maximum growth moves progressively downstream. For wavenumbers  $k_3\leq 4$, no downstream decay is observed within the computational domain of interest. 
According to the experimental observations from \citet{borovoy2022laminar}, the transition onset for this case occurs at approximately  ${x}\approx {129}$,  indicated by the vertical line in this panel. At this transition location, the optimal spanwise wavenumber is between $k_3=$ 4 and 5. The corresponding spanwise wavelengths of these optimal non-modal perturbations are comparable to the local boundary-layer thickness at the same location, consistent with previous findings in \citet{zhao2025excitation}.
{The resolution study for the SF-HLNS calculations is provided in figure \ref{fig:gridcheck}(b) in Appendix \ref{app:A}.}

In figure \ref{fig:linear_line_auat_k3k1k2_r1.2}(b), we select four representative streamwise positions and plot the dependence of $A_u$ and $A_T$ on $k_3$. Clearly, an optimal $k_3$ appears at each position, with the optimal value decreasing as ${x}$ increases. Additionally, panel (c) depicts the dependence of $A_u$ and $A_T$ on frequency, revealing that stationary non-modal perturbations experience the highest amplification. Finally, the variation of $A_u$ and $A_T$ with $\vartheta$ are shown in panel (d). A broad plateau appears between $\vartheta=$ {0$^{\circ}$} to 60$^\circ$, whereas amplitudes noticeably decrease for higher values of  $\vartheta$.

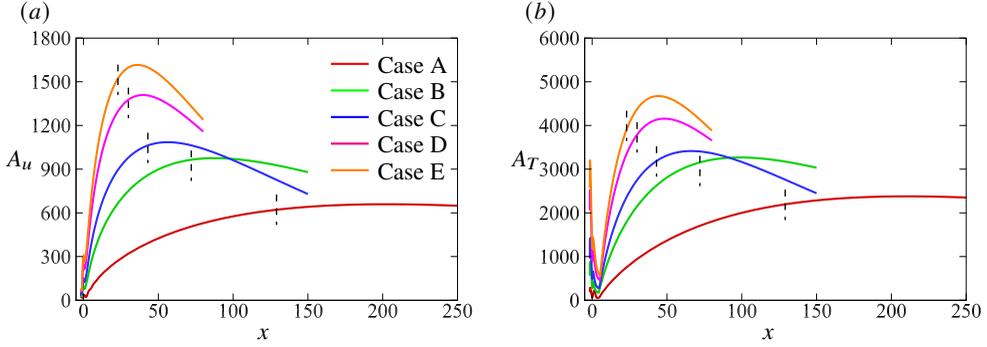
\begin{figure}
  \begin{center}
  \includegraphics[width = \textwidth]{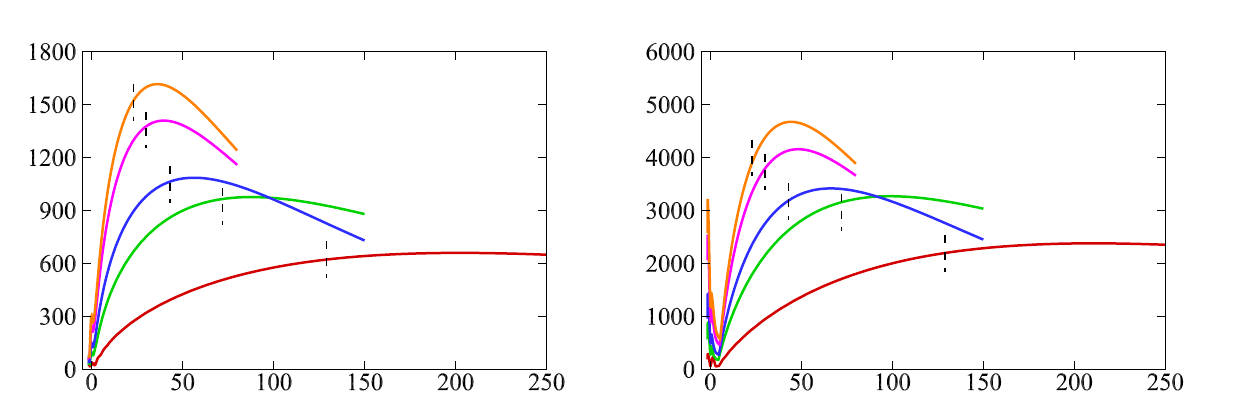}
  \put(-380,120){$(a)$}
  \put(-190,120){$(b)$}
  \put(-385,65){$A_u$}
  \put(-195,65){$A_T$}
  \put(-290,0){${x}$}
  \put(-100,0){${x}$} 
  \put(-263,63) {\begin{tikzpicture}
    \draw[red,thick]   (0,0) -- (0.5,0);
    \draw[green,thick] (0,-0.35) -- (0.5,-0.35);
    \draw[blue,thick]  (0,-0.7) -- (0.5,-0.7);
    \draw[magenta,thick] (0,-1.05) -- (0.5,-1.05);
    \draw[orange,thick]  (0,-1.4) -- (0.5,-1.4);
    \end{tikzpicture}}
  \put(-245,100) {Case A}
  \put(-245,90) {Case B}
  \put(-245,80) {Case C}
  \put(-245,70) {Case D}
  \put(-245,60) {Case E}
  \caption{Comparison of streamwise evolution of $A_u$ ($a$) and $A_T$ ($b$) across the five cases, where  $\omega=0$, $\vartheta=0^{\circ}$ and $k_3=$ 4, 7.5, 10, 14 and 16 for the five cases. The vertical lines represent the experimental transition onset.}
  \label{fig:linear_line_auat_zone_cases}
  \end{center}
\end{figure}
Repeating the same parametric study for other cases, we  find that $\omega=0$ and $\vartheta=0^\circ$ remain representative parameters for non-modal perturbations in all considered cases. However, the optimal spanwise wavenumber varies due to differences in local boundary-layer thickness. Specifically, the optimal  $k_3$ for the five cases are determined to be 4, 7.5, 10, 14 and 16. Using these chosen values, we compare the streamwise evolution of $A_u$ and $A_T$ across the five case studies, as shown in figure \ref{fig:linear_line_auat_zone_cases}. 
Clearly, as the Reynolds number (nose radius) increases, the non-modal perturbations attain their maximum amplitude farther upstream, and the peak amplification rate becomes more pronounced. This observation suggests that for larger bluntness, non-modal perturbations will likely reach nonlinear saturation earlier, potentially resulting in an earlier onset of bypass transition.
To confirm this, we will perform subsequent NPSE calculations, as presented in the following subsection.

\subsection{Nonlinear evolution of streaks}\label{subsec:streak_nonlinear}

\begin{figure}
  \begin{center}
  \includegraphics[width = \textwidth]{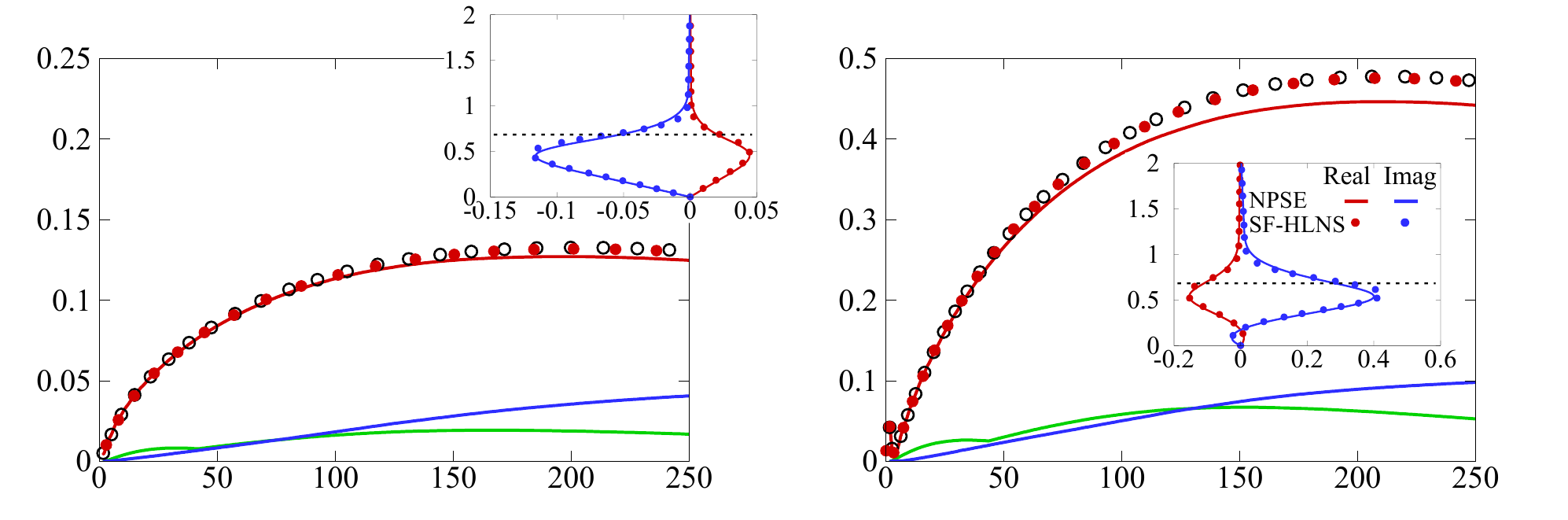}
  \put(-380,125){$(a)$}
  \put(-190,125){$(b)$}
  \put(-385,65){${A_{u}}$}
  \put(-195,65){${A_{T}}$}
  \put(-285,100){$y_n$}
  \put(-235,110){$\hat u_{0,1}$}
  \put(-260,98){\fontsize{6pt}{6pt}\selectfont $\delta_{BL}$}
  \put(-120,65){$y_n$}
  \put(-95,70){$\hat T_{0,1}$}
  \put(-48,62){\fontsize{6pt}{6pt}\selectfont $\delta_{BL}$}
  \put(-355,67) {\begin{tikzpicture}
    \draw[red,thick]   (0,0) -- (0.5,0);
    \draw[green,thick] (0,-0.35) -- (0.5,-0.35);
    \draw[blue,thick]  (0,-0.7) -- (0.5,-0.7);
    \node[draw,black,circle,inner sep=1pt, line width=0.8pt] at (0.25,-1.0) {};
    \node[draw,red,circle,fill,inner sep=1pt] at (0.25,-1.3) {};
    \end{tikzpicture}}
  \put(-337,103) {NPSE: (0,1)}
  \put(-337,93) {NPSE: (0,2)}
  \put(-337,83) {NPSE: (0,0)}
  \put(-337,73) {LPSE: ~(0,1)}
  \put(-337,63) {SF-HLNS}\\
  \includegraphics[width = \textwidth]{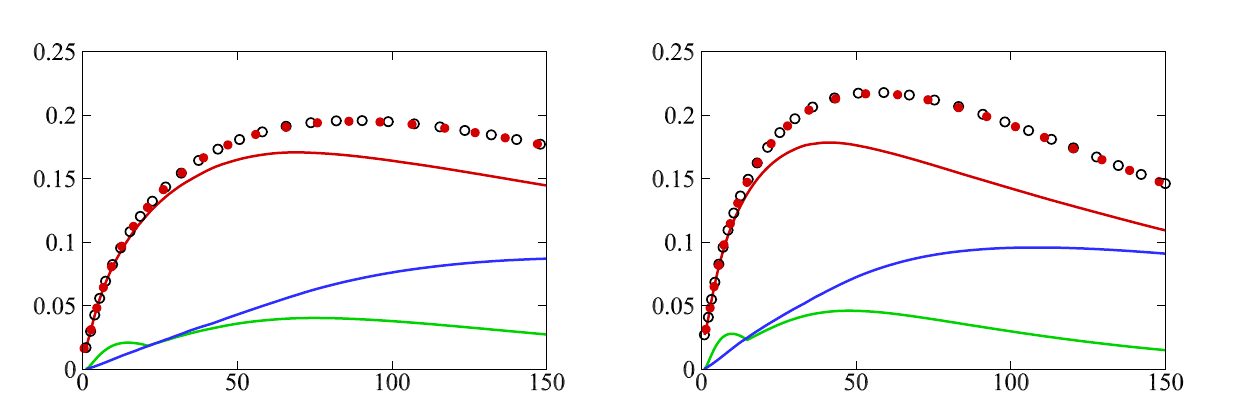}
  \put(-380,120){$(c)$}
  \put(-190,120){$(d)$}
  \put(-385,65){${A_{u}}$}
  \put(-195,65){${A_{u}}$}\\
  \includegraphics[width = \textwidth]{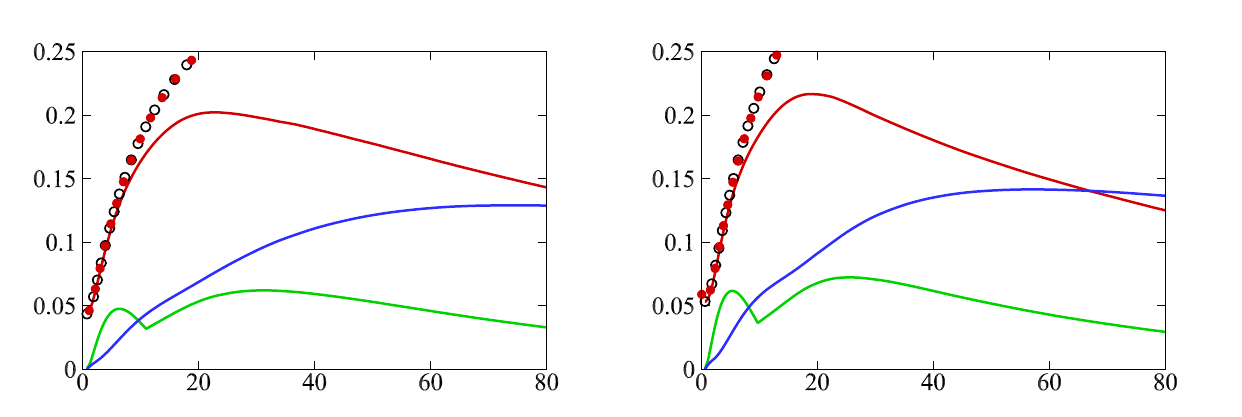}
  \put(-380,120){$(e)$}
  \put(-190,120){$(f)$}
  \put(-385,65){${A_{u}}$}
  \put(-195,65){${A_{u}}$}
  \put(-290,0){${x}$}
  \put(-100,0){${x}$} 
  \caption{Nonlinear evolution of the perturbation amplitudes {for noise level $Q=1.75\%$}, with results from the SF-HLNS and LPSE calculations included for comparison. ($a$,$b$): $A_u$ and $A_T$ for case A, with the inset figures illustrating the corresponding wall-normal profiles at ${x=146}$. ($c$-$f$): $A_u$ for cases B, C, D and E.}
  \label{fig:nonl_line_auat_cases}
  \end{center}
\end{figure}
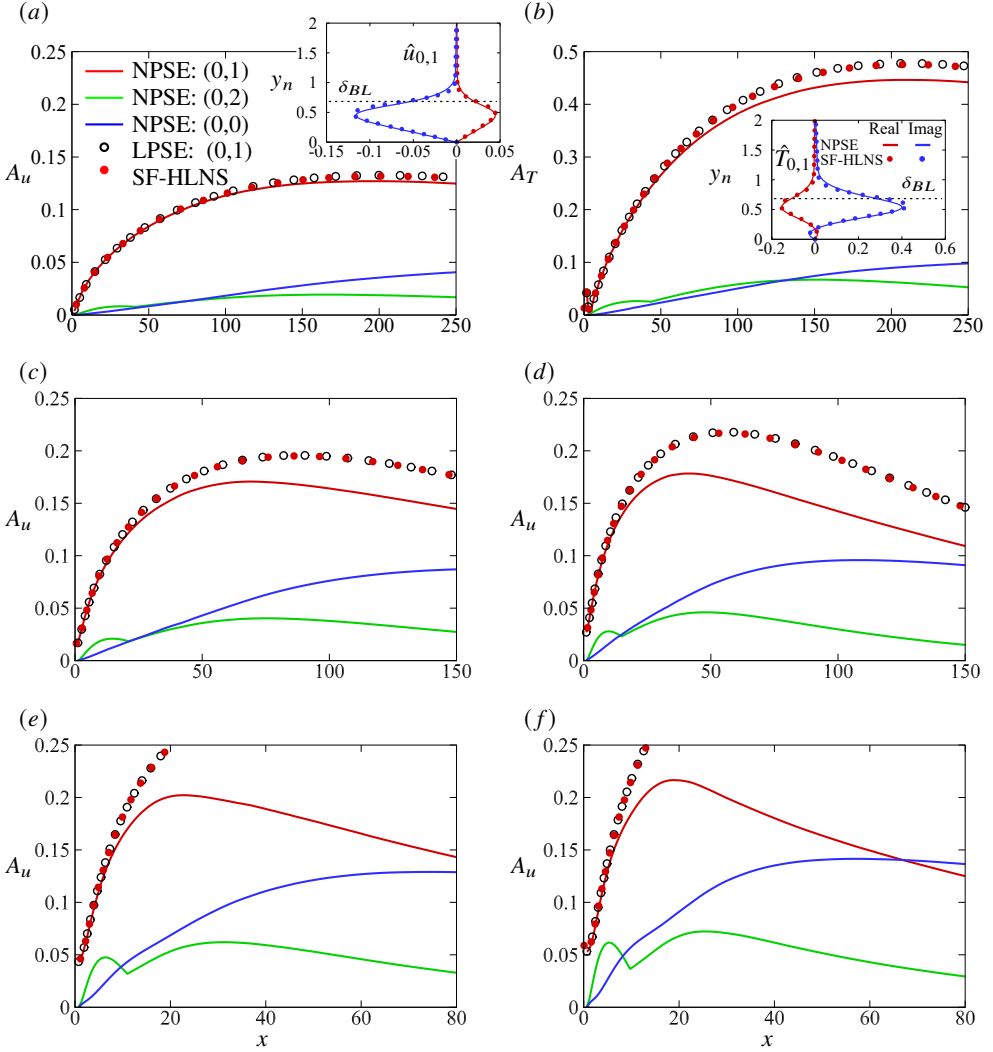
To predict the nonlinear evolution of non-modal perturbations, we perform NPSE calculations based on the non-modal receptivity results obtained via SF-HLNS. Given that non-modal perturbations exhibit stronger transient growth at higher Reynolds numbers, we opted for earlier starting positions for the NPSE calculations. Specifically, we select the dimensional starting position  $X_0^*=2$ mm, corresponding to $X_0=1.16\times 10^5/Re$ in the dimensionless form. For the five cases considered, the resulting values of $X_0$ are 1.67, 1.11, 1.00, 0.769 and 0.667, respectively. The mesh system for each case is the same as that for the SF-HLNS calculation, but with the nose region excluded.

The initial perturbation profiles are obtained from the SF-HLNS calculations, using the slow acoustic forcing that maximizes receptivity efficiency. In line with the findings in $\S$\ref{subsec:streak_excitation}, the forcing frequency is set to zero and the wave angle to {$\vartheta=0^\circ$} for all cases. The spanwise  wavenumbers are prescribed as $k_3=4$, 7.5, 10, 14 and 16 for cases  A, B, C, D and E, respectively. 
By setting the  amplitude of the oncoming pressure perturbation to  {$\varepsilon=0.0002$ ($Q=1.75\%$) or 0.0004 ($Q=3.5\%$)}, the corresponding initial amplitude for the NPSE calculation for each case is determined from the SF-HLNS calculation.

\begin{figure}
  \begin{center}
  \includegraphics[width = 0.95\textwidth]{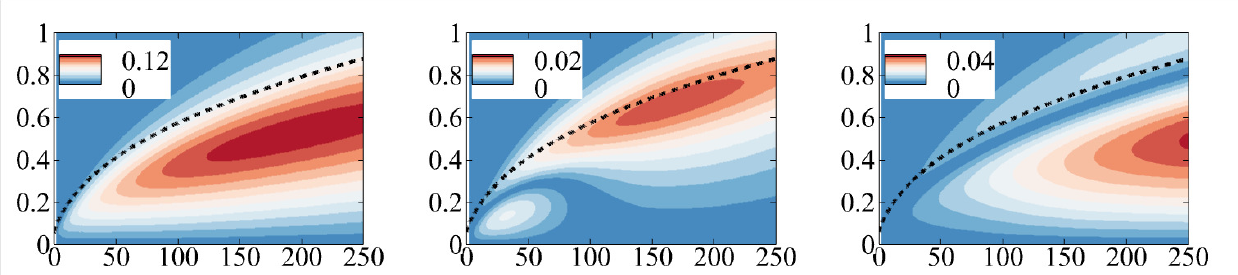}
  \put(-380,75){$(a)$}
  \put(-250,75){$(b)$}
  \put(-130,75){$(c)$}
  \put(-375,40){$y_n$}
  \put(-305,-5){${x}$}  
  \put(-185,-5){${x}$} 
  \put(-63,-5){${x}$} 
  \put(-345,45){{$|\hat u_{0,1}|$}}
  \put(-223,45){{$|\hat u_{0,2}|$}}
  \put(-102,45){{$|\hat u_{0,0}|$}}
  \caption{Contours of $|\hat u|$ {for components} ($a$) (0,1), ($b$) (0,2) and ($c$) (0,0) for case A obtained by NPSE calculations. The dashed line indicates the edge of the boundary layer.}
  \label{fig:nonl_cont_eigenf_r1.2}
  \end{center}
\end{figure}
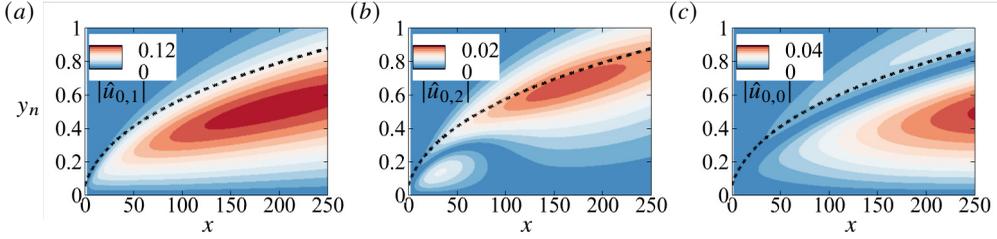

The solid lines in figure \ref{fig:nonl_line_auat_cases}$(a)$ illustrate the streamwise evolution of the velocity amplitude $A_u$ for the fundamental perturbation component (0,1), its second-order harmonic (0,2) and the mean-flow distortion (0,0) for case A under noise level {$Q=1.75\%$}. Note that both the fundamental and second-order harmonic components are stationary due to an initial fundamental frequency $\omega$ of zero. The fundamental component exhibits sustained growth downstream, ultimately reaching an amplitude of approximately {0.125} at the outlet of the computational domain. Such a trend generally matches the SF-HLNS calculation (red circles), with only a slightly lower amplitude observed in the downstream region due to weak nonlinearity. The amplitudes of the second-order harmonic and the mean-flow distortion at the downstream boundary of the computational domain are found to be approximately {0.02 and 0.04, respectively.}
The inset compares profiles of the real and imaginary parts of $\hat u_{0,1}$ obtained by NPSE and SF-HLNS calculations at ${x=146}$, demonstrating overall agreement.
Figures \ref{fig:nonl_cont_eigenf_r1.2}(a), (b) and (c) further illustrate the perturbation fields of $\hat u$ for (0,1), (0,2) and (0,0), respectively. 
The fundamental component (0,1) exhibits a single, well-defined peak in the bulk boundary-layer region. In contrast, the second-order harmonic (0,2) displays two peak regions, with its dominant peak occurring closer to the boundary-layer edge. The dominant peak of the mean-flow distortion (0,0) appears at a position lower than that of the fundamental component (0,1).
Figure \ref{fig:nonl_line_auat_cases} (b) presents  the evolution of the temperature amplitude $A_T$, with an inbox showing the wall-normal profiles at $x=146$. A similar trend is observed.  Notably, the temperature amplitudes for components  (0,1), (0,2) and (0,0) at the downstream boundary are {0.45, 0.05 and 0.1}, respectively, much greater than their velocity counterparts.

Figures \ref{fig:nonl_line_auat_cases}(c-f)  illustrate the streamwise evolution of the amplitude $A_u$ for cases B to E. In each case, the fundamental component (0,1) initially exhibits transient growth, followed by amplitude damping downstream after reaching saturation. As the Reynolds number increases, the saturation position shifts upstream, and the NPSE results deviate from the linear SF-HLNS predictions when $A_u$ exceeds approximately 0.15. Additionally, the amplitude of the fundamental component at its saturation position increases notably with increasing $Re$. The second-order harmonic component (0,2) also experiences transient growth for each case, but its saturation occurs farther downstream, with much smaller amplitudes compared to the (0,1) component. In contrast, the mean-flow distortion (0,0) increases gradually throughout the computational domain, exhibiting no damping behavior until the downstream outlet.

Additionally, the NPSE can be simplified to the linear parabolised stability equations (LPSE) by neglecting the nonlinear terms $f_{m,n}$ in (\ref{eq:NPSE}). In figure \ref{fig:nonl_line_auat_cases}, the LPSE calculations are also included (open circles) and agree closely with the SF-HLNS results. This excellent agreement further confirms the reliability of our numerical results.

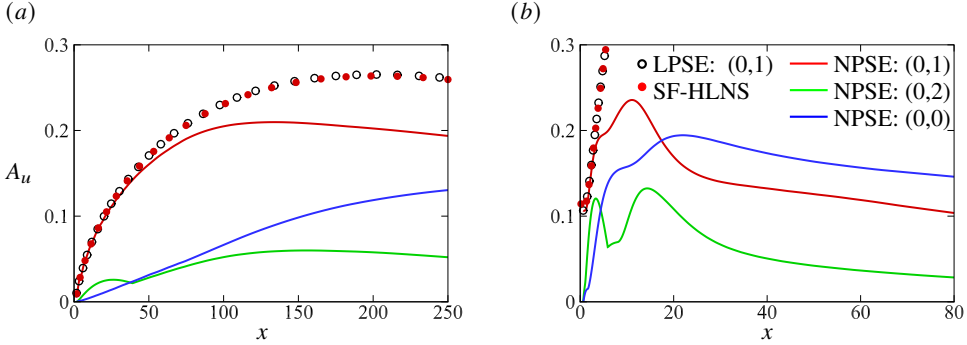
\begin{figure}
  \begin{center}
  \includegraphics[width = 0.49\textwidth]{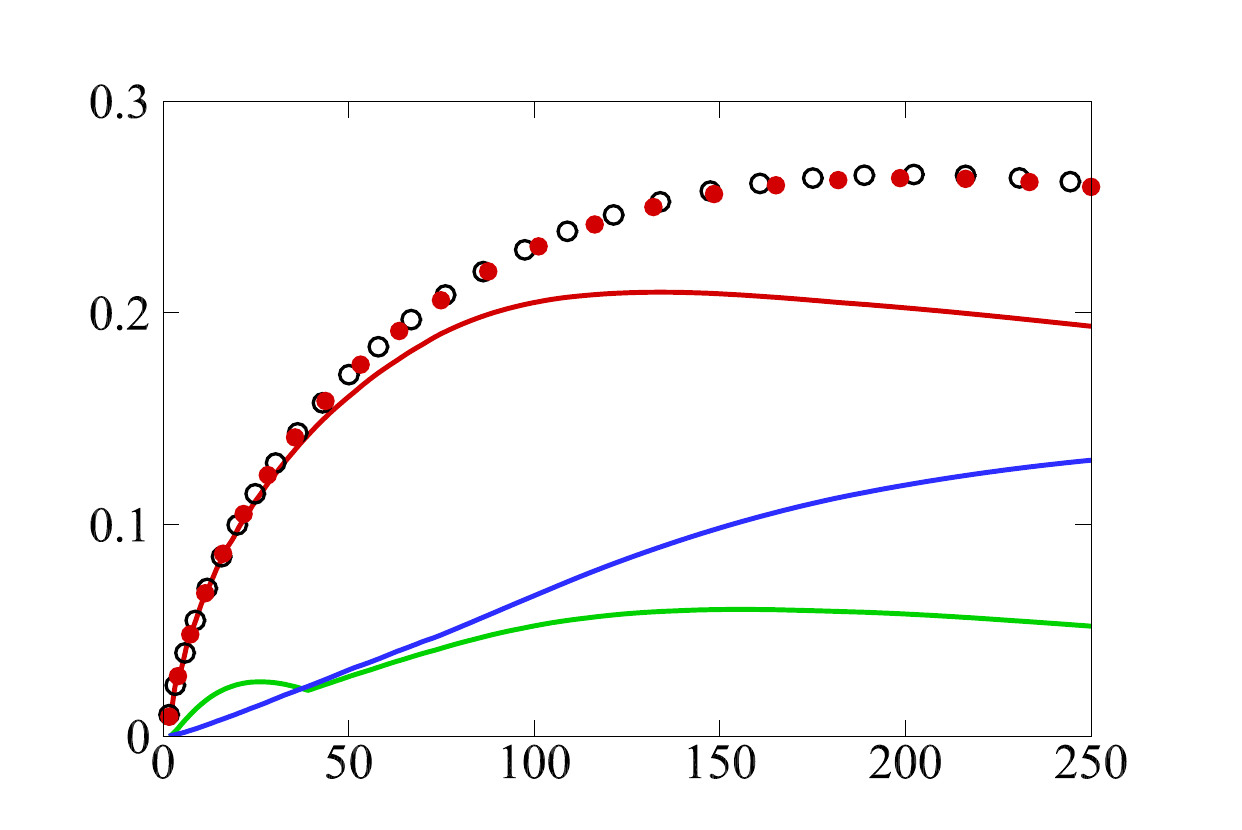}
  \includegraphics[width = 0.49\textwidth]{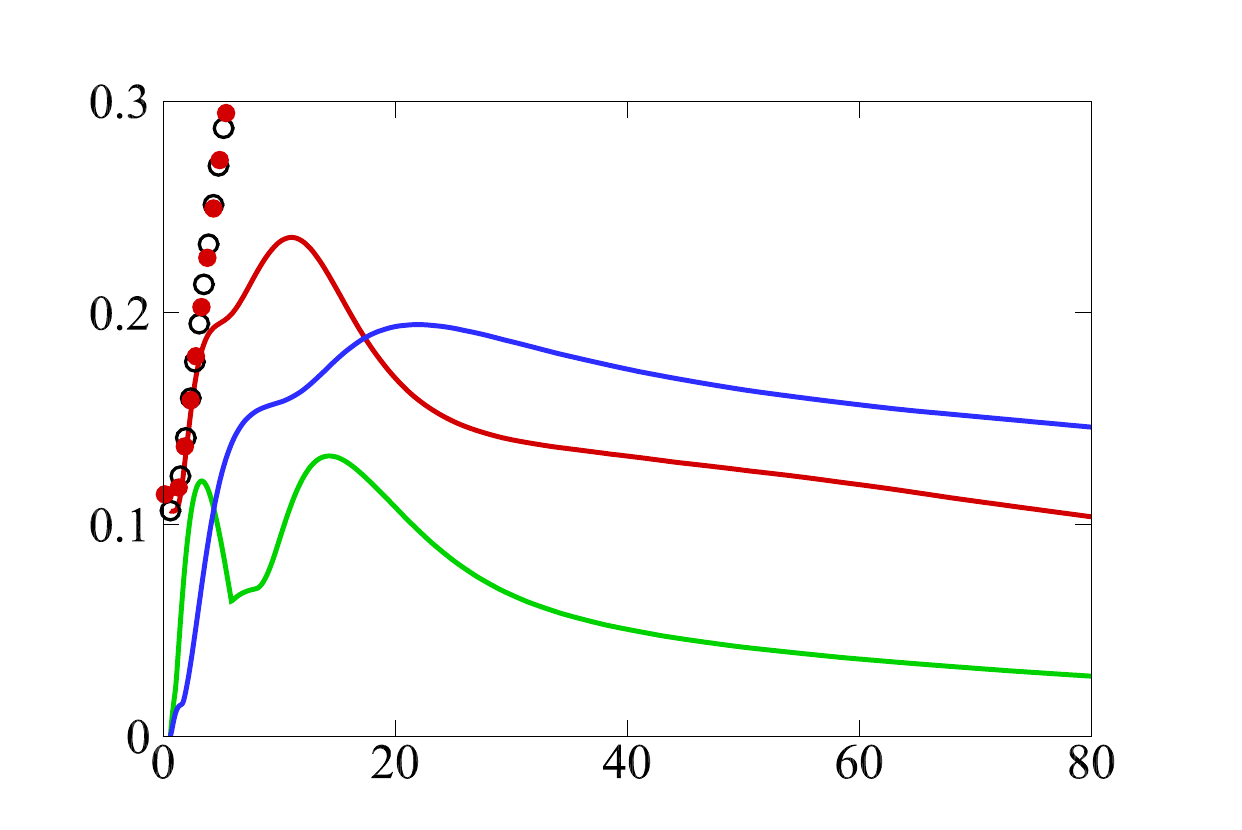}
  \put(-380,120){$(a)$}
  \put(-190,120){$(b)$}
  \put(-380,60){$A_{u}$}
  \put(-285,0){$x$}
  \put(-95,0){$x$}
  \put(-85,83) {\begin{tikzpicture}
    \draw[red,thick]   (0,0) -- (0.5,0);
    \draw[green,thick] (0,-0.35) -- (0.5,-0.35);
    \draw[blue,thick]  (0,-0.7) -- (0.5,-0.7);
    \end{tikzpicture}}
  \put(-142,93) {\begin{tikzpicture}
    \node[draw,black,circle,inner sep=1pt, line width=0.8pt] at (0.25,-1.0) {};
    \node[draw,red,circle,fill,inner sep=1pt] at (0.25,-1.3) {};
  \end{tikzpicture}}
  \put(-68,100) {NPSE: (0,1)}
  \put(-68,90) {NPSE: (0,2)}
  \put(-68,80) {NPSE: (0,0)}
  \put(-136,100) {LPSE: ~(0,1)}
  \put(-136,90) {SF-HLNS}
  \caption{{Nonlinear evolution of the perturbation amplitudes $A_u$ for noise level {$Q=3.5\%$}. ($a$) Case A; ($b$) case E. The line types are identical to those in figure \ref{fig:nonl_line_auat_cases}.}}
  \label{fig:nonl_line_auat_cases_2}
  \end{center}
\end{figure}
When the noise level is increased to {$Q=3.5\%$}, as shown in figure   \ref{fig:nonl_line_auat_cases_2}, nonlinear saturation occurs earlier in all cases, accompanied by a more pronounced deviation from linear predictions downstream. Notably, for case E, the mean-flow distortion  (0,0)  reaches saturation by {$x\approx20$}, potentially indicating much stronger secondary instability, which will be analysed in the next subsection.

\subsection{Secondary instability and BSA}\label{subsec:streak_sia}
\begin{figure}
  \begin{center}
  \includegraphics[width = \textwidth]{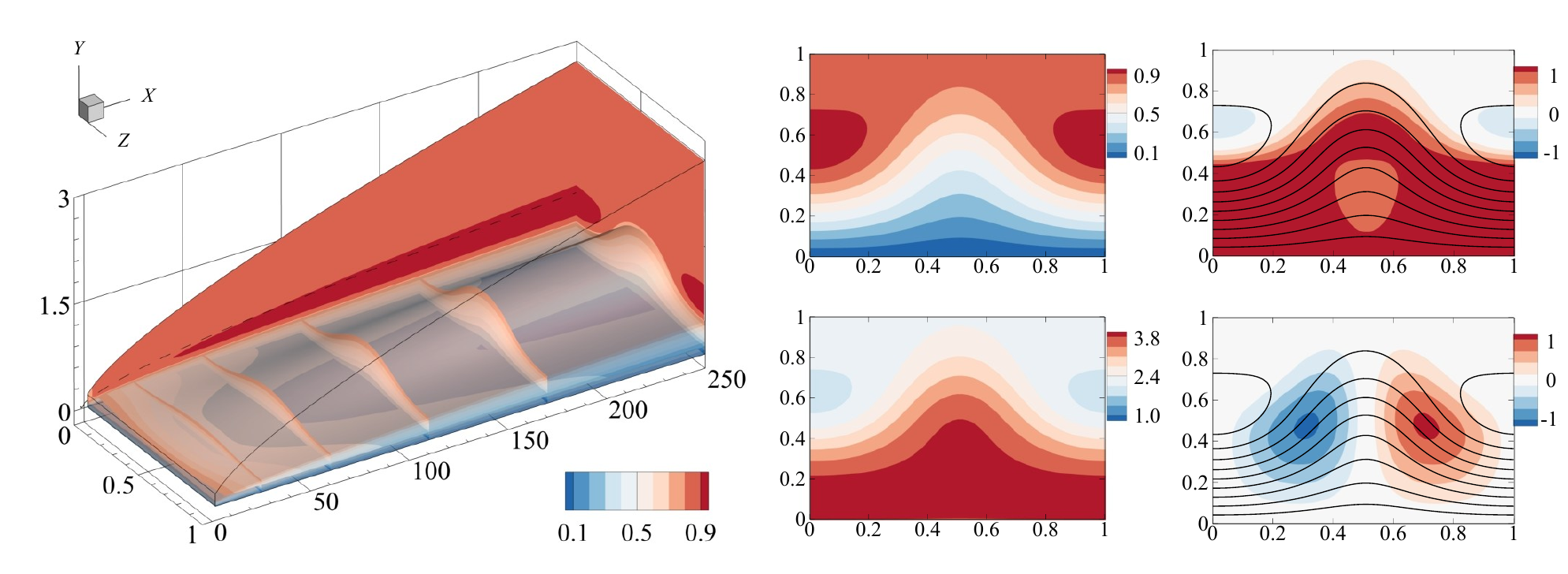}
  \put(-380,135){$(a)$}
  \put(-270,20){${x}$}
  \put(-380,72){$y_n$}
  \put(-370,10){{$z/z_l$}} 
  \put(-230,1){{$\bar{u}$}}
  \put(-205,135){$(b)$}
  \put(-203,100){$y_n$}
  \put(-203,35){$y_n$}
  \put(-155,1){{$z/z_l$}} 
  \put(-110,130){{$\bar{u}$}}
  \put(-110,65){{$\bar{T}$}}
  \put(-100,135){$(c)$}
  \put(-13,130){{$\frac{\partial \bar{u}}{\partial y_n} $}}
  \put(-13,65){{$\frac{\partial \bar{u}}{\partial z} $}}
  \put(-55,1){{$z/z_l$}} 
  \caption{Demonstration of the streaky base flow for case A with {$Q=1.75\%$}. $(a)$ Contours of $\bar{u}$ in the entire computational domain; $(b)$ contours of $\bar{u}$ and $\bar{T}$ at ${x}=146$; $(c)$ contours of ${\partial \bar{u}}/{\partial y_n} $ and ${\partial \bar{u}}/{\partial z} $ at ${x}=146$.}  
  \label{fig:sia_cont_streakyUBTB_r1.2}
  \end{center}
\end{figure}
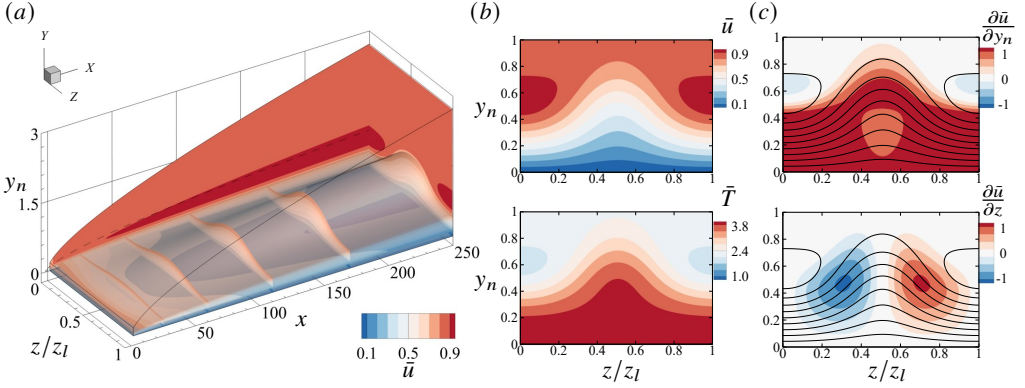
Once the non-modal perturbations reach the nonlinear saturated state, secondary instabilities characterised by smaller temporal and spatial scales would experience  rapid amplification, eventually accumulating enough amplitude to trigger bypass transition. 
Let us first consider the noise level of {$Q=1.75\%$}. Figure \ref{fig:sia_cont_streakyUBTB_r1.2} illustrates the distribution of the mean flow  $\bar u$ and $\bar T$, defined in (\ref{eq:SIA_baseflow}), obtained from the NPSE calculation for case A,  with only one spanwise period ${z_l=2\pi/k_3}$ displayed. The streaky structures are clearly visible, with their amplitude increasing downstream along ${x}$. At a downstream location $x=146$, pronounced gradients are apparent in both the $y_n$ and $z$ directions, suggesting the potential streak instabilities.

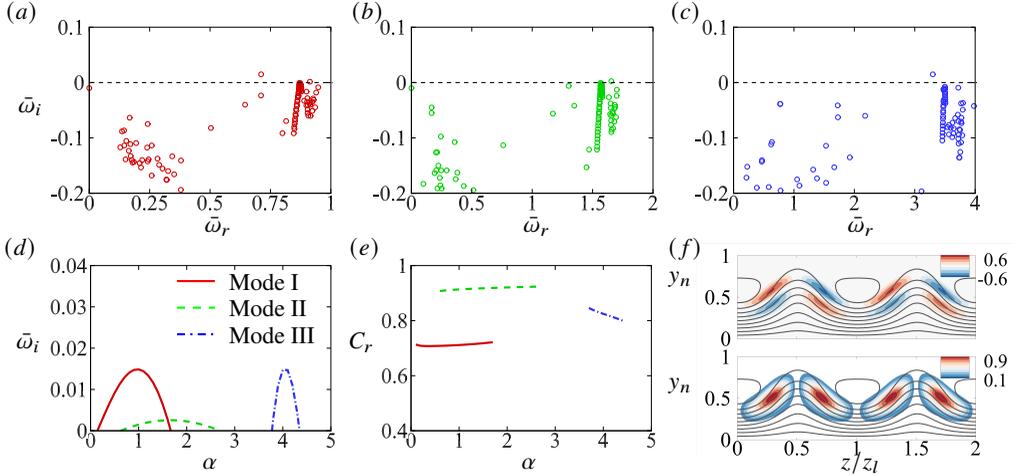
\begin{figure}
  \begin{center}
  \includegraphics[width = 0.95\textwidth]{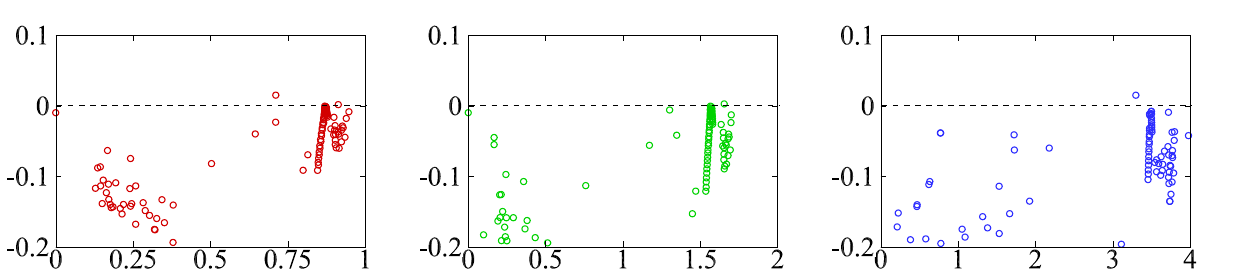}
  \put(-380,75){$(a)$}
  \put(-250,75){$(b)$}
  \put(-130,75){$(c)$}
  \put(-375,40){${\bar\omega_i}$}
  \put(-305,-5){${\bar\omega_r}$}  
  \put(-185,-5){${\bar\omega_r}$} 
  \put(-63,-5){${\bar\omega_r}$} \\
  \includegraphics[width = 0.63\textwidth]{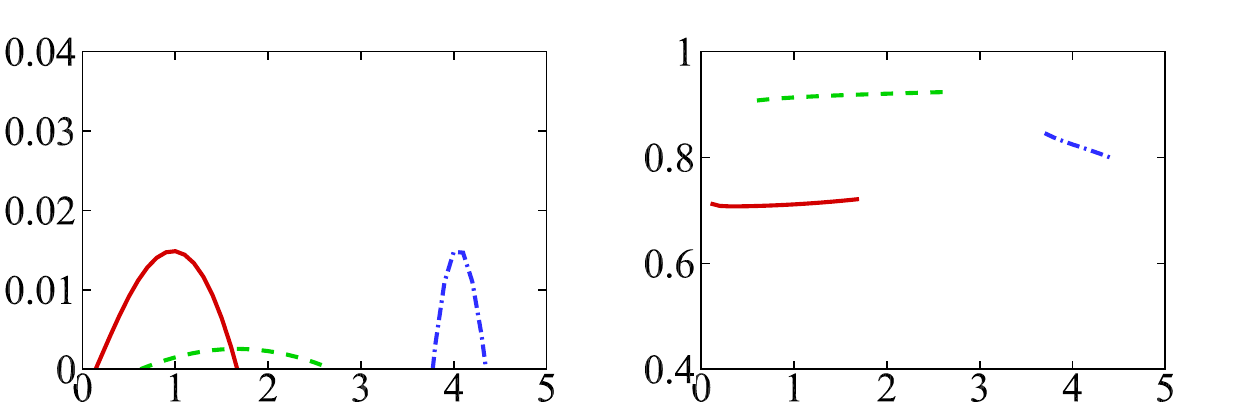}
  \includegraphics[width = 0.31\textwidth]{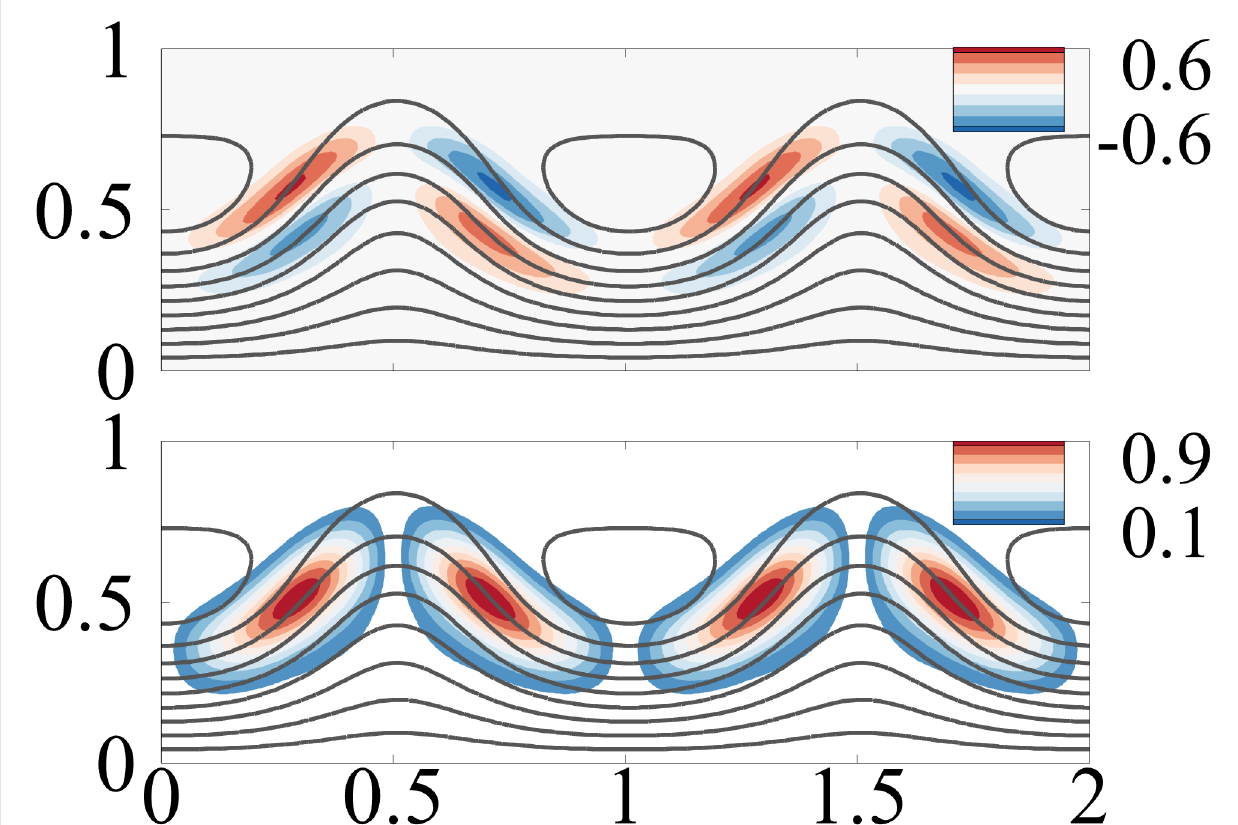}
  \put(-380,75){$(d)$}
  \put(-250,75){$(e)$}
  \put(-130,75){$(f)$}
  \put(-375,40){${\bar\omega_i}$}
  \put(-250,40){$C_r$}
  \put(-315,45) {\begin{tikzpicture}
    \draw[red,thick]   (0,0) -- (0.5,0);
    \draw[green,thick,dashed] (0,-0.35) -- (0.5,-0.35);
    \draw[blue,thick,dash dot]  (0,-0.7) -- (0.5,-0.7);
    \end{tikzpicture}}
  \put(-295,62) {Mode I}
  \put(-295,52) {Mode II}
  \put(-295,42) {Mode III}
  \put(-130,65){$y_n$}
  \put(-130,25){$y_n$}
  \put(-305,-5){$\alpha$}  
  \put(-185,-5){$\alpha$} 
  \put(-65,-5){{$z/z_l$}}
  \caption{{Temporal {BSA} of the fundamental $(\sigma_d=0)$ SI modes for case A with {$Q=1.75\%$} at $x=146$. $(a,b,c)$ Eigenvalue spectra in ${\bar\omega_i}-{\bar\omega_r}$ plane for $\alpha=$ 1, 1.8 and 4, respectively. $(d,e)$ Dependence on $\alpha$ of ${\bar\omega_i}$ and phase speed $C_r:=(\bar\omega/\alpha)_r$. $(f)$ The real part (top) and modulus (bottom) of the eigenfunction $\hat u_{SI}(y_n,z)$ of mode I with $\alpha=1$, for which ${\bar\omega_i}=0.015$. The eigenfunction is normalised by its maximum.}}
  \label{fig:sia_growth_local_r1.2_q=0}
  \end{center}
\end{figure}
\begin{figure}
  \begin{center}
  \includegraphics[width = 0.63\textwidth]{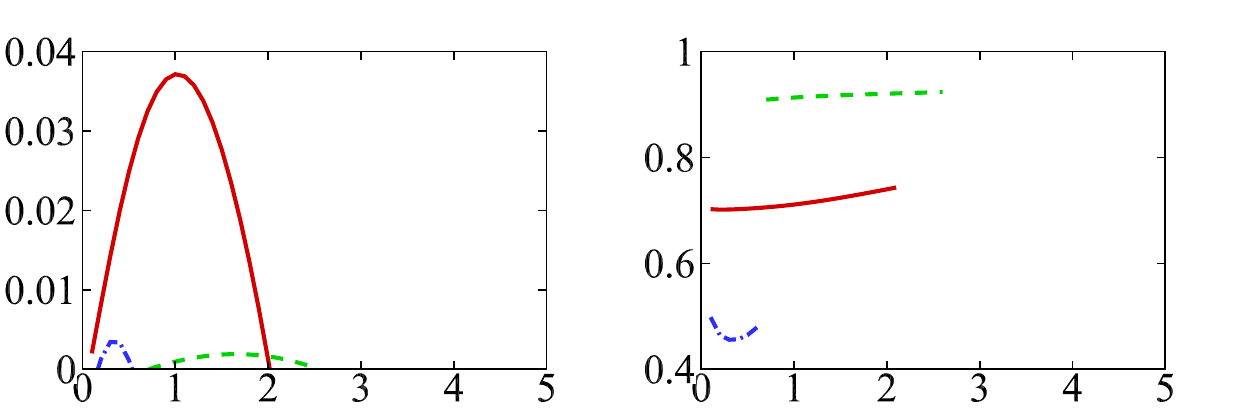}
  \includegraphics[width = 0.31\textwidth]{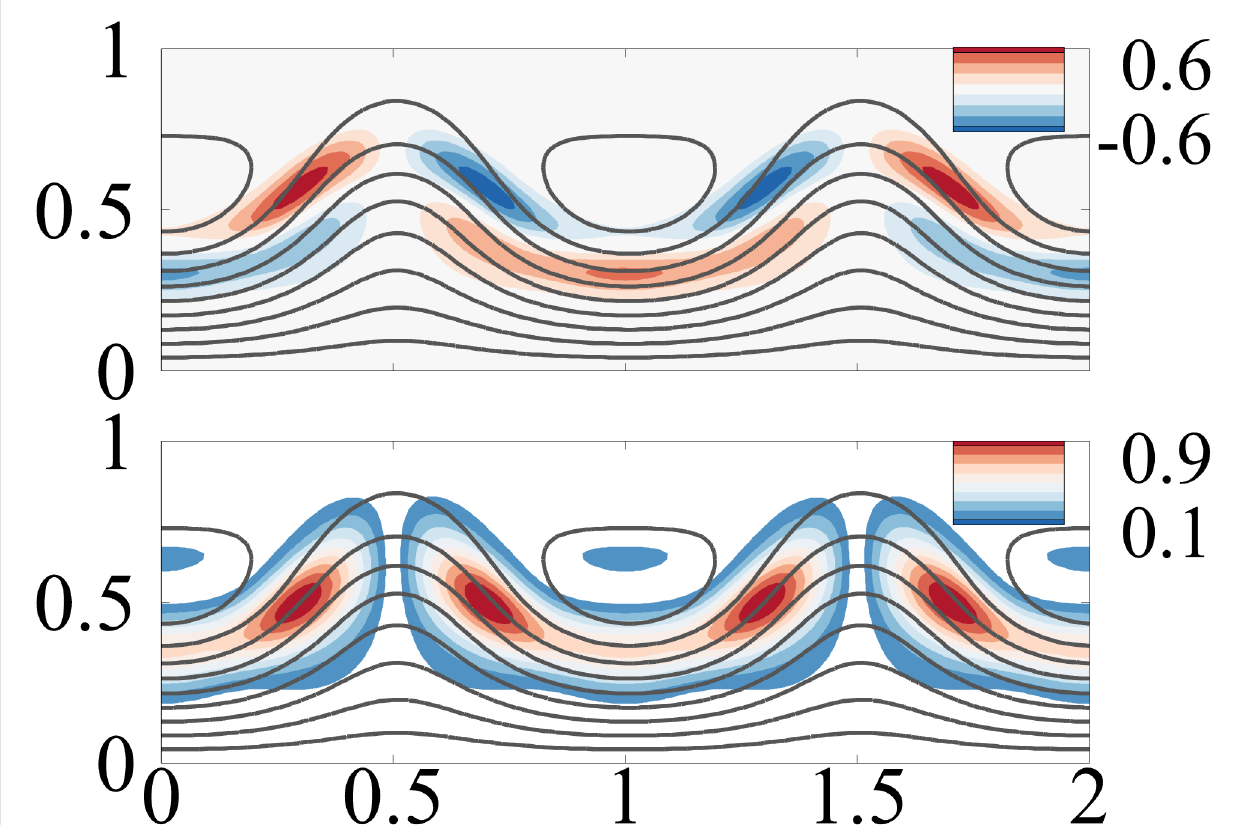}
  \put(-380,75){$(a)$}
  \put(-250,75){$(b)$}
  \put(-130,75){$(c)$}
  \put(-375,40){${\bar\omega_i}$}
   \put(-250,40){$C_r$}
  \put(-315,45) {\begin{tikzpicture}
    \draw[red,thick]   (0,0) -- (0.5,0);
    \draw[green,thick,dashed] (0,-0.35) -- (0.5,-0.35);
    \draw[blue,thick,dash dot]  (0,-0.7) -- (0.5,-0.7);
    \end{tikzpicture}}
  \put(-295,62) {Mode I}
  \put(-295,52) {Mode II}
  \put(-295,42) {Mode III}
  \put(-130,65){$y_n$}
  \put(-130,25){$y_n$}
  \put(-305,-5){$\alpha$}  
  \put(-185,-5){$\alpha$} 
  \put(-65,-5){{$z/z_l$}} 
  \caption{{Temporal BSA of the subharmonic $(\sigma_d=0.5)$ SI modes for case A with {$Q=1.75\%$} at $x=146$. $(a,b)$ Dependence on $\alpha$ of ${\bar\omega_i}$ and $C_r$. $(c)$ The real part (top) and modulus (bottom) of the eigenfunction $\re^{\ri \sigma_d k_3 z}\hat u_{SI}(y_n,z)$ of  mode I with $\alpha=1$, for which ${\bar\omega_i}=0.037$. The eigenfunction is normalised by its maximum.}}
  \label{fig:sia_growth_local_r1.2_q=0.5}
  \end{center}
\end{figure}

 Based on the streaky base flow at $x=146$ for case A, we perform temporal BSA of SI modes for two representative detuning parameters, $\sigma_d=0$ (fundamental) and 0.5 (subharmonic), as illustrated in figures \ref{fig:sia_growth_local_r1.2_q=0} and \ref{fig:sia_growth_local_r1.2_q=0.5}, respectively. For other values of the detuning parameter, the growth rates are expected to be of the same magnitude. Figures \ref{fig:sia_growth_local_r1.2_q=0}-(a,b,c) display the temporal eigenvalue spectra of the fundamental SI for three representative streamwise wavenumbers $\alpha$. In each panel, one distinct eigenmode with a positive growth rate $\bar \omega_i$ clearly appears, though at different real frequencies $\bar\omega_r$, where the subscript $r$ denotes the real part. By tracking these eigenvalues while gradually varying 
$\alpha$, we identify three distinct unstable SI mode branches. Figures \ref{fig:sia_growth_local_r1.2_q=0}-(d) and (e) respectively illustrate the dependence of growth rate $\bar\omega_i$ and the phase speed $C_r:=(\bar\omega/\alpha)_r$ on the wavenumber $\alpha$. These branches are denoted as modes I, II and III, ordered by ascending wavenumbers. Notably, mode I exists over a wide range of  $\alpha$ and attains the highest maximum growth rate among these modes. Its corresponding phase speed $C_r$ is approximately {0.71}. The contour plot of the eigenfunction $\hat u_{SI}(y_n,z)$ is shown in figure \ref{fig:sia_growth_local_r1.2_q=0}-(f), displaying two spanwise wavelengths. The extrema of the perturbation coincide with those of $\partial \bar u/\partial z$, depicted in figure\ref{fig:sia_cont_streakyUBTB_r1.2}-(c). Additionally, the perturbation displays an anti-symmetric structure about the axes defined by $z/z_l=0.5n$ with $n$ being an integer, indicating typical characteristics  of a sinuous mode.

By performing BSA for the subharmonic SI mode based on the same base flow, we have identified three unstable branches, whose growth rates and phase speeds are illustrated in figures \ref{fig:sia_growth_local_r1.2_q=0.5}-(a) and (b). 
Clearly, mode I emerges as the most unstable of these three branches. While the phase speeds of mode I in the subharmonic and fundamental SI modes are virtually identical, the maximum growth rate of the subharmonic mode is roughly twice that of its fundamental counterpart. The eigenfunction associated with the subharmonic mode I is presented in figure \ref{fig:sia_growth_local_r1.2_q=0.5}-(c), which is symmetric about $z/z_l=1$ yet anti-symmetric about $z/z_l=0.5$ or 1.5.

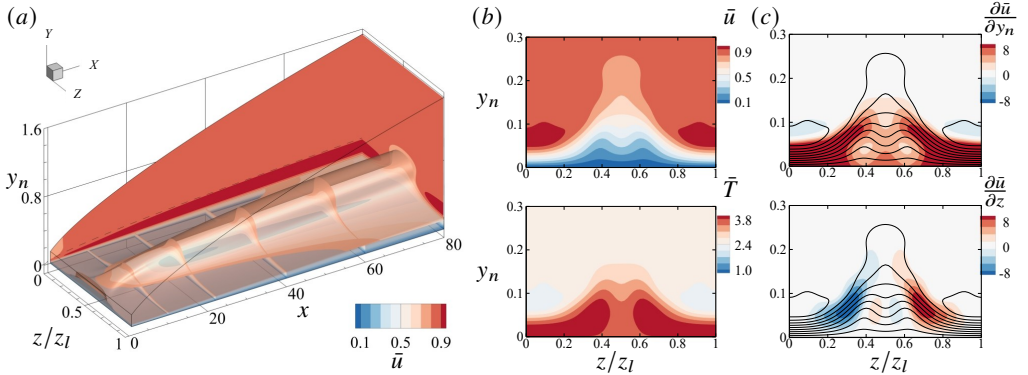
\begin{figure}
  \begin{center}
  \includegraphics[width = \textwidth]{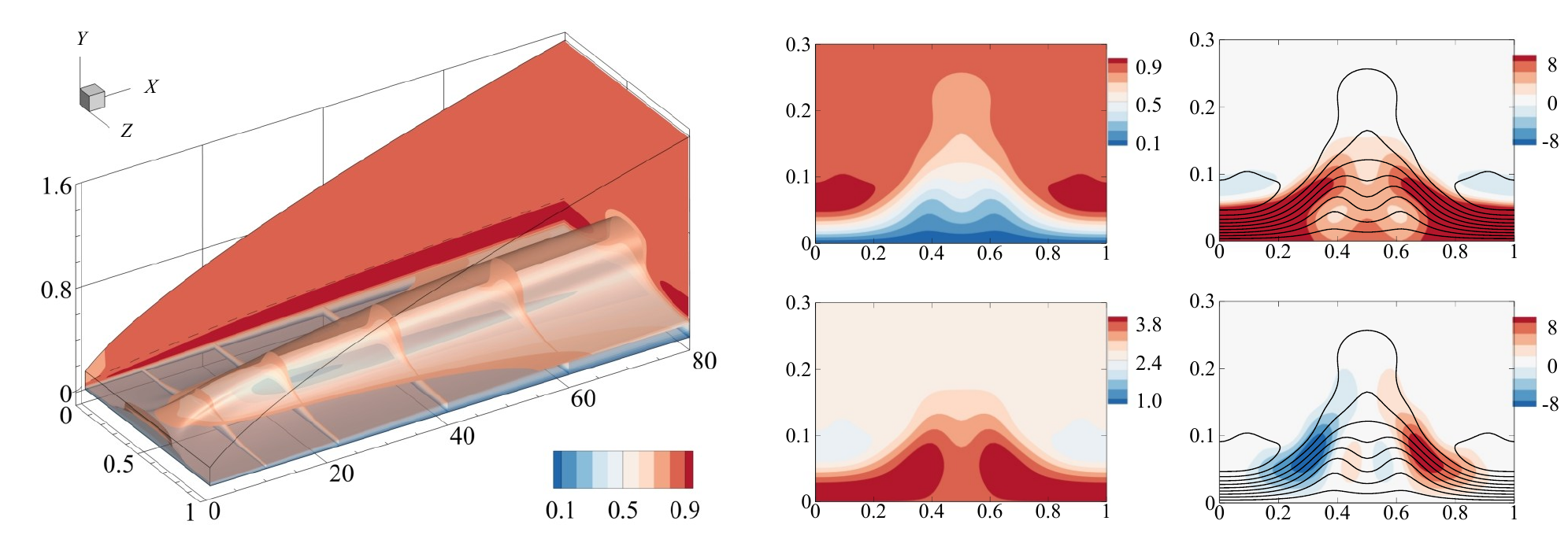}
  \put(-380,130){$(a)$}
  \put(-270,20){${x}$}
  \put(-380,70){$y_n$}
  \put(-370,10){{$z/z_l$}} 
  \put(-235,1){{$\bar{u}$}}
  \put(-205,130){$(b)$}
  \put(-203,100){$y_n$}
  \put(-203,35){$y_n$}
  \put(-158,1){{$z/z_l$}} 
  \put(-110,130){{$\bar{u}$}}
  \put(-110,65){{$\bar{T}$}}
  \put(-100,130){$(c)$}
  \put(-13,130){{$\frac{\partial \bar{u}}{\partial y_n} $}}
  \put(-13,65){{$\frac{\partial \bar{u}}{\partial z} $}}
  \put(-55,1){{$z/z_l$}}  
  \caption{{Demonstration of the streaky base flow for case E with {$Q=1.75\%$}. $(a)$ Contours of $\bar{u}$ in the entire computational domain; $(b)$ contours of $\bar{u}$ and $\bar{T}$ at $x=12$; $(c)$ contours of ${\partial \bar{u}}/{\partial y_n} $ and ${\partial \bar{u}}/{\partial z} $ at ${x}=12$.}}  
  \label{fig:sia_cont_streakyUBTB_r3.0}
  \end{center}
\end{figure}

For the bluntest configuration (case E) with {$Q=1.75\%$}, we demonstrate the streaky mean flow in figure \ref{fig:sia_cont_streakyUBTB_r3.0}.  Compared with case A, these streaks are far more tightly confined about the centreline ($z/z_l = 0.5$), yielding markedly steeper spanwise gradients in both $\bar u$ and $\bar T$.   These steep gradients may enhance the instability of the SI modes.
\begin{figure}
  \begin{center}
  \includegraphics[width = 0.63\textwidth]{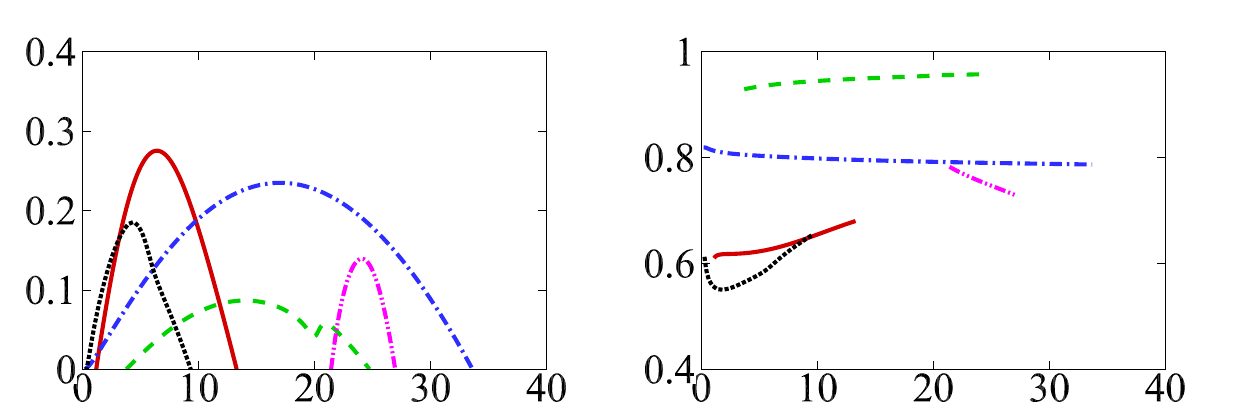}
  \includegraphics[width = 0.31\textwidth]{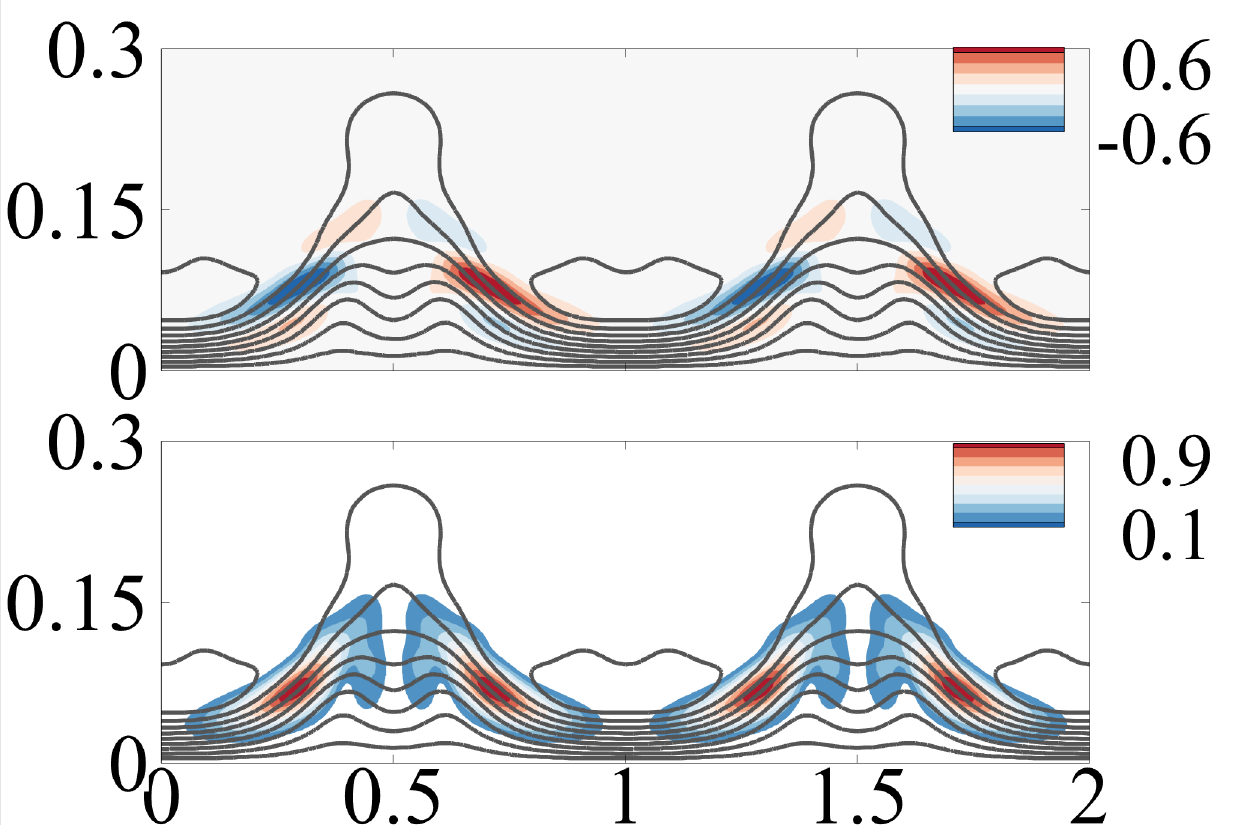}
  \put(-380,75){$(a)$}
  \put(-250,75){$(b)$}
  \put(-130,75){$(c)$}
  \put(-375,40){${\bar\omega_i}$}
  \put(-250,40){$C_r$}
  \put(-330,52) {\begin{tikzpicture}
    \draw[red,thick]   (0,0) -- (0.35,0);
    \draw[green,thick,dashed] (0,-0.25) -- (0.35,-0.25);
    \draw[blue,thick,dash dot]  (0,-0.5) -- (0.35,-0.5);
    \draw[magenta,thick,dash dot dot] (1.3,0) -- (1.65,0);
    \draw[black,thick,dotted]  (1.3,-0.25) -- (1.65,-0.25);
    \end{tikzpicture}}
  \put(-318,65) {\fontsize{6pt}{6pt}\selectfont Model I}
  \put(-318,57) {\fontsize{6pt}{6pt}\selectfont Model II}
  \put(-318,49) {\fontsize{6pt}{6pt}\selectfont Model III}
  \put(-283,65) {\fontsize{6pt}{6pt}\selectfont Model IV}
  \put(-283,57) {\fontsize{6pt}{6pt}\selectfont Model V}
  \put(-130,65){$y_n$}
  \put(-130,25){$y_n$}
  \put(-305,-5){$\alpha$}  
  \put(-185,-5){$\alpha$} 
  \put(-65,-5){{$z/z_l$}} 
  \caption{{Temporal BSA of the fundamental SI modes for case E with {$Q=1.75\%$} at $x=12$. {$(a,b)$ Dependence on $\alpha$ of $\bar\omega_i$ and $C_r$. $(c)$ The real part (top) and modulus (bottom) of the eigenfunction $\hat u_{SI}(y_n,z)$ of  mode I with $\alpha=6.4$, for which $\bar\omega_i=0.275$. The eigenfunction is normalised by its maximum.}}}
  \label{fig:sia_growth_local_r3.0}
  \end{center}
\end{figure}
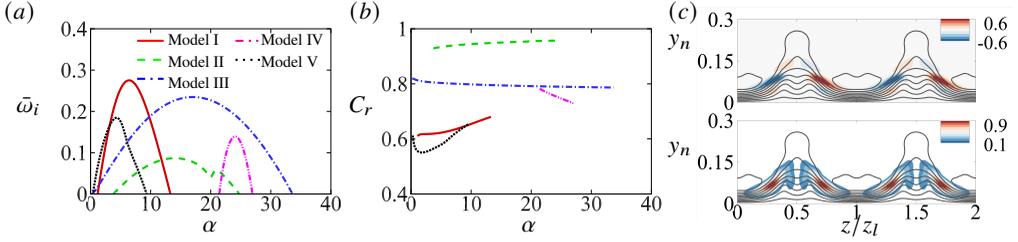
Figure \ref{fig:sia_growth_local_r3.0} displays the instability properties of the streaky base flow at $x=12$ for case E. Five unstable branches of the fundamental SI modes are identified, among which mode I is the most unstable. The phase speed of mode I ranges approximately from 0.6 to 0.7, and the eigenprofiles displayed in panel (c) are also anti-symmetric about the axes $z/z_l=0.5n$,
closely resembling those observed in figure \ref{fig:sia_growth_local_r1.2_q=0}. Remarkably, the maximum growth rate for the current profile is almost 0.3, significantly greater than that in case A. This indicates that increasing the nose radius not only enhances the receptivity of the non-modal perturbations but also promotes the amplification of secondary instability modes, consequently leading to an earlier onset of transition.

\begin{figure}
  \begin{center}
  \includegraphics[width = 0.95\textwidth]{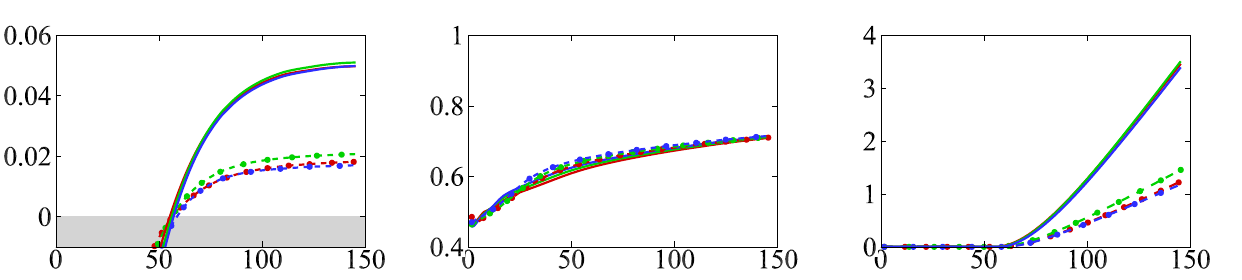}
  \put(-380,78) {$(ai)$}
  \put(-255,78) {$(aii)$}
  \put(-130,78) {$(aiii)$}
  \put(-380,40) {$-\alpha_i$}
  \put(-250,40) {$C_r$}
  \put(-125,40) {$N$}
  \put(-105,24) {\begin{tikzpicture}
    \draw[red,thick,dashed]   (0,0) -- (0.35,0);
    \node[draw,red,circle,fill,inner sep=0.4pt, line width=0.8pt] at (0.175,0) {};
    \draw[green,thick,dashed] (0,-0.25) -- (0.35,-0.25);
    \node[draw,green,circle,fill,inner sep=0.4pt, line width=0.8pt] at (0.175,-0.25) {};
    \draw[blue,thick,dashed]  (0,-0.5) -- (0.35,-0.5);
    \node[draw,blue,circle,fill,inner sep=0.4pt, line width=0.8pt] at (0.175,-0.5) {};
    \draw[red,thick]   (0,-0.75) -- (0.35,-0.75);
    \draw[green,thick] (0,-1) -- (0.35,-1);
    \draw[blue,thick]  (0,-1.25) -- (0.35,-1.25);
    \draw[decorate, decoration={brace, amplitude=3pt}, thick](0.75,0) -- (0.75,-0.5);
    \draw[decorate, decoration={brace, amplitude=3pt}, thick](0.75,-0.7) -- (0.75,-1.2);
    \end{tikzpicture}}
  \put(-105,66) {\fontsize{6pt}{6pt}\selectfont $\bar \omega=$}
  \put(-93,58) {\fontsize{6pt}{6pt}\selectfont $0.5$}
  \put(-93,51) {\fontsize{6pt}{6pt}\selectfont $0.7$}
  \put(-93,44) {\fontsize{6pt}{6pt}\selectfont $0.9$}
  \put(-93,37) {\fontsize{6pt}{6pt}\selectfont $0.6$}
  \put(-93,30) {\fontsize{6pt}{6pt}\selectfont $0.7$}
  \put(-93,23) {\fontsize{6pt}{6pt}\selectfont $0.8$}
  \put(-78,52) {\fontsize{6pt}{6pt}\selectfont $\sigma_d=0$}
  \put(-78,32) {\fontsize{6pt}{6pt}\selectfont $\sigma_d=0.5$}
  \\
  \includegraphics[width = 0.95\textwidth]{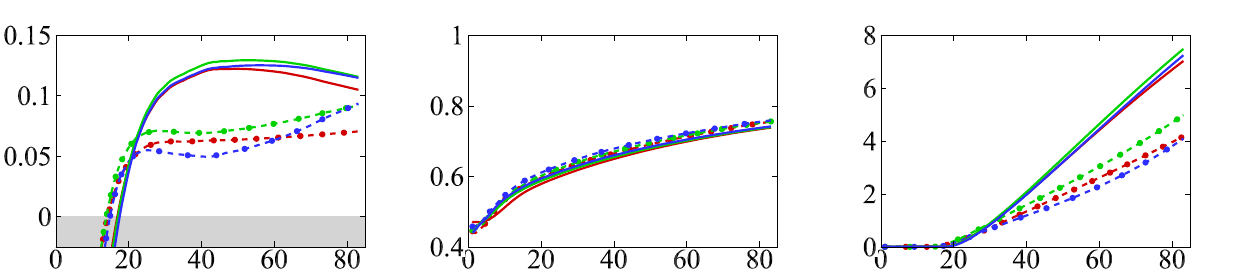}
  \put(-380,78) {$(bi)$}
  \put(-255,78) {$(bii)$}
  \put(-130,78) {$(biii)$}
  \put(-380,40) {$-\alpha_i$}
  \put(-250,40) {$C_r$}
  \put(-125,40) {$N$}
  \put(-105,24) {\begin{tikzpicture}
    \draw[red,thick,dashed]   (0,0) -- (0.35,0);
    \node[draw,red,circle,fill,inner sep=0.4pt, line width=0.8pt] at (0.175,0) {};
    \draw[green,thick,dashed] (0,-0.25) -- (0.35,-0.25);
    \node[draw,green,circle,fill,inner sep=0.4pt, line width=0.8pt] at (0.175,-0.25) {};
    \draw[blue,thick,dashed]  (0,-0.5) -- (0.35,-0.5);
    \node[draw,blue,circle,fill,inner sep=0.4pt, line width=0.8pt] at (0.175,-0.5) {};
    \draw[red,thick]   (0,-0.75) -- (0.35,-0.75);
    \draw[green,thick] (0,-1) -- (0.35,-1);
    \draw[blue,thick]  (0,-1.25) -- (0.35,-1.25);
    \draw[decorate, decoration={brace, amplitude=3pt}, thick](0.75,0) -- (0.75,-0.5);
    \draw[decorate, decoration={brace, amplitude=3pt}, thick](0.75,-0.7) -- (0.75,-1.2);
    \end{tikzpicture}}
  \put(-105,66) {\fontsize{6pt}{6pt}\selectfont $\bar \omega=$}
  \put(-93,58) {\fontsize{6pt}{6pt}\selectfont $1.0$}
  \put(-93,51) {\fontsize{6pt}{6pt}\selectfont $1.5$}
  \put(-93,44) {\fontsize{6pt}{6pt}\selectfont $2.0$}
  \put(-93,37) {\fontsize{6pt}{6pt}\selectfont $1.0$}
  \put(-93,30) {\fontsize{6pt}{6pt}\selectfont $1.3$}
  \put(-93,23) {\fontsize{6pt}{6pt}\selectfont $1.5$}
  \put(-78,52) {\fontsize{6pt}{6pt}\selectfont $\sigma_d=0$}
  \put(-78,40) {\fontsize{6pt}{6pt}\selectfont $\sigma_d=0.5$}
  \\
  \includegraphics[width = 0.95\textwidth]{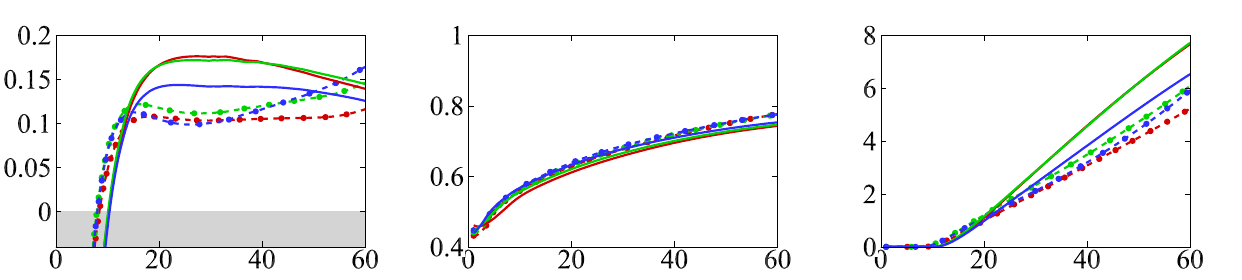}
  \put(-380,78) {$(ci)$}
  \put(-255,78) {$(cii)$}
  \put(-130,78) {$(ciii)$}
  \put(-380,40) {$-\alpha_i$}
  \put(-250,40) {$C_r$}
  \put(-125,40) {$N$}
  \put(-105,24) {\begin{tikzpicture}
    \draw[red,thick,dashed]   (0,0) -- (0.35,0);
    \node[draw,red,circle,fill,inner sep=0.4pt, line width=0.8pt] at (0.175,0) {};
    \draw[green,thick,dashed] (0,-0.25) -- (0.35,-0.25);
    \node[draw,green,circle,fill,inner sep=0.4pt, line width=0.8pt] at (0.175,-0.25) {};
    \draw[blue,thick,dashed]  (0,-0.5) -- (0.35,-0.5);
    \node[draw,blue,circle,fill,inner sep=0.4pt, line width=0.8pt] at (0.175,-0.5) {};
    \draw[red,thick]   (0,-0.75) -- (0.35,-0.75);
    \draw[green,thick] (0,-1) -- (0.35,-1);
    \draw[blue,thick]  (0,-1.25) -- (0.35,-1.25);
    \draw[decorate, decoration={brace, amplitude=3pt}, thick](0.75,0) -- (0.75,-0.5);
    \draw[decorate, decoration={brace, amplitude=3pt}, thick](0.75,-0.7) -- (0.75,-1.2);
    \end{tikzpicture}}
  \put(-105,66) {\fontsize{6pt}{6pt}\selectfont $\bar \omega=$}
  \put(-93,58) {\fontsize{6pt}{6pt}\selectfont $1.5$}
  \put(-93,51) {\fontsize{6pt}{6pt}\selectfont $2.0$}
  \put(-93,44) {\fontsize{6pt}{6pt}\selectfont $2.5$}
  \put(-93,37) {\fontsize{6pt}{6pt}\selectfont $1.5$}
  \put(-93,30) {\fontsize{6pt}{6pt}\selectfont $2.0$}
  \put(-93,23) {\fontsize{6pt}{6pt}\selectfont $2.5$}
  \put(-78,52) {\fontsize{6pt}{6pt}\selectfont $\sigma_d=0$}
  \put(-78,40) {\fontsize{6pt}{6pt}\selectfont $\sigma_d=0.5$}
  \\
  \includegraphics[width = 0.95\textwidth]{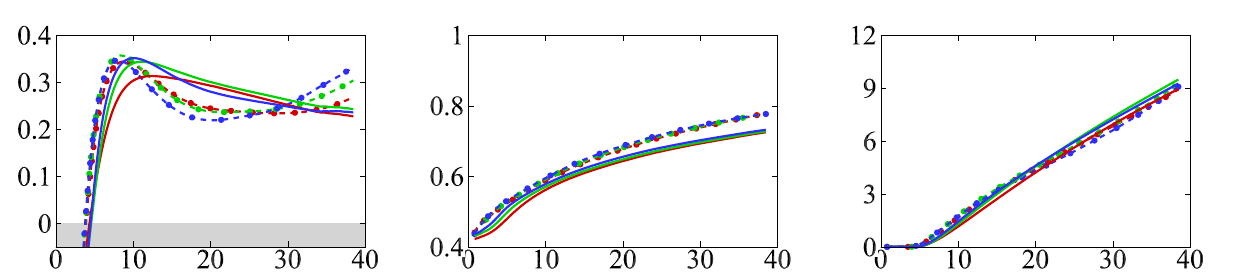}
  \put(-380,78) {$(di)$}
  \put(-255,78) {$(dii)$}
  \put(-130,78) {$(diii)$}
  \put(-380,40) {$-\alpha_i$}
  \put(-250,40) {$C_r$}
  \put(-125,40) {$N$}
  \put(-105,24) {\begin{tikzpicture}
    \draw[red,thick,dashed]   (0,0) -- (0.35,0);
    \node[draw,red,circle,fill,inner sep=0.4pt, line width=0.8pt] at (0.175,0) {};
    \draw[green,thick,dashed] (0,-0.25) -- (0.35,-0.25);
    \node[draw,green,circle,fill,inner sep=0.4pt, line width=0.8pt] at (0.175,-0.25) {};
    \draw[blue,thick,dashed]  (0,-0.5) -- (0.35,-0.5);
    \node[draw,blue,circle,fill,inner sep=0.4pt, line width=0.8pt] at (0.175,-0.5) {};
    \draw[red,thick]   (0,-0.75) -- (0.35,-0.75);
    \draw[green,thick] (0,-1) -- (0.35,-1);
    \draw[blue,thick]  (0,-1.25) -- (0.35,-1.25);
    \draw[decorate, decoration={brace, amplitude=3pt}, thick](0.75,0) -- (0.75,-0.5);
    \draw[decorate, decoration={brace, amplitude=3pt}, thick](0.75,-0.7) -- (0.75,-1.2);
    \end{tikzpicture}}
  \put(-105,66) {\fontsize{6pt}{6pt}\selectfont $\bar \omega=$}
  \put(-93,58) {\fontsize{6pt}{6pt}\selectfont $3.0$}
  \put(-93,51) {\fontsize{6pt}{6pt}\selectfont $3.5$}
  \put(-93,44) {\fontsize{6pt}{6pt}\selectfont $4.0$}
  \put(-93,37) {\fontsize{6pt}{6pt}\selectfont $2.0$}
  \put(-93,30) {\fontsize{6pt}{6pt}\selectfont $2.5$}
  \put(-93,23) {\fontsize{6pt}{6pt}\selectfont $3.0$}
  \put(-78,52) {\fontsize{6pt}{6pt}\selectfont $\sigma_d=0$}
  \put(-78,40) {\fontsize{6pt}{6pt}\selectfont $\sigma_d=0.5$}
  \\
  \includegraphics[width = 0.95\textwidth]{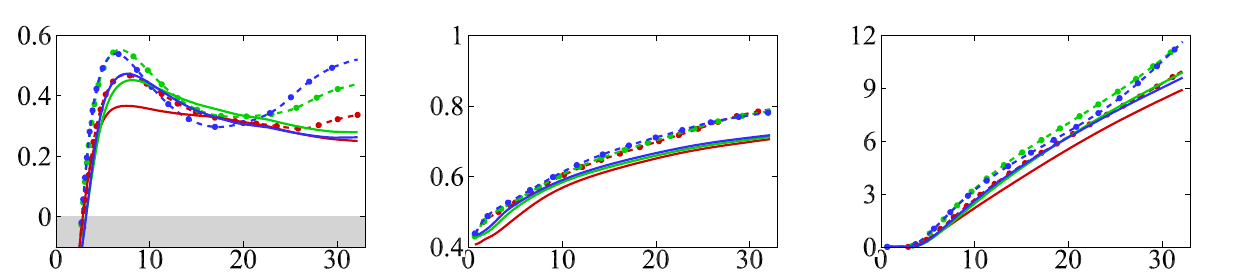}
  \put(-380,78) {$(ei)$}
  \put(-255,78) {$(eii)$}
  \put(-130,78) {$(eiii)$}
  \put(-380,40) {$-\alpha_i$}
  \put(-250,40) {$C_r$}
  \put(-125,40) {$N$}
  \put(-310,-3){$x$}  
  \put(-188,-3){$x$}
  \put(-68,-3){$x$}
  \put(-105,26) {\begin{tikzpicture}
    \draw[red,thick,dashed]   (0,0) -- (0.35,0);
    \node[draw,red,circle,fill,inner sep=0.4pt, line width=0.8pt] at (0.175,0) {};
    \draw[green,thick,dashed] (0,-0.25) -- (0.35,-0.25);
    \node[draw,green,circle,fill,inner sep=0.4pt, line width=0.8pt] at (0.175,-0.25) {};
    \draw[blue,thick,dashed]  (0,-0.5) -- (0.35,-0.5);
    \node[draw,blue,circle,fill,inner sep=0.4pt, line width=0.8pt] at (0.175,-0.5) {};
    \draw[red,thick]   (0,-0.75) -- (0.35,-0.75);
    \draw[green,thick] (0,-1) -- (0.35,-1);
    \draw[blue,thick]  (0,-1.25) -- (0.35,-1.25);
    \draw[decorate, decoration={brace, amplitude=3pt}, thick](0.75,0) -- (0.75,-0.5);
    \draw[decorate, decoration={brace, amplitude=3pt}, thick](0.75,-0.7) -- (0.75,-1.2);
    \end{tikzpicture}}
  \put(-105,66) {\fontsize{6pt}{6pt}\selectfont $\bar \omega=$}
  \put(-93,60) {\fontsize{6pt}{6pt}\selectfont $3.0$}
  \put(-93,53) {\fontsize{6pt}{6pt}\selectfont $4.0$}
  \put(-93,46) {\fontsize{6pt}{6pt}\selectfont $5.0$}
  \put(-93,39) {\fontsize{6pt}{6pt}\selectfont $3.0$}
  \put(-93,32) {\fontsize{6pt}{6pt}\selectfont $3.5$}
  \put(-93,25) {\fontsize{6pt}{6pt}\selectfont $4.0$}
  \put(-78,54) {\fontsize{6pt}{6pt}\selectfont $\sigma_d=0$}
  \put(-80,44) {\fontsize{6pt}{6pt}\selectfont $\sigma_d=0.5$}  
  \caption{Streamwise evolution of the spatial instability properties obtained from fundamental {$(\sigma_d=0)$} and subharmonic {$(\sigma_d=0.5)$} SI analyses {for noise level {$Q=1.75\%$}}, where representative frequencies {$\bar \omega$} of the SI modes are considered. Left column: spatial growth rate $-\alpha_i$ of the most unstable mode; middle column:   phase speed $C_r$; right column: $N$-factor. From top row to bottom row: case A to case  E. The {gray shaded} area in the left column indicates the stable area $-\alpha_i<0$.} 
  \label{fig:sia_growth_modeI_xs_rs}
  \end{center}
\end{figure}
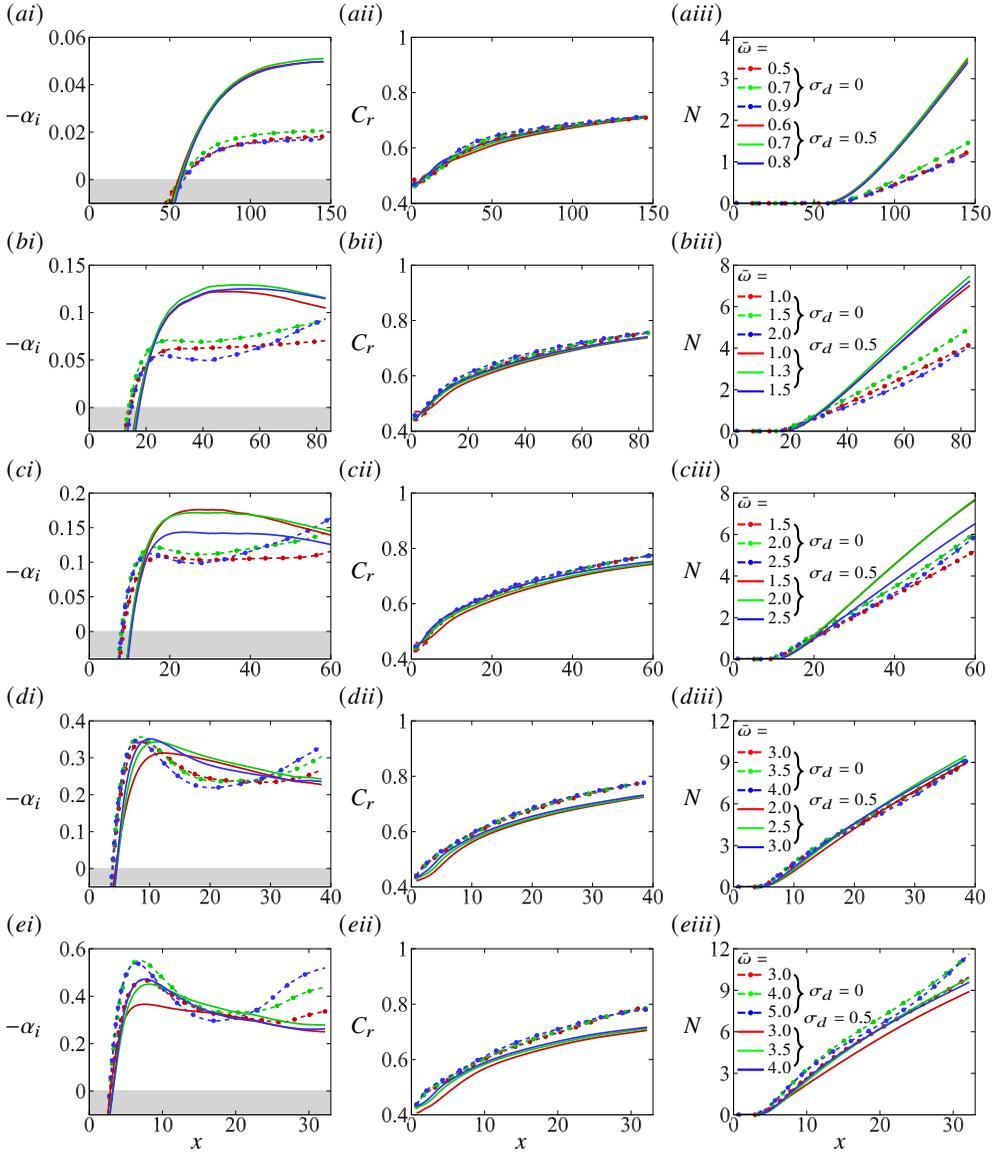
Using the looping procedure introduced in $\S$\ref{subsubsec:SIA} to map our temporal-growth results into the spatial-growth results, the left column of figure \ref{fig:sia_growth_modeI_xs_rs} plots the streamwise evolution of the spatial growth rates $-\alpha_i$ for all five cases with {$Q=1.75\%$}.  We show both the fundamental and subharmonic SI modes, each evaluated at three representative frequencies.  Every curve features a neutral point, indicating the onset of SI, followed by a rapid increase in growth rate immediately downstream.  After reaching a peak, $-\alpha_i$ decreases gradually with $x$; for the fundamental mode, it then exhibits a second rise farther downstream.
For the smallest bluntness (cases A and B), subharmonic modes remain more amplified than the fundamentals throughout the domain of interest.  In cases C and D, however, the  growth rates of the fundamental modes overtake those of the subharmonic modes in the far-downstream region, and in the bluntest case E, the fundamental SI modes are dominant almost everywhere.
The middle column of figure \ref{fig:sia_growth_modeI_xs_rs} shows the corresponding phase speeds $C_r$, which increase monotonically with $x$ from about 0.4 to 0.8.  Finally, the right column presents the evolution of the amplification factor $N$ defined in (\ref{eq:N_steady}). For each case, the $N$ factor grows monotonically in the region downstream of the neutral position. These $N$-$x$ curves form the basis for our bypass-transition predictions in the next subsection.

\begin{figure}
  \begin{center}
  \includegraphics[width = 0.95\textwidth]{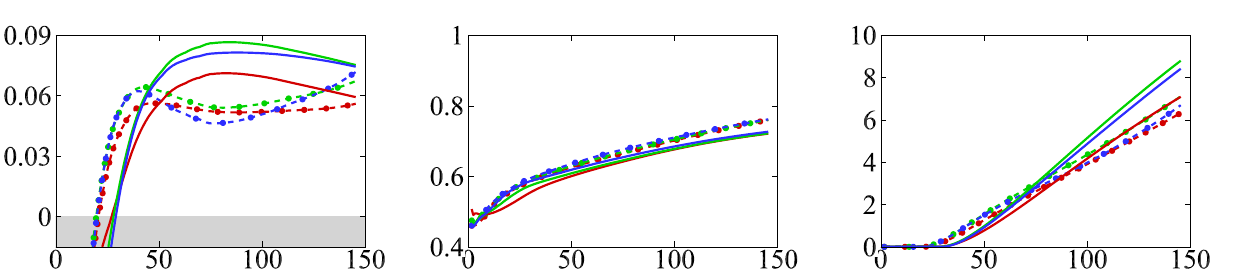}
  \put(-380,78) {$(ai)$}
  \put(-255,78) {$(aii)$}
  \put(-130,78) {$(aiii)$}
  \put(-380,40) {$-\alpha_i$}
  \put(-250,40) {$C_r$}
  \put(-125,40) {$N$}
  \put(-105,26) {\begin{tikzpicture}
    \draw[red,thick,dashed]   (0,0) -- (0.35,0);
    \node[draw,red,circle,fill,inner sep=0.4pt, line width=0.8pt] at (0.175,0) {};
    \draw[green,thick,dashed] (0,-0.25) -- (0.35,-0.25);
    \node[draw,green,circle,fill,inner sep=0.4pt, line width=0.8pt] at (0.175,-0.25) {};
    \draw[blue,thick,dashed]  (0,-0.5) -- (0.35,-0.5);
    \node[draw,blue,circle,fill,inner sep=0.4pt, line width=0.8pt] at (0.175,-0.5) {};
    \draw[red,thick]   (0,-0.75) -- (0.35,-0.75);
    \draw[green,thick] (0,-1) -- (0.35,-1);
    \draw[blue,thick]  (0,-1.25) -- (0.35,-1.25);
    \draw[decorate, decoration={brace, amplitude=3pt}, thick](0.75,0) -- (0.75,-0.5);
    \draw[decorate, decoration={brace, amplitude=3pt}, thick](0.75,-0.7) -- (0.75,-1.2);
    \end{tikzpicture}}
  \put(-105,66) {\fontsize{6pt}{6pt}\selectfont $\bar \omega=$}
  \put(-93,60) {\fontsize{6pt}{6pt}\selectfont $0.6$}
  \put(-93,53) {\fontsize{6pt}{6pt}\selectfont $0.8$}
  \put(-93,46) {\fontsize{6pt}{6pt}\selectfont $1.0$}
  \put(-93,39) {\fontsize{6pt}{6pt}\selectfont $0.4$}
  \put(-93,32) {\fontsize{6pt}{6pt}\selectfont $0.6$}
  \put(-93,25) {\fontsize{6pt}{6pt}\selectfont $0.8$}
  \put(-78,54) {\fontsize{6pt}{6pt}\selectfont $\sigma_d=0$}
  \put(-80,40) {\fontsize{6pt}{6pt}\selectfont $\sigma_d=0.5$}
  \\
  \includegraphics[width = 0.95\textwidth]{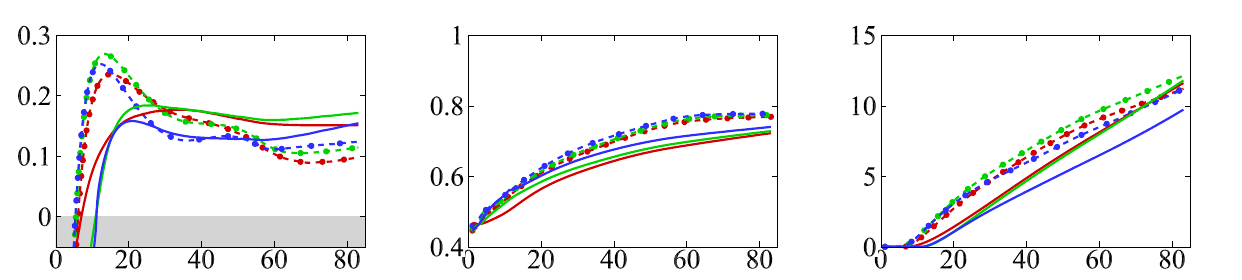}
  \put(-380,78) {$(bi)$}
  \put(-255,78) {$(bii)$}
  \put(-130,78) {$(biii)$}
  \put(-380,40) {$-\alpha_i$}
  \put(-250,40) {$C_r$}
  \put(-125,40) {$N$}
  \put(-105,27) {\begin{tikzpicture}
    \draw[red,thick,dashed]   (0,0) -- (0.35,0);
    \node[draw,red,circle,fill,inner sep=0.4pt, line width=0.8pt] at (0.175,0) {};
    \draw[green,thick,dashed] (0,-0.25) -- (0.35,-0.25);
    \node[draw,green,circle,fill,inner sep=0.4pt, line width=0.8pt] at (0.175,-0.25) {};
    \draw[blue,thick,dashed]  (0,-0.5) -- (0.35,-0.5);
    \node[draw,blue,circle,fill,inner sep=0.4pt, line width=0.8pt] at (0.175,-0.5) {};
    \draw[red,thick]   (0,-0.75) -- (0.35,-0.75);
    \draw[green,thick] (0,-1) -- (0.35,-1);
    \draw[blue,thick]  (0,-1.25) -- (0.35,-1.25);
    \draw[decorate, decoration={brace, amplitude=3pt}, thick](0.75,0) -- (0.75,-0.5);
    \draw[decorate, decoration={brace, amplitude=3pt}, thick](0.75,-0.7) -- (0.75,-1.2);
    \end{tikzpicture}}
  \put(-105,66) {\fontsize{6pt}{6pt}\selectfont $\bar \omega=$}
  \put(-93,60) {\fontsize{6pt}{6pt}\selectfont $1.5$}
  \put(-93,53) {\fontsize{6pt}{6pt}\selectfont $2.0$}
  \put(-93,46) {\fontsize{6pt}{6pt}\selectfont $2.5$}
  \put(-93,39) {\fontsize{6pt}{6pt}\selectfont $1.0$}
  \put(-93,32) {\fontsize{6pt}{6pt}\selectfont $1.5$}
  \put(-93,25) {\fontsize{6pt}{6pt}\selectfont $2.0$}
  \put(-78,54) {\fontsize{6pt}{6pt}\selectfont $\sigma_d=0$}
  \put(-82,44) {\fontsize{6pt}{6pt}\selectfont $\sigma_d=0.5$}
  \\
  \includegraphics[width = 0.95\textwidth]{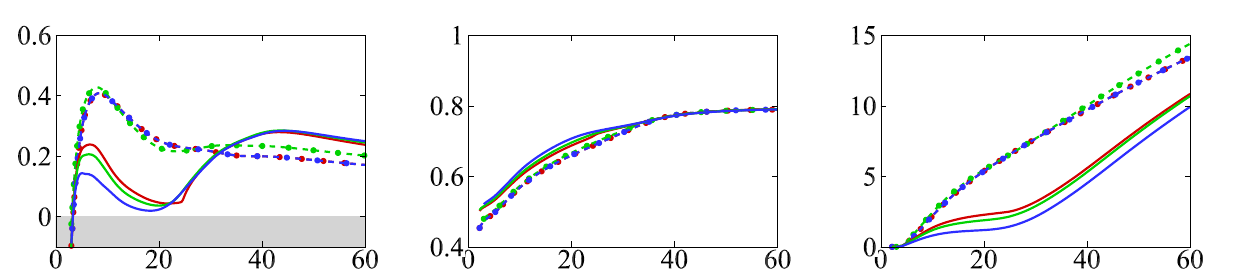}
  \put(-380,78) {$(ci)$}
  \put(-255,78) {$(cii)$}
  \put(-130,78) {$(ciii)$}
  \put(-380,40) {$-\alpha_i$}
  \put(-250,40) {$C_r$}
  \put(-125,40) {$N$}
  \put(-105,27) {\begin{tikzpicture}
    \draw[red,thick,dashed]   (0,0) -- (0.35,0);
    \node[draw,red,circle,fill,inner sep=0.4pt, line width=0.8pt] at (0.175,0) {};
    \draw[green,thick,dashed] (0,-0.25) -- (0.35,-0.25);
    \node[draw,green,circle,fill,inner sep=0.4pt, line width=0.8pt] at (0.175,-0.25) {};
    \draw[blue,thick,dashed]  (0,-0.5) -- (0.35,-0.5);
    \node[draw,blue,circle,fill,inner sep=0.4pt, line width=0.8pt] at (0.175,-0.5) {};
    \draw[red,thick]   (0,-0.75) -- (0.35,-0.75);
    \draw[green,thick] (0,-1) -- (0.35,-1);
    \draw[blue,thick]  (0,-1.25) -- (0.35,-1.25);
    \draw[decorate, decoration={brace, amplitude=3pt}, thick](0.75,0) -- (0.75,-0.5);
    \draw[decorate, decoration={brace, amplitude=3pt}, thick](0.75,-0.7) -- (0.75,-1.2);
    \end{tikzpicture}}
  \put(-105,66) {\fontsize{6pt}{6pt}\selectfont $\bar \omega=$}
  \put(-93,61) {\fontsize{6pt}{6pt}\selectfont $2.5$}
  \put(-93,54) {\fontsize{6pt}{6pt}\selectfont $3.0$}
  \put(-93,47) {\fontsize{6pt}{6pt}\selectfont $3.5$}
  \put(-93,40) {\fontsize{6pt}{6pt}\selectfont $4.0$}
  \put(-93,33) {\fontsize{6pt}{6pt}\selectfont $4.5$}
  \put(-93,26) {\fontsize{6pt}{6pt}\selectfont $5.0$}
  \put(-78,55) {\fontsize{6pt}{6pt}\selectfont $\sigma_d=0$}
  \put(-82,44) {\fontsize{6pt}{6pt}\selectfont $\sigma_d=0.5$}
  \\
  \includegraphics[width = 0.95\textwidth]{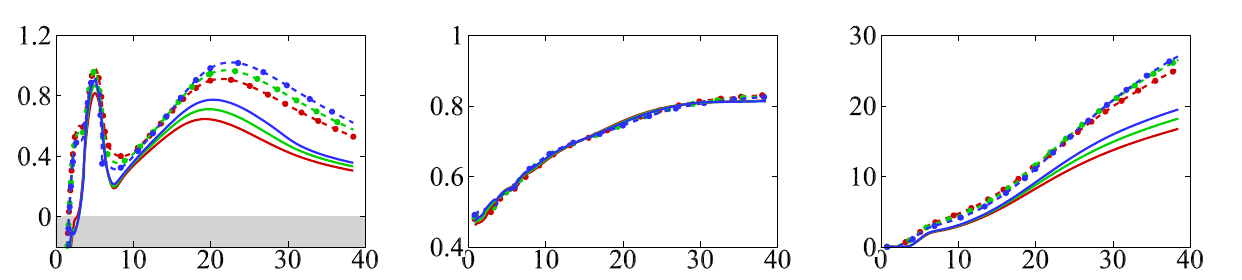}
  \put(-380,78) {$(di)$}
  \put(-255,78) {$(dii)$}
  \put(-130,78) {$(diii)$}
  \put(-380,40) {$-\alpha_i$}
  \put(-250,40) {$C_r$}
  \put(-125,40) {$N$}
  \put(-105,26) {\begin{tikzpicture}
    \draw[red,thick,dashed]   (0,0) -- (0.35,0);
    \node[draw,red,circle,fill,inner sep=0.4pt, line width=0.8pt] at (0.175,0) {};
    \draw[green,thick,dashed] (0,-0.25) -- (0.35,-0.25);
    \node[draw,green,circle,fill,inner sep=0.4pt, line width=0.8pt] at (0.175,-0.25) {};
    \draw[blue,thick,dashed]  (0,-0.5) -- (0.35,-0.5);
    \node[draw,blue,circle,fill,inner sep=0.4pt, line width=0.8pt] at (0.175,-0.5) {};
    \draw[red,thick]   (0,-0.75) -- (0.35,-0.75);
    \draw[green,thick] (0,-1) -- (0.35,-1);
    \draw[blue,thick]  (0,-1.25) -- (0.35,-1.25);
    \draw[decorate, decoration={brace, amplitude=3pt}, thick](0.75,0) -- (0.75,-0.5);
    \draw[decorate, decoration={brace, amplitude=3pt}, thick](0.75,-0.7) -- (0.75,-1.2);
    \end{tikzpicture}}
  \put(-105,66) {\fontsize{6pt}{6pt}\selectfont $\bar \omega=$}
  \put(-93,60) {\fontsize{6pt}{6pt}\selectfont $6.0$}
  \put(-93,53) {\fontsize{6pt}{6pt}\selectfont $6.5$}
  \put(-93,46) {\fontsize{6pt}{6pt}\selectfont $7.0$}
  \put(-93,39) {\fontsize{6pt}{6pt}\selectfont $4.5$}
  \put(-93,32) {\fontsize{6pt}{6pt}\selectfont $5.0$}
  \put(-93,25) {\fontsize{6pt}{6pt}\selectfont $5.5$}
  \put(-78,54) {\fontsize{6pt}{6pt}\selectfont $\sigma_d=0$}
  \put(-80,42) {\fontsize{6pt}{6pt}\selectfont $\sigma_d=0.5$}
  \\
  \includegraphics[width = 0.95\textwidth]{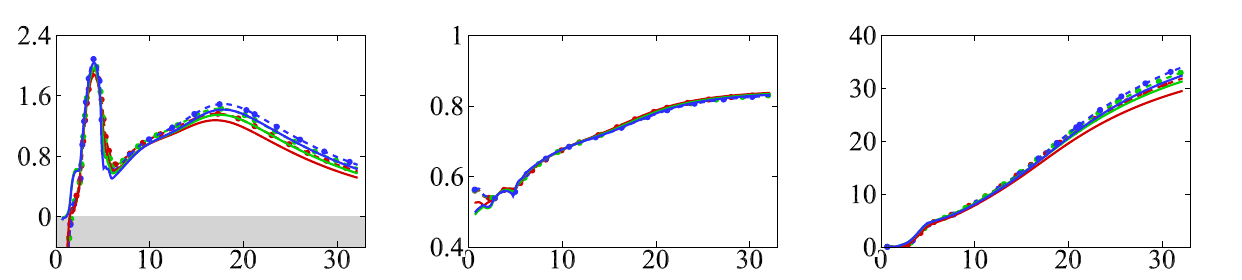}
  \put(-380,78) {$(ei)$}
  \put(-255,78) {$(eii)$}
  \put(-130,78) {$(eiii)$}
  \put(-380,40) {$-\alpha_i$}
  \put(-250,40) {$C_r$}
  \put(-125,40) {$N$}
  \put(-310,-3){$x$}  
  \put(-188,-3){$x$}
  \put(-68,-3){$x$}
  \put(-105,26) {\begin{tikzpicture}
    \draw[red,thick,dashed]   (0,0) -- (0.35,0);
    \node[draw,red,circle,fill,inner sep=0.4pt, line width=0.8pt] at (0.175,0) {};
    \draw[green,thick,dashed] (0,-0.25) -- (0.35,-0.25);
    \node[draw,green,circle,fill,inner sep=0.4pt, line width=0.8pt] at (0.175,-0.25) {};
    \draw[blue,thick,dashed]  (0,-0.5) -- (0.35,-0.5);
    \node[draw,blue,circle,fill,inner sep=0.4pt, line width=0.8pt] at (0.175,-0.5) {};
    \draw[red,thick]   (0,-0.75) -- (0.35,-0.75);
    \draw[green,thick] (0,-1) -- (0.35,-1);
    \draw[blue,thick]  (0,-1.25) -- (0.35,-1.25);
    \draw[decorate, decoration={brace, amplitude=3pt}, thick](0.75,0) -- (0.75,-0.5);
    \draw[decorate, decoration={brace, amplitude=3pt}, thick](0.75,-0.7) -- (0.75,-1.2);
    \end{tikzpicture}}
  \put(-105,66) {\fontsize{6pt}{6pt}\selectfont $\bar \omega=$}
  \put(-93,60) {\fontsize{6pt}{6pt}\selectfont $7.5$}
  \put(-93,53) {\fontsize{6pt}{6pt}\selectfont $8.0$}
  \put(-93,46) {\fontsize{6pt}{6pt}\selectfont $8.5$}
  \put(-93,39) {\fontsize{6pt}{6pt}\selectfont $7.5$}
  \put(-93,32) {\fontsize{6pt}{6pt}\selectfont $8.0$}
  \put(-93,25) {\fontsize{6pt}{6pt}\selectfont $8.5$}
  \put(-78,54) {\fontsize{6pt}{6pt}\selectfont $\sigma_d=0$}
  \put(-80,42) {\fontsize{6pt}{6pt}\selectfont $\sigma_d=0.5$}  
  \caption{Streamwise evolution of the spatial instability properties obtained from fundamental {$(\sigma_d=0)$} and subharmonic {$(\sigma_d=0.5)$} SI analyses {for noise level $Q=3.5\%$} , where the line types are identical to figure \ref{fig:sia_growth_modeI_xs_rs}.}  \label{fig:sia_growth_modeI_xs_rs_2}
  \end{center}
\end{figure}
Increasing the noise level to {$Q=3.5\%$}, we perform the same {BSA} and obtain the streamwise evolution of the instability and {$N$} factor shown in figure \ref{fig:sia_growth_modeI_xs_rs_2}. While the phase speeds of the secondary instability modes remain almost identical to those for {$Q=1.75\%$}, the growth rate and {$N$} factor are remarkably enhanced.

\subsection{Comparison with experimental data}\label{subsec:comparison}
\begin{figure}
  \begin{center}
    \includegraphics[width = 0.9\textwidth]{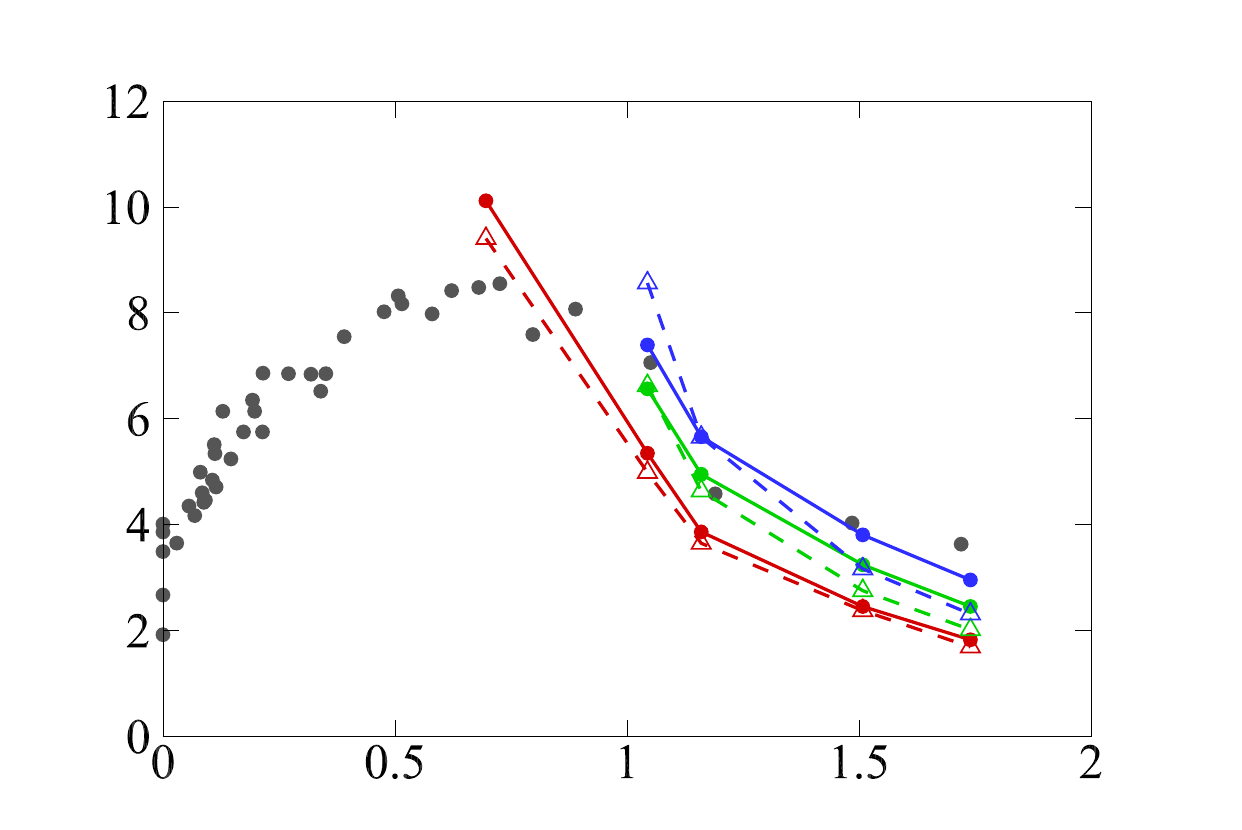}
  \put(-330,90){{\rotatebox{90}{$Re_{\mathrm{tr}}(\times 10^6)$}}}
  \put(-185,0){{$Re(\times 10^5)$}}
  \put(-150,133) {\begin{tikzpicture}
    \node[draw,gray,circle,fill,inner sep=1.2pt, line width=0.8pt] at (0.4,0) {};
    \draw[red,thick]   (0,-0.35) -- (0.8,-0.35);
    \node[draw,red,circle,fill,inner sep=1.2pt, line width=0.8pt] at (0.4,-0.35) {};
    \draw[green,thick] (0,-0.7) -- (0.8,-0.7);
    \node[draw,green,circle,fill,inner sep=1.2pt, line width=0.8pt] at (0.4,-0.7) {};
    \draw[blue,thick]  (0,-1.05) -- (0.8,-1.05);
    \node[draw,blue,circle,fill,inner sep=1.2pt, line width=0.8pt] at (0.4,-1.05) {};
    \draw[red,dashed,thick]   (0,-1.4) -- (0.8,-1.4);
    \draw[green,dashed,thick] (0,-1.75) -- (0.8,-1.75);
    \draw[blue,dashed,thick]  (0,-2.1) -- (0.8,-2.1);
    \end{tikzpicture}}
  \put(-90,129) {\begin{tikzpicture}
    \draw[decorate, decoration={brace, amplitude=3pt}, thick](2.3,0) -- (2.3,-0.9);
    \draw[decorate, decoration={brace, amplitude=3pt}, thick](2.3,-1.1) -- (2.3,-2.0);
    \end{tikzpicture}}
  \put(-142,151) {\begin{tikzpicture}
    \path[draw=red, line width=0.6pt] (0, 0) -- (-3pt, -5.2pt) -- (3pt, -5.2pt) -- cycle;
    \end{tikzpicture}}  
  \put(-142,141) {\begin{tikzpicture}
    \path[draw=green, line width=0.6pt] (0, 0) -- (-3pt, -5.2pt) -- (3pt, -5.2pt) -- cycle;
    \end{tikzpicture}} 
  \put(-142,131) {\begin{tikzpicture}
    \path[draw=blue, line width=0.6pt] (0, 0) -- (-3pt, -5.2pt) -- (3pt, -5.2pt) -- cycle;
    \end{tikzpicture}} 
  \put(-125,190){Borovoy et al. (2022)}
  \put(-125,180){$N_{\mathrm{tr}}=3.5$}
  \put(-125,170){$N_{\mathrm{tr}}=5$}
  \put(-125,160){$N_{\mathrm{tr}}=6$}
  \put(-125,150){$N_{\mathrm{tr}}=8$}
  \put(-125,140){$N_{\mathrm{tr}}=10$}
  \put(-125,130){$N_{\mathrm{tr}}=12$}
  \put(-85,174){$Q=1.75\%$}
  \put(-85,142){$Q=3.5\%$}
  \caption{Comparison of the transition Reynolds number $Re_{\mathrm {tr}}$ predicted by the e-N approach (continuous curves) with the experimental data (black circles).}
  \label{fig:sia_transition_eN}
  \end{center}
\end{figure}
Based on amplitude amplification factor $N$ predicted  in the right columns of {figures \ref{fig:sia_growth_modeI_xs_rs} and \ref{fig:sia_growth_modeI_xs_rs_2}}, the onset of bypass transition can be estimated using a suitable  threshold {$N_{\mathrm {tr}}$}. Since direct calculation of the receptivity of the secondary instabilities in a streaky boundary layer remains infeasible, we assume, on the basis of nearly identical wind‐tunnel background noise levels for different runs, that the SI‐mode receptivity efficiency is comparable across all nose‐radius cases.  While this assumption inevitably introduces some uncertainty,  it allows us to use a single threshold $N_{\rm tr}$ for all nose radii with the same noise level. 

To validate our prediction of  bypass transition, the threshold of $N$ factor $N_{\rm tr}$ is calibrated using the experimental data \citep{borovoy2022laminar} for a moderate-bluntness configuration (case C), yielding values of 5 and 10 for {$Q=1.75\%$ and  $Q=3.5\%$}, respectively. Using these thresholds,  the predicted transition Reynolds numbers are plotted as the solid and dashed green curves in figure \ref{fig:sia_transition_eN}. Note that for case A, the maximum $N$ factor throughout the computational domain remains below $N_{\rm tr}$, and thus no transition location is displayed. In both curves, increasing the  Reynolds number or nose bluntness leads to an upstream shift of the transition location, resembling the experimental observation for large bluntness configurations. 
To assess the robustness of the prediction framework, two alternative $N_{\rm tr}$ values are tested for each noise level:  $N_{\rm tr}=$ 3.5 and 6 for {$Q=1.75\%$} and $N_{\rm tr}=$ 8 and 12 for {$Q=3.5\%$}. Due to the strong amplification rates of the SI modes, a $\pm 20\%$ variation in   $N_{\rm tr}$ leads to only limited changes in the predicted transition location. All  curves show general agreement with experimental measurements for  $Re\gtrsim0.7\times10^5$, the regime in which transition reversal is observed. Notably, even when the assumed background noise level is doubled from {$1.75\%$ to $3.5\%$}, our predictive framework remains valid once an appropriate $N_{\rm tr}$ is selected.
The remaining quantitative discrepancies may stem from slight variations in SI receptivity that is not accounted for in the model, or from the uncertainties in the estimated spectral distribution of the background noise.

\section{Concluding remarks and discussions}\label{sec:Conclusion}
Accurate prediction of laminar–turbulent transition remains a fundamental challenge, owing to various influencing factors such as the properties of  the dominant disturbances, their receptivity into the boundary layer, and the ensuing nonlinear interactions. In natural transitions, the laminar phase is dominated by the exponential growth of normal instability modes, whose growth rates, as predicted by  linear stability analysis, govern their amplitude evolution. Neglecting the receptivity and nonlinear saturation processes, the conventional e-N method introduced an empirical threshold, the $N$ factor, to model the entire laminar phase, thus providing a convenient framework for  predicting natural transition. By contrast, bypass transition follows a more intricate sequence: (1) freestream disturbances excite non-modal, streaky perturbations in the boundary layer (receptivity); (2)
these perturbations undergo transient growth and reach a nonlinear saturation phase (nonlinear evolution); (3) high-growth-rate secondary instability modes then develop on the saturated streaks (secondary instability), spawning turbulence spots and leading to streak breakdown.
Because bypass transition is not driven solely by the linear amplification of primary  perturbations, the classical e-N method by itself cannot reliably predict the transition onset. To address this issue, we propose a hybrid prediction framework that explicitly accounts for all three stages of bypass transition.

In hypersonic boundary‐layer transition over a blunt body, the initial stage, the receptivity of non‐modal perturbations, is particularly complex because of the impact of the detached bow shock and the  entropy layer. This complexity arises from two main challenges: (i) perturbation development in the nose-region entropy layer exhibits strong non‐parallel effects, rendering conventional linear stability analysis and parabolised stability equations inapplicable; (2)
quantifying the interaction between freestream disturbances and the bow shock is inherently difficult.
Thanks to the recent advances by \cite{zhao2025excitation}, we can now employ the high-efficient, high‐accuracy SF-HLNS approach to quantify this receptivity process and to identify the key parameters of the oncoming forcing that maximize the energy amplification of linear non-modal perturbations for a given blunt body. Notably, increased bluntness enhances receptivity efficiency, indicating a shortened streamwise distance required  to reach nonlinear saturation.

Since the SF-HLNS framework neglects nonlinear effects, we extend  \cite{zhao2025excitation} by applying the NPSE to calculate the downstream evolution of non-modal perturbations in regions where non-parallelism of the base flow becomes weak. We validate this approach by comparing LPSE predictions with SF-HLNS results in the same downstream domain. While both methods agree on linear amplification, NPSE uniquely captures the ensuing nonlinear interactions among different Fourier components. As a result, a distinct mushroom  structure develops during the nonlinear phase, characterised by an earlier onset and a thinner neck (indicating strong effects of high-order harmonics) as the nose radius increases.

Predicting bypass transition requires more than tracking the low‐frequency evolution of non‐modal perturbations. As shown in  many simulations \citep{Wu2023new}, small‐scale turbulent spots emerge  in the late-phase of the streaky structures, driven by rapidly growing secondary instabilities. To capture this, we carry out the BSA on the streaky base flow, revealing that stronger velocity and temperature gradients significantly boost secondary‐instability growth rates. By integrating these growth rates downstream from their neutral points, we obtain the amplification factor $N$, and the transition onset is determined when $N$ reaches a certain threshold.

The three‐stage hybrid framework thus provides a complete methodology for predicting bypass transition in hypersonic boundary layers over blunt bodies. To validate it, we selected five cases from the experimental conditions of \cite{borovoy2022laminar} and applied our hybrid approach. Qualitatively, the numerical results reproduce the transition reversal phenomenon: beyond a certain nose radius, increasing bluntness causes transition to move upstream. More importantly, the calculated transition onsets agree overall with the experimental data, as shown in Figure \ref{fig:sia_transition_eN}.

Finally, we note that the transition predictions in this study are subject to uncertainties stemming from two aspects: (1) the input amplitude of the freestream  forcing, and (2) the criterion of the $N$ factor threshold used to determine the transition onset. To address the first uncertainty, two representative noise levels,  {$Q=1.75\%$ and $3.5\%$}, are adopted. The predicted transition Reynolds numbers (or transition locations) remain in reasonable agreement with experimental measurements once an appropriate $N_{\rm tr}$ threshold is selected. To assess the influence of the second uncertainty, the threshold $N_{\rm tr}$ is varied by $\pm 20\%$ for each noise level. The resulting predictions continue to align satisfactorily with experimental trends, supporting the robustness of our predictive framework. Nevertheless, we emphasize that direct experimental characterization of the freestream perturbation spectrum is essential for reliable interpretation of   transition experiments. In  conventional wind tunnels, where the freestream perturbations are typically dominated by slow acoustic waves radiated from the turbulent boundary layers on the tunnel walls, measurements of spatiotemporal spectra of pressure perturbation  alone are often sufficient. In contrast, in quiet wind tunnels, where the acoustic waves are significantly suppressed by  suction at the nozzle, the type of the dominant perturbations must be identified by measuring not only pressure perturbation, but also density and/or velocity perturbations.  The second uncertainty, related to the threshold, calls for a dedicated  investigation of the receptivity  of the secondary instabilities, which represents a clear direction for our subsequent  work.

\backsection[Funding]{QS and LZ are supported by NSFC (no. 12372222). MD is supported by NSFC (nos. 12588201,92371104), the CAS Strategic Priority Research Program (no. XDB0620102) and CAS project for Young Scientists in Basic Research (YSBR-087).}

\backsection[Declaration of interests]{The authors report no conflict of interest.}

\appendix
\section{Resolution study}\label{app:A}
\begin{figure}
  \begin{center}
  \includegraphics[width = 0.49\textwidth]{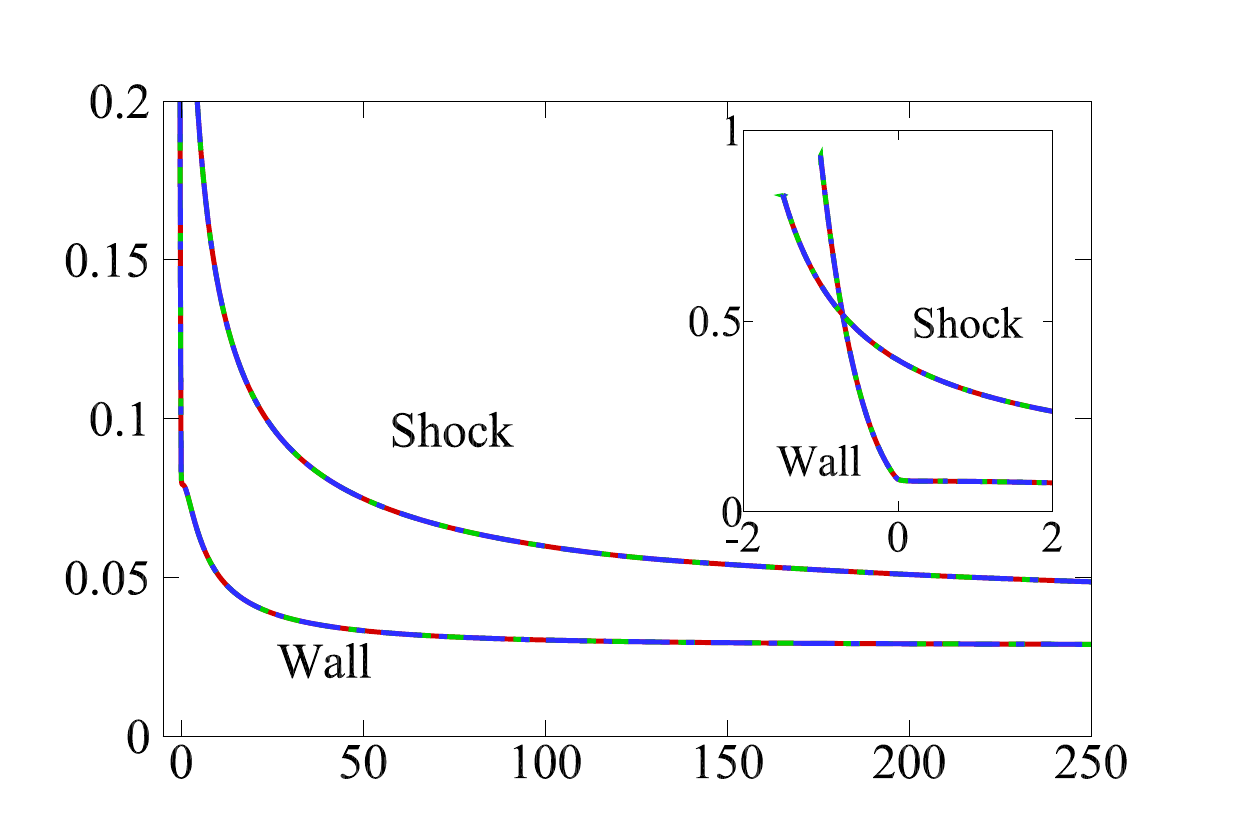}
  \includegraphics[width = 0.49\textwidth]{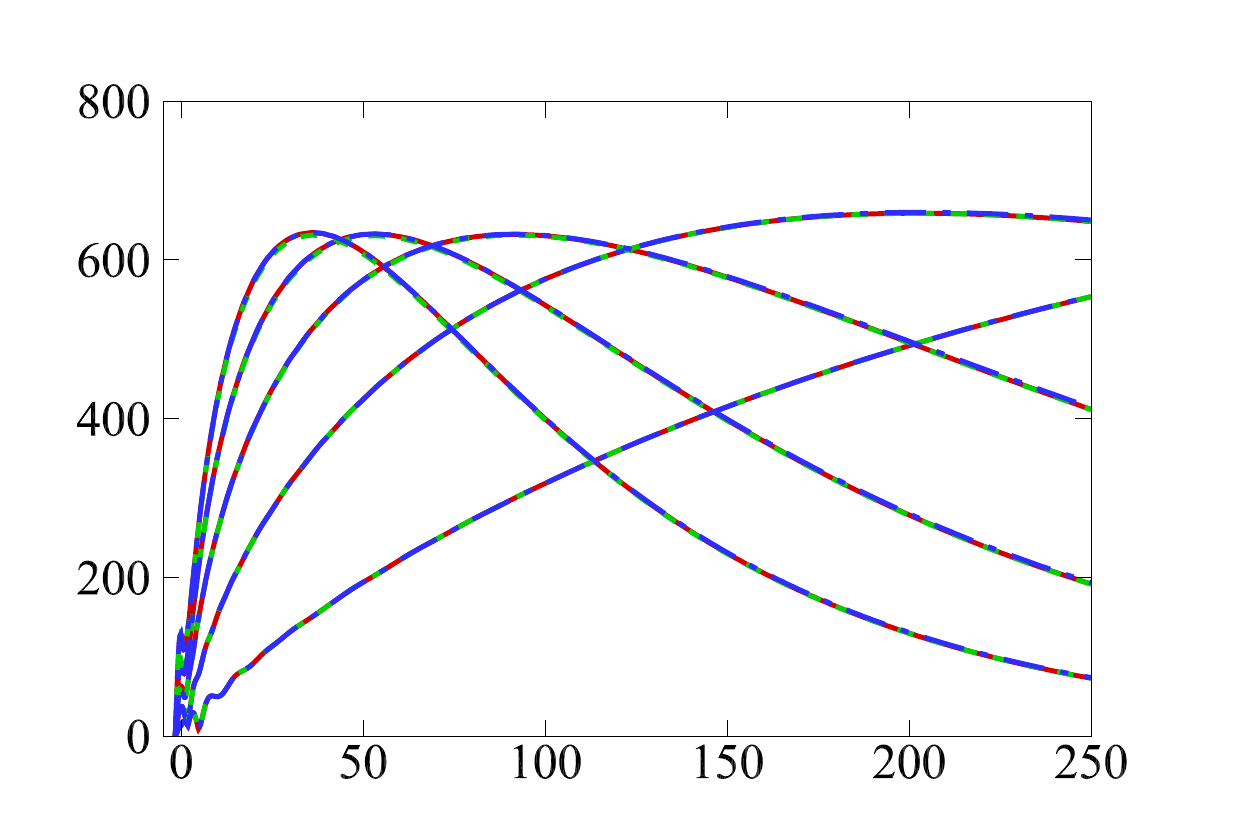}
  \put(-380,120){$(a)$}
  \put(-190,120){$(b)$}
  \put(-380,60) {$p_B$} 
  \put(-190,60) {$A_u$}
  \put(-285,2)  {$x$}
  \put(-95,2)   {$x$}
  \put(-340,83) {\begin{tikzpicture}
    \draw[red,thick]   (0,0) -- (0.5,0);
    \draw[green,thick,dashed] (0,-0.35) -- (0.5,-0.35);
    \draw[blue,thick,dash dot]  (0,-0.7) -- (0.5,-0.7);
    \end{tikzpicture}}
  \put(-320,100){$551~~\times301$}
  \put(-320,90) {$1101\times301$}
  \put(-320,80) {$551~~\times601$}
  \put(-125,34) {$k_3=2$}
  \put(-60,85)  {$k_3=4$}
  \put(-55,60)  {$k_3=6$}
  \put(-55,34)  {$k_3=8$}
  \put(-60,18)  {$k_3=10$}
  \caption{Resolution study for case A. ($a$) Streamwise evolution of the base-flow pressure {$p_B$} at the shock and the wall; ($b$) streamwise evolution of the perturbation amplitude $A_u$  for $\omega=0$, $\vartheta=0^{\circ}$ and various $k_3$. The baseline results are depicted in red, while the results from refining the streamwise and wall-normal grid points are shown in green and blue, respectively. }
  \label{fig:gridcheck}
  \end{center}
\end{figure}
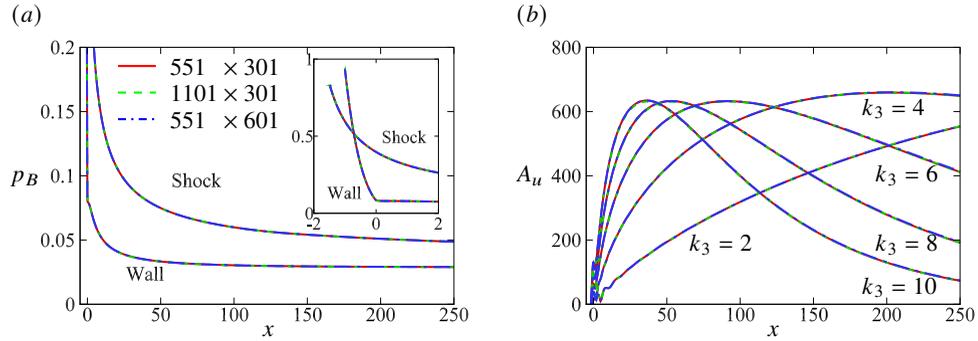
To assess grid convergence, we repeated the base‐flow and SF‐HLNS computations on meshes refined in both the streamwise and wall‐normal directions, starting from our baseline grid of 551$\times$301 points.  Figure \ref{fig:gridcheck}-(a) shows the streamwise variation of the base‐flow pressure $p_B$ at the wall and the shock, and figure \ref{fig:gridcheck}-(b) plots the evolution of perturbation amplitudes for $\omega=0$, $\vartheta=0^\circ$, with $k_3$ varies from 2 to 10.  The virtually perfect collapse of all curves demonstrates that our results are mesh‐independent and confirms the accuracy of our calculations.

\bibliographystyle{jfm}
\bibliography{Bypass_Transition}

\begin{thebibliography}{38}
\expandafter\ifx\csname natexlab\endcsname\relax\def\natexlab#1{#1}\fi
\def\au#1{#1} \def\ed#1{#1} \def\yr#1{#1}\def\at#1{#1}\def\jt#1{\textit{#1}} \def\bt#1{#1}\def\bvol#1{\textbf{#1}} \def\vol#1{#1} \def\pg#1{#1} \def\publ#1{#1}\def\arxiv#1{#1}\def\org#1{#1}\def\st#1{\textit{#1}}

\bibitem[Andersson {\em et~al.\/}(1999)Andersson, Berggren \& Henningson]{andersson1999optimal}
{\sc \au{Andersson, P.}, \au{Berggren, M.} \& \au{Henningson, D.~S.}} \yr{1999}  \at{Optimal disturbances and bypass transition in boundary layers}.  \jt{Phys. Fluids}  \bvol{11}~(1),  \pg{134--150}.

\bibitem[Andersson {\em et~al.\/}(2001)Andersson, Brandt, Bottaro \& Henningson]{andersson2001breakdown}
{\sc \au{Andersson, P.}, \au{Brandt, L.}, \au{Bottaro, A.} \& \au{Henningson, D.~S.}} \yr{2001}  \at{On the breakdown of boundary layer streaks}.  \jt{J. Fluid Mech.}  \bvol{428},  \pg{29--60}.

\bibitem[Borovoy {\em et~al.\/}(2022)Borovoy, Radchenko, Aleksandrov \& Mosharov]{borovoy2022laminar}
{\sc \au{Borovoy, V.~Y}, \au{Radchenko, V.~N}, \au{Aleksandrov, S.~V} \& \au{Mosharov, V.~E}} \yr{2022}  \at{Laminar-turbulent transition reversal on a blunted plate with various leading-edge shapes}.  \jt{AIAA J.}  \bvol{60}~(1),  \pg{497--507}.

\bibitem[Brandt {\em et~al.\/}(2004)Brandt, Schlatter \& Henningson]{brandt2004transition}
{\sc \au{Brandt, L.}, \au{Schlatter, P.} \& \au{Henningson, D.~S.}} \yr{2004}  \at{Transition in boundary layers subject to free-stream turbulence}.  \jt{J. Fluid Mech.}  \bvol{517},  \pg{167--198}.

\bibitem[Casper {\em et~al.\/}(2016)Casper, Beresh, Henfling, Spillers, Pruett \& Schneider]{casper2016hypersonic}
{\sc \au{Casper, K.~M.}, \au{Beresh, S.~J.}, \au{Henfling, J.~F.}, \au{Spillers, R.~W.}, \au{Pruett, B.~OM} \& \au{Schneider, S.~P.}} \yr{2016}  \at{Hypersonic wind-tunnel measurements of boundary-layer transition on a slender cone}.  \jt{AIAA J.}  \bvol{54}~(4),  \pg{1250--1263}.

\bibitem[Dietz \& Hein(1999)]{dietz1999entropy}
{\sc \au{Dietz, G.} \& \au{Hein, S.}} \yr{1999}  \at{Entropy-layer instabilities over a blunted flat plate in supersonic flow}.  \jt{Phys. Fluids}  \bvol{11}~(1),  \pg{7--9}.

\bibitem[Dong {\em et~al.\/}(2020)Dong, Liu \& Wu]{Dong2020receptivity}
{\sc \au{Dong, M.}, \au{Liu, Y.} \& \au{Wu, X.}} \yr{2020}  \at{Receptivity of inviscid modes in supersonic boundary layers due to scattering of freestream sound by wall roughness}.  \jt{J. Fluid Mech.}  \bvol{896},  \pg{A23}.

\bibitem[Duan {\em et~al.\/}(2019)Duan, Choudhari, Chou, Munoz, Radespiel, Schilden, Schr{\"o}der, Marineau, Casper, Chaudhry, Candler, Gray \& Schneider]{duan2019characterization}
{\sc \au{Duan, L.}, \au{Choudhari, M.~M.}, \au{Chou, A.}, \au{Munoz, F.}, \au{Radespiel, R.}, \au{Schilden, T.}, \au{Schr{\"o}der, W.}, \au{Marineau, E.~C.}, \au{Casper, K.~M.}, \au{Chaudhry, R.~S.}, \au{Candler, G.~V.}, \au{Gray, K.~A.} \& \au{Schneider, S.~P.}} \yr{2019}  \at{Characterization of freestream disturbances in conventional hypersonic wind tunnels}.  \jt{J. Spacecr. Rockets}  \bvol{56}~(2),  \pg{357--368}.

\bibitem[Fedorov \& Tumin(2004)]{Fedorov2004evolution}
{\sc \au{Fedorov, A.~V.} \& \au{Tumin, A.}} \yr{2004}  \at{Evolution of disturbances in entropy layer on blunted plate in supersonic flow}.  \jt{AIAA J.}  \bvol{42}~(1),  \pg{89--94}.

\bibitem[Guo {\em et~al.\/}(2025)Guo, Hao \& Wen]{guo2025transition}
{\sc \au{Guo, P.}, \au{Hao, J.} \& \au{Wen, C.~Y.}} \yr{2025}  \at{Transition reversal over a blunt plate at {M}ach 5}.  \jt{J. Fluid Mech.}  \bvol{1005},  \pg{A5}.

\bibitem[Han {\em et~al.\/}(2021)Han, Yuan, Dong \& Fan]{han2021secondary}
{\sc \au{Han, L.}, \au{Yuan, J.}, \au{Dong, M.} \& \au{Fan, Z.}} \yr{2021}  \at{Secondary instability of the spike-bubble structures induced by nonlinear {Rayleigh-Taylor} instability with a diffuse interface}.  \jt{Phys. Rev. E}  \bvol{104}~(3),  \pg{035108}.

\bibitem[Herbert(1988)]{herbert1988secondary}
{\sc \au{Herbert, T.}} \yr{1988}  \at{Secondary instability of boundary layers}.  \jt{Annu. Rev. Fluid Mech.}  \bvol{20}~(1),  \pg{487--526}.

\bibitem[Herbert(1997)]{Herbert1997parabolized}
{\sc \au{Herbert, T.}} \yr{1997}  \at{Parabolized stability equations}.  \jt{Annu. Rev. Fluid Mech.}  \bvol{29}~(1),  \pg{245--283}.

\bibitem[Jewell {\em et~al.\/}(2018)Jewell, Kennedy, Laurence \& Kimmel]{jewell2018transition}
{\sc \au{Jewell, J.~S.}, \au{Kennedy, R.~E.}, \au{Laurence, S.~J.} \& \au{Kimmel, R.~L.}} \yr{2018} {Transition on a variable bluntness 7-degree cone at high Reynolds number}.  \bt{In {\em 2018 AIAA Aerospace Sciences Meeting\/}},  \pg{pp. AIAA Paper 2018--1822}.  \publ{Kissimmee, Florida, U.S.A.: AIAA}.

\bibitem[Jewell \& Kimmel(2017)]{jewell2017boundary}
{\sc \au{Jewell, J.~S.} \& \au{Kimmel, R.~L.}} \yr{2017}  \at{{Boundary-layer stability analysis for Stetson's Mach 6 blunt-cone experiments}}.  \jt{J. Spacecr. Rockets}  \bvol{54}~(1),  \pg{258--265}.

\bibitem[Lei \& Zhong(2012)]{lei2012linear}
{\sc \au{Lei, J.} \& \au{Zhong, X.}} \yr{2012}  \at{Linear stability analysis of nose bluntness effects on hypersonic boundary layer transition}.  \jt{J. Spacecr. Rockets}  \bvol{49}~(1),  \pg{24--37}.

\bibitem[Leib {\em et~al.\/}(1999)Leib, Wundrow \& Goldstein]{Leib1999}
{\sc \au{Leib, S.}, \au{Wundrow, D.} \& \au{Goldstein, M.}} \yr{1999}  \at{Effect of free-stream turbulence and other vortical disturbances on a laminar boundary layer}.  \jt{J. Fluid Mech.}  \bvol{380},  \pg{169--203}.

\bibitem[Liu {\em et~al.\/}(2022)Liu, Schuabb, Duan, Paredes \& Choudhari]{liu2022interaction}
{\sc \au{Liu, Y.}, \au{Schuabb, M.}, \au{Duan, L.}, \au{Paredes, P.} \& \au{Choudhari, M.~M.}} \yr{2022} Interaction of a tunnel-like acoustic disturbance field with a blunt cone boundary layer at {Mach 8}.  \bt{In {\em AIAA Aviation 2022 Forum\/}},  \pg{pp. AIAA Paper 2022--3250}.  \publ{Chicago, IL \& Virtual, U.S.A.: AIAA}.

\bibitem[Lysenko(1990)]{lysenko1990influence}
{\sc \au{Lysenko, V.~I.}} \yr{1990}  \at{Influence of the entropy layer on the stability of a supersonic shock layer and transition of the laminar boundary layer to turbulence}.  \jt{J. Appl. Mech. Tech. Phys.}  \bvol{31}~(6),  \pg{868--873}.

\bibitem[Malik {\em et~al.\/}(1990)Malik, Spall \& Chang]{malik1990effect}
{\sc \au{Malik, M.~R.}, \au{Spall, R.~E.} \& \au{Chang, C.~L.}} \yr{1990} Effect of nose bluntness on boundary layer stability and transition.  \bt{In {\em 28th Aerospace Sciences Meeting\/}},  \pg{pp. AIAA Paper 1990--0112}.  \publ{Reno, Nevada, U.S.A.: AIAA}.

\bibitem[Paredes {\em et~al.\/}(2018)Paredes, Choudhari \& Li]{paredes2018blunt}
{\sc \au{Paredes, P.}, \au{Choudhari, M.~M.} \& \au{Li, F.}} \yr{2018}  \at{Blunt-body paradox and improved application of transient-growth framework}.  \jt{AIAA J.}  \bvol{56}~(7),  \pg{2604--2614}.

\bibitem[Quintanilha {\em et~al.\/}(2022)Quintanilha, Paredes, Hanifi \& Theofilis]{quintanilha2022transient}
{\sc \au{Quintanilha, H.~JR.}, \au{Paredes, P.}, \au{Hanifi, A.} \& \au{Theofilis, V.}} \yr{2022}  \at{Transient growth analysis of hypersonic flow over an elliptic cone}.  \jt{J. Fluid Mech.}  \bvol{935},  \pg{A40}.

\bibitem[Reshotko \& Khan(1980)]{reshotko1980stability}
{\sc \au{Reshotko, E.} \& \au{Khan, M. M.~S.}} \yr{1980}  \at{Stability of the laminar boundary layer on a blunted plate in supersonic flow}.  \bt{In {\em Symposium on laminar-turbulent transition\/}},  \pg{pp. 185--200}.  \publ{Stuttgart, Germany, F.R.: Springer}.

\bibitem[Ricco {\em et~al.\/}(2011)Ricco, Luo \& Wu]{ricco2011evolution}
{\sc \au{Ricco, P.}, \au{Luo, J.} \& \au{Wu, X.}} \yr{2011}  \at{Evolution and instability of unsteady nonlinear streaks generated by free-stream vortical disturbances}.  \jt{J. Fluid Mech.}  \bvol{677},  \pg{1--38}.

\bibitem[Song {\em et~al.\/}(2024)Song, Dong, Zhao, Chu \& Wu]{song2024influence}
{\sc \au{Song, Q.}, \au{Dong, M.}, \au{Zhao, L.}, \au{Chu, X.} \& \au{Wu, N.}} \yr{2024}  \at{Influence of spanwise wall vibration on non-modal perturbations subject to freestream vortical disturbances in hypersonic boundary layers}.  \jt{J. Fluid Mech.}  \bvol{999},  \pg{A57}.

\bibitem[Song {\em et~al.\/}(2023{\natexlab{{\em a\/}}})Song, Zhao \& Dong]{song2023effect}
{\sc \au{Song, Q.}, \au{Zhao, L.} \& \au{Dong, M.}} \yr{2023{\natexlab{{\em a\/}}}}  \at{Effect of porous coatings on the nonlinear evolution of {M}ack modes in hypersonic boundary layers}.  \jt{Phys. Fluids}  \bvol{35}~(5),  \pg{054115}.

\bibitem[Song {\em et~al.\/}(2023{\natexlab{{\em b\/}}})Song, Dong \& Zhao]{songrj2023effect}
{\sc \au{Song, R.}, \au{Dong, M.} \& \au{Zhao, L.}} \yr{2023{\natexlab{{\em b\/}}}}  \at{Effect of cone rotation on the nonlinear evolution of {Mack} modes in supersonic boundary layers}.  \jt{J. Fluid Mech.}  \bvol{971},  \pg{A4}.

\bibitem[Stetson(1967)]{stetson1967shock}
{\sc \au{Stetson, K.}} \yr{1967}  \at{Shock tunnel investigation of boundary-layer transition at {M = 5.5}}.  \jt{AIAA J.}  \bvol{5}~(5),  \pg{899--906}.

\bibitem[Stetson(1983)]{stetson1983nosetip}
{\sc \au{Stetson, K.}} \yr{1983} Nosetip bluntness effects on cone frustum boundary layer transition in hypersonic flow.  \bt{In {\em 16th Fluid and Plasmadynamics Conference\/}},  \pg{pp. AIAA Paper 1983--1763}.  \publ{Danvers, Massachusett, U.S.A.: AIAA}.

\bibitem[Towne {\em et~al.\/}(2022)Towne, Rigas, Kamal, Pickering \& Colonius]{Towne2022}
{\sc \au{Towne, A.}, \au{Rigas, G.}, \au{Kamal, O.}, \au{Pickering, E.} \& \au{Colonius, T.}} \yr{2022}  \at{Efficient global resolvent analysis via the one-way {Navier–Stokes} equations}.  \jt{J. Fluid Mech.}  \bvol{948},  \pg{A9}.

\bibitem[Wagner {\em et~al.\/}(2018)Wagner, Sch{\"u}lein, Petervari, Hannemann, Ali, Cerminara \& Sandham]{wagner2018combined}
{\sc \au{Wagner, A.}, \au{Sch{\"u}lein, E.}, \au{Petervari, R.}, \au{Hannemann, K.}, \au{Ali, S.~RC.}, \au{Cerminara, A.} \& \au{Sandham, N.~D.}} \yr{2018}  \at{Combined free-stream disturbance measurements and receptivity studies in hypersonic wind tunnels by means of a slender wedge probe and direct numerical simulation}.  \jt{J. Fluid Mech.}  \bvol{842},  \pg{495--531}.

\bibitem[Wan {\em et~al.\/}(2023)Wan, Chen, Tu, Xiang, Yuan \& Duan]{wan2023effects}
{\sc \au{Wan, B.}, \au{Chen, J.}, \au{Tu, G.}, \au{Xiang, X.}, \au{Yuan, X.} \& \au{Duan, M.}} \yr{2023}  \at{Effects of nose bluntness on entropy-layer stabilities over cones and wedges}.  \jt{Acta Mech. Sin.}  \bvol{39}~(1),  \pg{122176}.

\bibitem[Wu(2023)]{Wu2023new}
{\sc \au{Wu, X.}} \yr{2023}  \at{New insights into turbulent spots}.  \jt{Annu. Rev. Fluid Mech.}  \bvol{55}~(1),  \pg{45--75}.

\bibitem[Xu {\em et~al.\/}(2024)Xu, Ricco \& Marensi]{xu2024excitation}
{\sc \au{Xu, D.}, \au{Ricco, P.} \& \au{Marensi, E.}} \yr{2024}  \at{Excitation and stability of nonlinear compressible g{\"o}rtler vortices and streaks induced by free-stream vortical disturbances}.  \jt{J. Fluid Mech.}  \bvol{1000},  \pg{A93}.

\bibitem[Zhang {\em et~al.\/}(2018)Zhang, Dong \& Zhang]{Zhang2018}
{\sc \au{Zhang, A.}, \au{Dong, M.} \& \au{Zhang, Y.}} \yr{2018}  \at{Receptivity of secondary instability modes in streaky boundary layers}.  \jt{Phys. Fluids}  \bvol{30}~(11),  \pg{114102}.

\bibitem[Zhao \& Dong(2025)]{zhao2025excitation}
{\sc \au{Zhao, L.} \& \au{Dong, M.}} \yr{2025}  \at{Excitation of non-modal perturbations in hypersonic boundary layers by free stream forcing: shock-fitting harmonic linearised {Navier--Stokes} approach}.  \jt{J. Fluid Mech.}  \bvol{1013},  \pg{A44}.

\bibitem[Zhao {\em et~al.\/}(2016)Zhao, Zhang, Liu \& Luo]{zhao2016improved}
{\sc \au{Zhao, L.}, \au{Zhang, C.}, \au{Liu, J.} \& \au{Luo, J.}} \yr{2016}  \at{Improved algorithm for solving nonlinear parabolized stability equations}.  \jt{Chin. Phys. B}  \bvol{25}~(8),  \pg{084701}.

\bibitem[Zhong \& Wang(2012)]{Zhong2012direct}
{\sc \au{Zhong, X.} \& \au{Wang, X.}} \yr{2012}  \at{Direct numerical simulation on the receptivity, instability and transition of hypersonic boundary layers}.  \jt{Annu. Rev. Fluid Mech.}  \bvol{44}~(1),  \pg{527--561}.

\end{thebibliography}
\end{document}